\documentclass[aps,prd,twocolumn,showpacs,nofootinbib,floatfix,superscriptaddress]{revtex4-1}

\usepackage{amsmath,bm}
\usepackage{amssymb}
\usepackage{mathrsfs}

\usepackage{color}
\usepackage{dcolumn}
\usepackage{graphicx}
\usepackage{multirow}
\usepackage{rotating}
\usepackage{afterpage}
\usepackage{enumerate}
\usepackage{xspace}
\usepackage{url}

\usepackage{fancyhdr}
\usepackage{float}

\newcommand{\herwig}    {\textsc{herwig}\xspace}
\newcommand{\pythia}    {\textsc{pythia}\xspace}
\newcommand{\alpgen}    {\textsc{alpgen}\xspace}
\newcommand{\tmva}    {\textsc{tmva}\xspace}

\newcommand{\geant}     {\textsc{geant}\xspace}
\newcommand{\mcatnlo}   {\textsc{mc@nlo}\xspace}
\newcommand{\resbos}   {\textsc{resbos}\xspace}

\newcommand{\alppyt}    {{\sc alpgen+pythia}\xspace}
\newcommand{\alpher}    {{\sc alpgen+herwig}\xspace}
\newcommand{\mcherwig}  {{\sc mc@nlo+herwig}\xspace}
\newcommand{\comphep}   {{\sc comphep}\xspace}
\newcommand{\mcfm}   {{\sc mcfm}\xspace}

\newcommand{\toppp}   {{\sc top++}\xspace}
\newcommand{\collie}   {{\sc collie}\xspace}

\newcommand{\met}       {\mbox{$\not\!\!E_T$}\xspace}
\newcommand{\metNo}       {\mbox{$\not\!\!E_T$}}

\newcommand{\wplus}     {\ensuremath{W+}jets\xspace}
\newcommand{\wplusTw}{\ensuremath{W+\ge 2} jets\xspace}
\newcommand{\zplus}     {\ensuremath{Z/\gamma^{*}+}jets\xspace}
\newcommand{\ptmiss}    {\ensuremath{{p\kern-0.5em\slash}_{T}}\xspace}
\newcommand{\muplus}    {\ensuremath{\mu +}jets\xspace}
\newcommand{\eplus}     {$e +$jets\xspace}
\newcommand{\lplus}     {$\ell +$jets\xspace}
\newcommand{\lplustw}     {$\ell +2$\,jets\xspace}
\newcommand{\lplusth}     {$\ell +3$\,jets\xspace}

\newcommand{\lplusgefo}     {$\ell + \ge 4$\,jets\xspace}

\newcommand{\ppbar}{\ensuremath{p\bar{p}}\xspace}
\newcommand{\qqbar}{\ensuremath{q\bar{q}}\xspace}
\newcommand{\ttbar}{\ensuremath{t\bar{t}}\xspace}
\newcommand{\mTT}{\ensuremath{m(t\bar{t})}\xspace}

\newcommand{\mm}{\mathrm}

\newcommand{\ljets}{$\ell +$jets\xspace}
\newcommand{\dilep}{$\ell\ell$\xspace}
\newcommand{\dzero}     {D0\xspace}

\newcommand{\mmax}{\ensuremath{j_{\mm{b-ID}}^{\mm{max}}}\xspace}

\newcommand{\wjets}{\ensuremath{W}+\rm jets\xspace}

\newcommand{\wlp}{\ensuremath{Wlp} + jets\xspace}
\newcommand{\zlp}{\ensuremath{Z/\gamma^{*}lp}+\rm jets\xspace}
\newcommand{\zlllp}{\ensuremath{Z/\gamma^{*} \rightarrow \ell\ell}+\rm jets\xspace}
\newcommand{\whf}       {\ensuremath{(Wc\bar{c}+Wb\bar{b}) +}jets\xspace}
\newcommand{\zhf}       {\ensuremath{(Z/\gamma^{*}c\bar{c}+Z/\gamma^{*}b\bar{b}) +}jets\xspace}
\newcommand{\xsttbar}   {\ensuremath{\sigma(t\bar{t})}\xspace}

\newcommand\T{\rule{0pt}{2.6ex}}       
\newcommand{\mvaM}{$b$-ID MVA\xspace}
\newcommand{\topoM}{combined MVA\xspace}
\newcommand{\mvaME}{$b$-ID MVA}
\newcommand{\topoME}{combined MVA}
\newcommand{\topoMnoB}{topological MVA\xspace}
\newcommand{\bidM}{$b$-ID MVA\xspace}
\newcommand{\mMC}{$m_t^{\mathrm{MC}}$\xspace}

\newcommand{\xsecLJ}{$ \sigma_{\ttbar} = 7.33 \pm 0.14\,(\mm{stat.})\,^{+0.71}_{-0.61}\,(\mm{syst.})$\xspace}
\newcommand{\xsecDL}{$ \sigma_{\ttbar} = 7.58 \pm 0.35\,(\mm{stat.})\,^{+0.69}_{-0.58}\,(\mm{syst.})$\xspace}
\newcommand{\xsecComb}{$ \sigma_{\ttbar} = 7.26 \pm 0.13\,(\mm{stat.})\,^{+0.57}_{-0.50}\,(\mm{syst.})$\xspace}

\newcolumntype{d}[1]{D{.}{\,\pm\,}{#1}}
\newcolumntype{e}[1]{D{.}{}{#1}}
\newcolumntype{z}[1]{D{,}{\,\pm\,}{#1}}

\begin{document}

\widetext

\hspace{5.2in} \mbox{FERMILAB-PUB-16-180-E}

\title{Measurement of the inclusive {\boldmath \ttbar} production cross section in \\ {\boldmath \ppbar} collisions at {\boldmath $\sqrt{s}=1.96$} TeV and determination of the top quark pole mass}

\affiliation{LAFEX, Centro Brasileiro de Pesquisas F\'{i}sicas, Rio de Janeiro, Brazil}
\affiliation{Universidade do Estado do Rio de Janeiro, Rio de Janeiro, Brazil}
\affiliation{Universidade Federal do ABC, Santo Andr\'e, Brazil}
\affiliation{University of Science and Technology of China, Hefei, People's Republic of China}
\affiliation{Universidad de los Andes, Bogot\'a, Colombia}
\affiliation{Charles University, Faculty of Mathematics and Physics, Center for Particle Physics, Prague, Czech Republic}
\affiliation{Czech Technical University in Prague, Prague, Czech Republic}
\affiliation{Institute of Physics, Academy of Sciences of the Czech Republic, Prague, Czech Republic}
\affiliation{Universidad San Francisco de Quito, Quito, Ecuador}
\affiliation{LPC, Universit\'e Blaise Pascal, CNRS/IN2P3, Clermont, France}
\affiliation{LPSC, Universit\'e Joseph Fourier Grenoble 1, CNRS/IN2P3, Institut National Polytechnique de Grenoble, Grenoble, France}
\affiliation{CPPM, Aix-Marseille Universit\'e, CNRS/IN2P3, Marseille, France}
\affiliation{LAL, Universit\'e Paris-Sud, CNRS/IN2P3, Orsay, France}
\affiliation{LPNHE, Universit\'es Paris VI and VII, CNRS/IN2P3, Paris, France}
\affiliation{CEA, Irfu, SPP, Saclay, France}
\affiliation{IPHC, Universit\'e de Strasbourg, CNRS/IN2P3, Strasbourg, France}
\affiliation{IPNL, Universit\'e Lyon 1, CNRS/IN2P3, Villeurbanne, France and Universit\'e de Lyon, Lyon, France}
\affiliation{III. Physikalisches Institut A, RWTH Aachen University, Aachen, Germany}
\affiliation{Physikalisches Institut, Universit\"at Freiburg, Freiburg, Germany}
\affiliation{II. Physikalisches Institut, Georg-August-Universit\"at G\"ottingen, G\"ottingen, Germany}
\affiliation{Institut f\"ur Physik, Universit\"at Mainz, Mainz, Germany}
\affiliation{Ludwig-Maximilians-Universit\"at M\"unchen, M\"unchen, Germany}
\affiliation{Panjab University, Chandigarh, India}
\affiliation{Delhi University, Delhi, India}
\affiliation{Tata Institute of Fundamental Research, Mumbai, India}
\affiliation{University College Dublin, Dublin, Ireland}
\affiliation{Korea Detector Laboratory, Korea University, Seoul, Korea}
\affiliation{CINVESTAV, Mexico City, Mexico}
\affiliation{Nikhef, Science Park, Amsterdam, the Netherlands}
\affiliation{Radboud University Nijmegen, Nijmegen, the Netherlands}
\affiliation{Joint Institute for Nuclear Research, Dubna, Russia}
\affiliation{Institute for Theoretical and Experimental Physics, Moscow, Russia}
\affiliation{Moscow State University, Moscow, Russia}
\affiliation{Institute for High Energy Physics, Protvino, Russia}
\affiliation{Petersburg Nuclear Physics Institute, St. Petersburg, Russia}
\affiliation{Instituci\'{o} Catalana de Recerca i Estudis Avan\c{c}ats (ICREA) and Institut de F\'{i}sica d'Altes Energies (IFAE), Barcelona, Spain}
\affiliation{Uppsala University, Uppsala, Sweden}
\affiliation{Taras Shevchenko National University of Kyiv, Kiev, Ukraine}
\affiliation{Lancaster University, Lancaster LA1 4YB, United Kingdom}
\affiliation{Imperial College London, London SW7 2AZ, United Kingdom}
\affiliation{The University of Manchester, Manchester M13 9PL, United Kingdom}
\affiliation{University of Arizona, Tucson, Arizona 85721, USA}
\affiliation{University of California Riverside, Riverside, California 92521, USA}
\affiliation{Florida State University, Tallahassee, Florida 32306, USA}
\affiliation{Fermi National Accelerator Laboratory, Batavia, Illinois 60510, USA}
\affiliation{University of Illinois at Chicago, Chicago, Illinois 60607, USA}
\affiliation{Northern Illinois University, DeKalb, Illinois 60115, USA}
\affiliation{Northwestern University, Evanston, Illinois 60208, USA}
\affiliation{Indiana University, Bloomington, Indiana 47405, USA}
\affiliation{Purdue University Calumet, Hammond, Indiana 46323, USA}
\affiliation{University of Notre Dame, Notre Dame, Indiana 46556, USA}
\affiliation{Iowa State University, Ames, Iowa 50011, USA}
\affiliation{University of Kansas, Lawrence, Kansas 66045, USA}
\affiliation{Louisiana Tech University, Ruston, Louisiana 71272, USA}
\affiliation{Northeastern University, Boston, Massachusetts 02115, USA}
\affiliation{University of Michigan, Ann Arbor, Michigan 48109, USA}
\affiliation{Michigan State University, East Lansing, Michigan 48824, USA}
\affiliation{University of Mississippi, University, Mississippi 38677, USA}
\affiliation{University of Nebraska, Lincoln, Nebraska 68588, USA}
\affiliation{Rutgers University, Piscataway, New Jersey 08855, USA}
\affiliation{Princeton University, Princeton, New Jersey 08544, USA}
\affiliation{State University of New York, Buffalo, New York 14260, USA}
\affiliation{University of Rochester, Rochester, New York 14627, USA}
\affiliation{State University of New York, Stony Brook, New York 11794, USA}
\affiliation{Brookhaven National Laboratory, Upton, New York 11973, USA}
\affiliation{Langston University, Langston, Oklahoma 73050, USA}
\affiliation{University of Oklahoma, Norman, Oklahoma 73019, USA}
\affiliation{Oklahoma State University, Stillwater, Oklahoma 74078, USA}
\affiliation{Brown University, Providence, Rhode Island 02912, USA}
\affiliation{University of Texas, Arlington, Texas 76019, USA}
\affiliation{Southern Methodist University, Dallas, Texas 75275, USA}
\affiliation{Rice University, Houston, Texas 77005, USA}
\affiliation{University of Virginia, Charlottesville, Virginia 22904, USA}
\affiliation{University of Washington, Seattle, Washington 98195, USA}
\author{V.M.~Abazov} \affiliation{Joint Institute for Nuclear Research, Dubna, Russia}
\author{B.~Abbott} \affiliation{University of Oklahoma, Norman, Oklahoma 73019, USA}
\author{B.S.~Acharya} \affiliation{Tata Institute of Fundamental Research, Mumbai, India}
\author{M.~Adams} \affiliation{University of Illinois at Chicago, Chicago, Illinois 60607, USA}
\author{T.~Adams} \affiliation{Florida State University, Tallahassee, Florida 32306, USA}
\author{J.P.~Agnew} \affiliation{The University of Manchester, Manchester M13 9PL, United Kingdom}
\author{G.D.~Alexeev} \affiliation{Joint Institute for Nuclear Research, Dubna, Russia}
\author{G.~Alkhazov} \affiliation{Petersburg Nuclear Physics Institute, St. Petersburg, Russia}
\author{A.~Alton$^{a}$} \affiliation{University of Michigan, Ann Arbor, Michigan 48109, USA}
\author{A.~Askew} \affiliation{Florida State University, Tallahassee, Florida 32306, USA}
% ...
%
%
\collaboration{The D0 Collaboration\footnote{With visitors from
$^{a}$Augustana College, Sioux Falls, South Dakota, USA,
$^{b}$The University of Liverpool, Liverpool, United Kingdom,
$^{c}$DESY, Hamburg, Germany,
$^{d}$CONACyT, Mexico City, Mexico,
$^{e}$SLAC, Menlo Park, California, USA,
$^{f}$University College London, London, United Kingdom,
$^{g}$Centro de Investigacion en Computacion - IPN, Mexico City, Mexico,
$^{h}$Universidade Estadual Paulista, S\~ao Paulo, Brazil,
$^{i}$Karlsruher Institut f\"ur Technologie (KIT) - Steinbuch Centre for Computing (SCC),
D-76128 Karlsruhe, Germany,
$^{j}$Office of Science, U.S. Department of Energy, Washington, D.C. 20585, USA,
$^{k}$American Association for the Advancement of Science, Washington, D.C. 20005, USA,
$^{l}$Kiev Institute for Nuclear Research, Kiev, Ukraine,
$^{m}$University of Maryland, College Park, Maryland 20742, USA
$^{n}$European Orgnaization for Nuclear Research (CERN), Geneva, Switzerland
and
$^{o}$Purdue University, West Lafayette, Indiana 47907, USA.
}} \noaffiliation
\vskip 0.25cm

\vskip 0.25cm

\date{May 20, 2016}

\begin{abstract}

The inclusive cross section of top quark-antiquark pairs produced in \ppbar collisions at \mbox{$\sqrt{s}=1.96$ TeV} is measured in the lepton$+$jets and dilepton decay channels. The data sample corresponds to 9.7 fb${}^{-1}$ of integrated luminosity recorded with the \dzero detector during Run II of the Fermilab Tevatron Collider. Employing multivariate analysis techniques we measure the cross section in the two decay channels and we perform a combined cross section measurement. For a top quark mass of 172.5 GeV, we measure a combined inclusive top quark-antiquark pair production cross section of  \mbox{\xsecComb$\mm{pb}$} which is consistent with standard model predictions. We also perform a likelihood fit to the measured and predicted top quark mass dependence of the inclusive cross section, which yields a measurement of the pole mass of the top quark. The extracted value is $m_t = 172.8 \pm 1.1\,(\mm{theo.})\,^{+3.3}_{-3.1}\,(\mm{exp.})$ GeV.
\end{abstract}

\maketitle

%
%
%
%
\section{Introduction} 
\label{toc:Intro}
The top quark, discovered by the CDF and \dzero experiments in 1995 \cite{top_disc1,top_disc2}, is the heaviest of all elementary particles in the standard model (SM). The production of top quark-antiquark pairs (\ttbar) at the Fermilab Tevatron Collider is dominated by the quark-antiquark (\qqbar) annihilation process. The measurement of the inclusive \ttbar production cross section provides a direct test of quantum chromodynamics (QCD), the theory of the strong interaction. Inclusive \ttbar production cross sections have been previously measured at the Tevatron \cite{Publ54_xsec,cdf_incl} and the LHC \cite{cms_incl,atlas_incl,lhcb_incl}. In this article we present a measurement using a refined analysis technique, which is optimized to be less dependent on the top quark mass. Compared to the previous \dzero result \cite{Publ54_xsec} we employ nearly a factor of 2 more data, which allows for higher precision tests of perturbative QCD (pQCD).

The mass of the top quark has been directly measured with a precision of less than 0.43\% in a single measurement \cite{d0mass}. The Tevatron combination currently yields a top quark mass of $174.34 \pm 0.64$ GeV \cite{TevatronMassCombo}. The direct measurements employed for the Tevatron combination are based on analysis techniques which use \ttbar events provided by Monte Carlo (MC) simulation for different assumed values of the top quark mass $m_t$. Applying these techniques to data yields a mass quantity corresponding to the top quark mass scheme implemented in the MC and we refer to that quantity as the ``MC mass" or $m_t^{\mathrm{MC}}$. Theoretical arguments suggest that \mMC is within about 1 GeV of the well-defined top quark pole mass \cite{poleToMCmassDiff}. An alternative measurement approach employs the inclusive \ttbar cross section to extract the mass of the top quark $m_t$. We assume the SM cross section dependence on $m_t$ as provided by the highest order of pQCD available at this time, namely a next-to-next-to-leading-order (NNLO) calculation. Comparing the dependence of the inclusive cross section on $m_t$, as calculated in pQCD, with the experimental measurement, accounting for the variation of the acceptance with $m_t$, yields a theoretically well-defined top quark pole mass. We employ this approach to extract a top quark pole mass with reduced experimental uncertainties due to our optimized analysis technique.

Events are selected in the lepton+jets (\ljets) and dilepton (\dilep) top quark decay channels, where the lepton ($\ell$) refers to either an electron or a muon. These channels correspond to $t\bar{t} \rightarrow W^{+}b W^{-}\bar{b}$ decays, where in the \ljets channel one of the two $W$ bosons decays leptonically ($W \rightarrow \ell \nu$), while the other $W$ boson decays hadronically ($W \rightarrow q\bar{q}'$). In the dilepton decay channel both $W$ bosons decay leptonically. Both decay channels include small contributions from electrons and muons stemming from the decay of $\tau$ leptons ($t \rightarrow Wb \rightarrow \tau \nu_{\tau} b \rightarrow \ell \nu_{\ell} \nu_{\tau} b$). 

\section{Measurement strategy and outline} 
\label{toc:strategy} 
This measurement uses various multivariate analysis (MVA) techniques \cite{mva,mva2,mva3}, as implemented in \tmva \cite{tmva}, to measure the inclusive cross section in the \ljets and \dilep decay channels. For the dilepton decay channel we use a discriminant solely based on the output distribution of the MVA employed to identify jets that are likely to originate from $b$ quarks ($b$-tagged jets) \cite{bid_nim}. This method is superior to a simple cut-and-count analysis since each \ttbar event contains two $b$-quarks from the decays of top quarks. We refer to this method in the following as ``\mvaME." We construct a combined discriminant for events in the \ljets decay channel to make the best use of the distinct topological signature of top quark events along with $b$-tagging information. We refer to this method in the following as ``\topoME." We use the entire distribution of the MVA discriminants in each decay channel to build MC templates. We use nuisance parameters to profile systematic uncertainties and to constrain their impact using data. For a combined inclusive \ttbar cross section measurement we simultaneously employ the discriminant distribution of the \mvaM in the dilepton decay channel and the \topoM in the \ljets decay channel in a nuisance-parameter-based profiling method. This combination benefits from the cross-calibration of the two different decay channels, leading to reduced systematic uncertainties.

This article is organized as follows. In Sec.\ \ref{toc:detector} we provide a brief review of the relevant aspects of the \dzero detector and object reconstruction. A brief description of our event simulation approach, the QCD predictions employed, and a discussion of the assumptions for the modeling of the signal and background contributions follows in Sec.\ \ref{toc:generators}. The selection requirements for \ttbar events in the \ljets and \dilep decay channels are discussed in \mbox{Sec.\ \ref{toc:data_samples}}. The determination of the sample composition in the two decay channels, the resulting event yields, and distributions of the data compared to MC are discussed in Sec.\ \ref{toc:sampleComp}. The details of the MVA techniques employed in this measurement are described in Sec.\ \ref{toc:methods_intro}. The methodology of the inclusive \ttbar production cross section measurement is described in Sec.\ \ref{toc:fitting_collie}, and the systematic uncertainties relevant for this measurement are discussed in Sec.\ \ref{toc:xsec_sys}. The results of the cross section measurement are presented in Sec.\ \ref{toc:results}, followed by the extraction of the top quark pole mass given in Sec.\ \ref{toc:tmass}, and we conclude in Sec.\ \ref{toc:conclusion}.

%
%
%
%
\section{The \dzero Detector and object reconstruction}
\label{toc:detector}
The \dzero detector \cite{d0detector} consists of several subdetectors designed for identification and reconstruction of the products of \ppbar collisions. A silicon microstrip tracker (SMT) \cite{smt,smt_l0} and central fiber tracker surround the interaction region for pseudorapidities\footnote{The pseudorapidity $\eta=-\ln\left[\tan(\theta/2)\right]$ is measured relative to the center of the detector, and $\theta$ is the polar angle with respect to the proton beam direction. The azimuthal angle $\phi$ is orthogonal to $\theta$. The $z$ axis is pointing along the proton beam direction.} $|\eta| < 3$ and $|\eta| < 2.5$, respectively. These elements of the central tracking system are located within a superconducting solenoidal magnet generating a 1.9 T field, providing measurements for reconstructing event vertices and trajectories of charged particles. The SMT allows for a precision of $40~\mm{\mu m}$ or better for the reconstructed primary \ppbar interaction vertex (PV) in the plane transverse to the beam direction. The impact parameter of typical charged-particle trajectories relative to the PV is determined with a precision between 20 and 50 $\mu$m depending on the number of SMT hits and particle momentum. The impact parameter and its measurement uncertainty are key components of the lifetime-based identification of jets containing $b$ quarks. Particle energies are measured using a liquid argon sampling calorimeter that is segmented into a central calorimeter covering $|\eta| < 1.1$, and two end calorimeters extending the coverage to $|\eta| = 4.2$. Outside of the calorimetry, trajectories of muons are measured using three layers of tracking detectors and scintillation trigger counters, and an iron toroidal magnet generating a 1.8 T field between the first two layers \cite{muonPaper}. Plastic scintillator arrays are located in front of the end calorimeter cryostats to measure the luminosity \cite{lumi_nim,lumiTM}. The trigger and data acquisition systems are designed to accommodate the high luminosities provided by the Tevatron \cite{triggerSystem}.

\subsection{Object reconstruction}
\label{toc:objectReco}
The object reconstruction is based on events identified by the D0 trigger system in which we require at least one lepton or at least one lepton and a jet. Since electrons mostly deposit energy in the electromagnetic (EM) calorimeter, the reconstruction and identification (ID) of electrons \cite{elec_nim} is based on clusters in the EM calorimeter with an associated track. Such a track, as reconstructed by the central tracking detector, is required to have a minimum transverse momentum, $p_T$, of 5 GeV that points to the EM cluster within a window of $\Delta \eta \times \Delta \phi = 0.05 \times 0.05$. We define an angular separation $\Delta R = \sqrt{(\Delta\eta)^2 + (\Delta\phi)^2}$ based on the distance $R = \sqrt{\eta^2+\phi^2}$ in the $\eta$-$\phi$ plane. Electron candidates are required to be isolated by only accepting events with $\Delta R(e,\mm{jet}) > 0.5$ (the definition and reconstruction of a jet is discussed below). Further selection requirements on these electron candidates are applied by means of a multivariate analysis of the calorimeter shower profiles and tracking information. MC efficiencies are adjusted to match data efficiencies measured in electron enriched data samples.

The identification of muons \cite{muon_nim} begins with a candidate formed using information from the muon system. Such a candidate is required to have a track, as reconstructed by the central tracking devices, associated with it. This association employs a $\chi^2$ measure to match muon tracks provided by the muon detector with a track from the central tracking detector, taking into account effects from multiple scattering and energy loss, as well as the inhomogeneous magnetic field. Isolation criteria are applied based on the information from the hadronic and electromagnetic calorimeters and the central tracking devices. MC efficiencies are adjusted to match data efficiencies measured in muon enriched data samples.

Jets are reconstructed from energy depositions in the calorimeter using a midpoint cone algorithm \cite{runIICone} employing a cone size of 0.5. Jets containing a muon within an angular separation of $\Delta R(\mu,\mm{jet}) < 0.5$ are considered to originate from a semileptonic $b$-quark decay and are corrected for the momentum carried away by the muon and the neutrino. For this correction, it is assumed that the neutrino carries the same momentum as the muon.

The jet energy scale (JES) \cite{jescorrection} corrects the measured energy of the jet to the energy of its constituent particles. The JES is derived using a quark-jet-dominated \mbox{$\gamma$ + jet} sample, and corrects data and MC for the difference in detector responses between jets and electromagnetic showers. An additional correction based on the single-particle response accounts for the different characteristics of quark and gluon jets. This correction implements a calibration of the simulated response to single particles inside a jet using data \cite{jescorrection}. Jets in MC simulations have their transverse momenta smeared so that the simulated resolution matches that observed in data. Calibrations of the jet reconstruction and identification efficiency in MC simulations are determined using \zplus data events. Jets are required to contain at least two tracks (see Sec.~\ref{toc:data_samples}), and in MC simulations the corresponding efficiency is adjusted to match that derived in dijet data.

The presence of a neutrino in the final state of the top quark decay can be detected only from the energy imbalance in the transverse plane, denoted by \met. This is reconstructed from the vector sum of the transverse energies of all calorimeter cells above a certain threshold. The vector opposite to this total visible momentum vector is denoted the raw missing energy vector. The fully corrected \met is obtained after correcting for the effects of JES, muon momenta, and muon minimally ionizing deposition in the calorimeter.

%
%
%
%
\section{Monte Carlo Simulations and QCD predictions}
\label{toc:generators}
We use MC simulations to simulate physics processes, to model the reconstruction of the observables, and to estimate systematic uncertainties associated with the measurements. Different MC event generators are used to implement hard scattering processes based on leading-order (LO) and next-to-leading-order (NLO) QCD calculations, and are complemented with parton shower evolution programs. To simulate detector effects, generated events are passed through a detailed simulation of the \dzero detector based on \geant~\cite{geant}. To account for effects from detector noise and additional overlapping \ppbar interactions, events are randomly recorded in \ppbar collisions and overlaid on the fully simulated MC events with the same instantaneous luminosity distribution as for data.

The \ttbar samples are generated with \mcatnlo version 3.4 \cite{mcatnlo} or with \alpgen version 2.11 \cite{alpgen}, which both  produce only on-shell top quarks.  For events generated with \mcatnlo, the parton showering is performed with \herwig version 6.510 \cite{herwig}. Events generated with \alpgen employ parton showering as implemented by \pythia version 6.409 \cite{pythia} or \herwig. We use the \alppyt signal sample as our default to measure the \ttbar cross section and the alternative \mcherwig, or \alpher, signal samples to estimate systematic uncertainties related to effects of NLO corrections or parton showering (see Sec.\ \ref{toc:xsec_sys}), respectively. Single top quark production $(q\bar{q}' \rightarrow t\bar{b}, q'g \rightarrow tq\bar{b})$ is modeled using \comphep \cite{comphep,comphepMan}. For events generated with \comphep, parton showering is implemented by \pythia.  The choice of the parton distribution functions (PDFs) made in generating MC events is CTEQ6L1 \cite{cteq6l}, with the exception of \mcatnlo and \comphep (for the $t$-channel single top quark production), where CTEQ6M \cite{cteq6m} PDFs are used. For all the MC simulations involving the generation of top quarks, a top quark mass of $m_t = 172.5$ GeV is used. The difference with the current Tevatron top quark mass combination of 174.34 GeV \cite{TevatronMassCombo} has negligible impact on the analysis. For the \ttbar $\to $ \lplus (\dilep) decay channel the branching fraction $B$ of $0.342 \pm 0.004$ ($0.04 \pm 0.001$) \cite{pdg} is used. These values include electrons and muons originating from the leptonic decay of $\tau$ leptons ($\tau \rightarrow \ell \nu_{\ell} \nu_{\tau}$).

Several QCD predictions for inclusive \ttbar cross sections have been calculated at higher orders than those included in the MC generators: approximate NNLO \cite{mochUwer}, fully resummed NNLO \cite{nnloInclXsec}, and an approximate next-to-NNLO \cite{kidonakis}. The scale used to calculate the inclusive \ttbar cross sections is set to $m_t$. For normalization of our MC events, we employ the approximate NNLO QCD calculation (using $m_t = 172.5$ GeV and the CTEQ6M PDF), which yields $7.48^{+0.48}_{-0.67}\thinspace(\mathrm{scale} + \mathrm{pdf})~\mathrm{pb}$. The result of this approximate NNLO calculation is close to the fully resummed NNLO QCD calculation (using $m_t = 172.5$ GeV), which finds $\sigma_{\mathrm{tot}}^{\mathrm{res}} = 7.35 ^{+0.23}_{-0.27}\thinspace(\mathrm{scale} + \mathrm{pdf})$ pb. The result of an approximate next-to-NNLO order calculation for \mbox{$m_t = 173$ GeV} finds $\sigma_{\mathrm{tot}}^{\mathrm{res}} = 7.37 \pm 0.39\thinspace(\mathrm{scale} + \mathrm{pdf})$ pb, very close to the fully resummed NNLO calculation. Both use the MSTW2008 NNLO PDF \cite{mstw2008nnlo}.

We use the fully resummed NNLO QCD calculation as implemented in \toppp \cite{toppp_prg} to derive the theoretical \ttbar cross section dependence as a function of the top quark mass (see Sec.\ \ref{toc:tmass}). The theoretical calculations use $\sqrt{s}=1.96$ TeV as input parameter which is known at the Tevatron to a precision of 0.1\%. This beam energy uncertainty yields a negligible  0.3\% effect on the fully resummed NNLO \ttbar cross section value.

\subsection{Modeling of background contributions in the $\bm{\ell+}$jets decay channel}
\label{toc:backgroundModel_ljets}
The main background to \ttbar production in the \ljets decay channel is the production of \wplus, including jets originating from heavy quarks. These events are generated with \alpgen interfaced to \pythia for showering and hadronization. The \wplus final state can be split into four subsamples according to parton flavor: $Wb\bar{b}+\mm{jets}$, $Wc\bar{c}+\mm{jets}$, $Wc+\mm{jets}$, and $W$ light partons $+$ jets (\wlp), where light refers to gluons, $u$, $d$ or $s$ quarks. The additional ``jets'' in these \wplus final states originate dominantly from gluon radiation. The \wplus contribution dominates especially at the lower jet multiplicities. The LO \alpgen cross sections are corrected for NLO effects as provided by \mcfm \cite{mcfm}: the $W+\mm{jets}$ cross section is multiplied by 1.30, and the cross sections of $W+$ heavy flavor (WHF) processes are additionally multiplied by a scale factor $s^{\mm{WHF}}$ of 1.47 for $Wb\bar{b}+\mm{jets}$ and $Wc\bar{c}+\mm{jets}$ and 1.27 for $Wc+\mm{jets}$. Apart from these theoretical corrections we constrain the absolute background normalization by employing the data as described below in Sec.\ \ref{toc:sampleComp}. The $p_T$ distribution of the $W$ boson in MC simulation is reweighted to match the $p_T$ distribution of $Z$ bosons measured in \dzero data \cite{zbosonpt} multiplied by the SM ratio of these two distributions calculated at NLO using \resbos \cite{resbos}.

The second most dominant background contribution is due to multijet processes where a jet is misidentified as an electron in the \eplus channel, or where a muon originating from the semileptonic decay of a heavy hadron appears to be isolated in the \muplus channel. More details and a brief discussion on the determination of the multijet background are given in Sec.\ \ref{toc:sampleComp}.

Other backgrounds include events from \zplus production, which includes $Z$ bosons and virtual photons ($\gamma^*$) decaying to electron, muon, or tau pairs. These events are generated with \alpgen interfaced to \pythia for showering and hadronization. The LO \alpgen predictions are corrected using the NLO calculation of \mcfm. The \zplus cross section is multiplied by 1.30. The heavy flavor components of the \zplus cross sections, $Z/\gamma^{*}c\bar{c}+$jets, $Z/\gamma^{*}b\bar{b} +$jets, are multiplied by an additional 1.67 and 1.52, respectively. The simulated $p_T$ distribution of the $Z$ boson is reweighted to match the measured $p_T$ distribution in $Z \rightarrow \ell \ell$ data \cite{zbosonpt}.

The single top quark background originates from $s$- and $t$-channel production, which are normalized to the NLO cross sections of 1.04 and 2.26 pb \cite{singleTopXsec}, respectively. As the single top quark background yields only a few events passing all selection criteria described later, we do not consider the dependence of this background on $m_t$.

Diboson production ($WW$, $WZ$, and $ZZ$ bosons) processes are another source of background and normalized to their NLO cross sections, calculated with \mcfm, of 11.62 pb, 3.25 pb, and 1.33 pb, respectively.

\subsection{Modeling of background contributions in the $\bm{\ell \ell}$ decay channel}
\label{toc:backgroundModel_ll}
The backgrounds in the dilepton decay channel are smaller than in the \ljets decay channel. The dominant source is \zplus production, followed by diboson production. For both processes the modeling employs the same implementation as described above for the \ljets decay channel.

The third most dominant source of background is multijet events, with the determination summarized in Sec.\ \ref{toc:sampleComp}.

%
%
%
%
\section{Event Selection}
\label{toc:data_samples}
This analysis is based upon the full Tevatron data sample recorded by the \dzero detector at $\sqrt{s}=1.96$ TeV and, after applying data quality requirements, corresponding to an integrated luminosity of 9.7 fb${}^{-1}$ \cite{lumi_nim}. The general selection criteria applied to both the \ljets and dilepton decay channels are summarized in the following:
\begin{enumerate}
\item \label{sel:vertexzpv} Accepted events have a PV within $|z_{\mm{PV}}|<60$~cm of the center of the detector along the beam axis.
\item \label{sel:vertextrk} The number of tracks associated with the PV is greater or equal three.
\item \label{sel:jets} After correcting the jet energy to the particle level, only jets with a transverse momentum $p_{T}>20$~GeV and $\left|\eta\right|<2.5$ are selected. 
\item \label{sel:jetsVC} Jets which satisfy the $b$-tagging requirement are required to have at least two tracks coming from the PV. More details on jet requirements for the individual decay channels are provided below.
\item \label{sel:leptons} Identified leptons are required to originate from the PV by demanding $|\Delta z(\ell, \mathrm{PV})| < 1~\mathrm{cm}$. These $z$ values correspond to the point of closest approach to the beam line of these tracks. 
\item \label{sel:leptonsdR} To ensure that electrons are isolated, an angular separation in $\Delta R$ of at least 0.5 between an electron and the closest jet is required.
\end{enumerate}

The measurements in both decay channels employ the $b$-tagging discriminant output distribution as provided by the $b$-ID MVA. The discriminant combines variables that characterize the presence and properties of secondary vertices and tracks within jets \cite{bid_nim}. We do not impose any requirements on this discriminant; instead we employ the entire distribution to measure the inclusive cross section as described in Sec.~\ref{toc:methods_intro}. 

The specific selection requirements for \ljets and \dilep events are described below; the requirements are chosen such that the selections are mutually exclusive.

\begin{figure*}[ht]
  \begin{center}
   \includegraphics[width=0.675\columnwidth,angle=0]{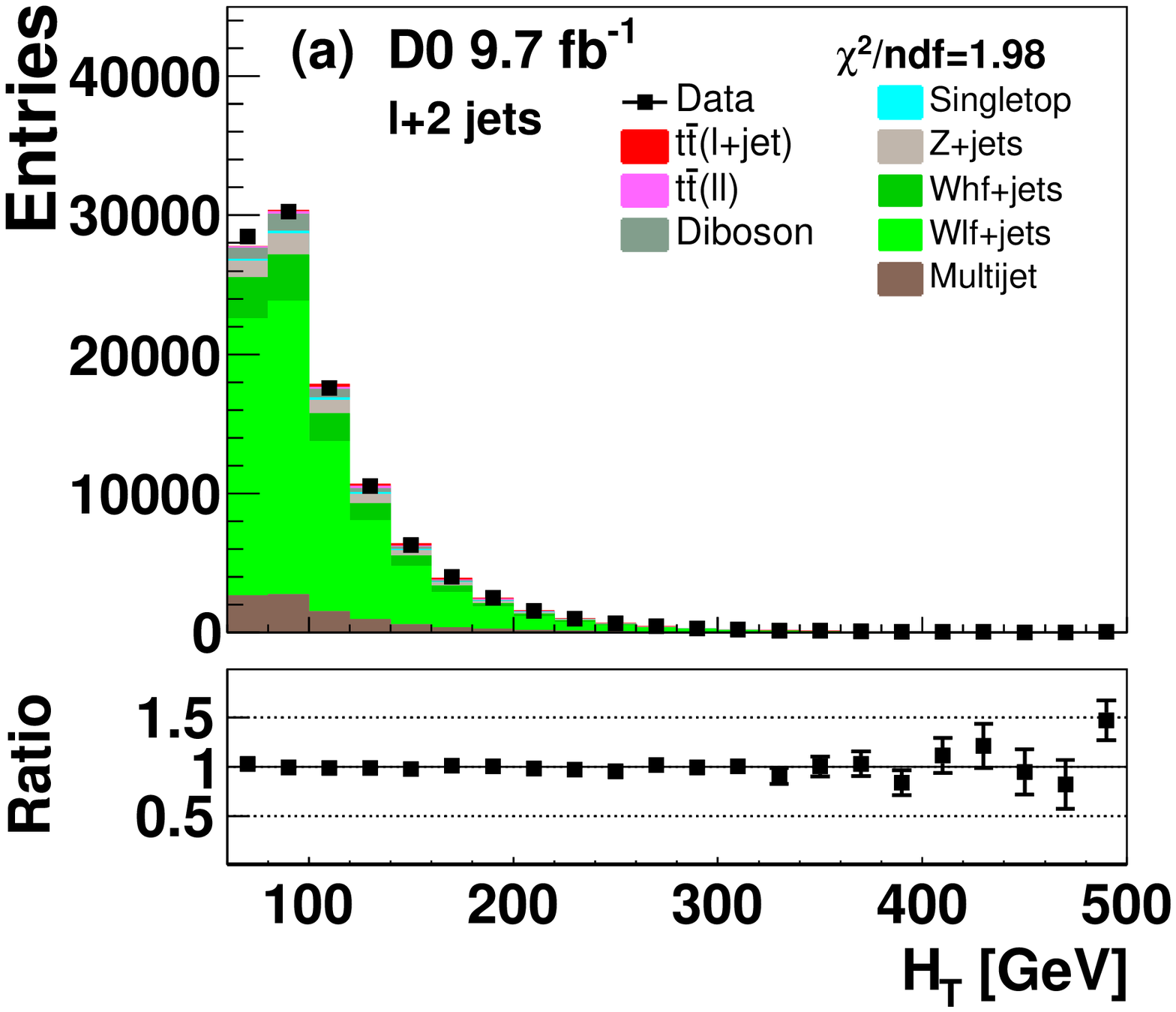}
   \includegraphics[width=0.675\columnwidth,angle=0]{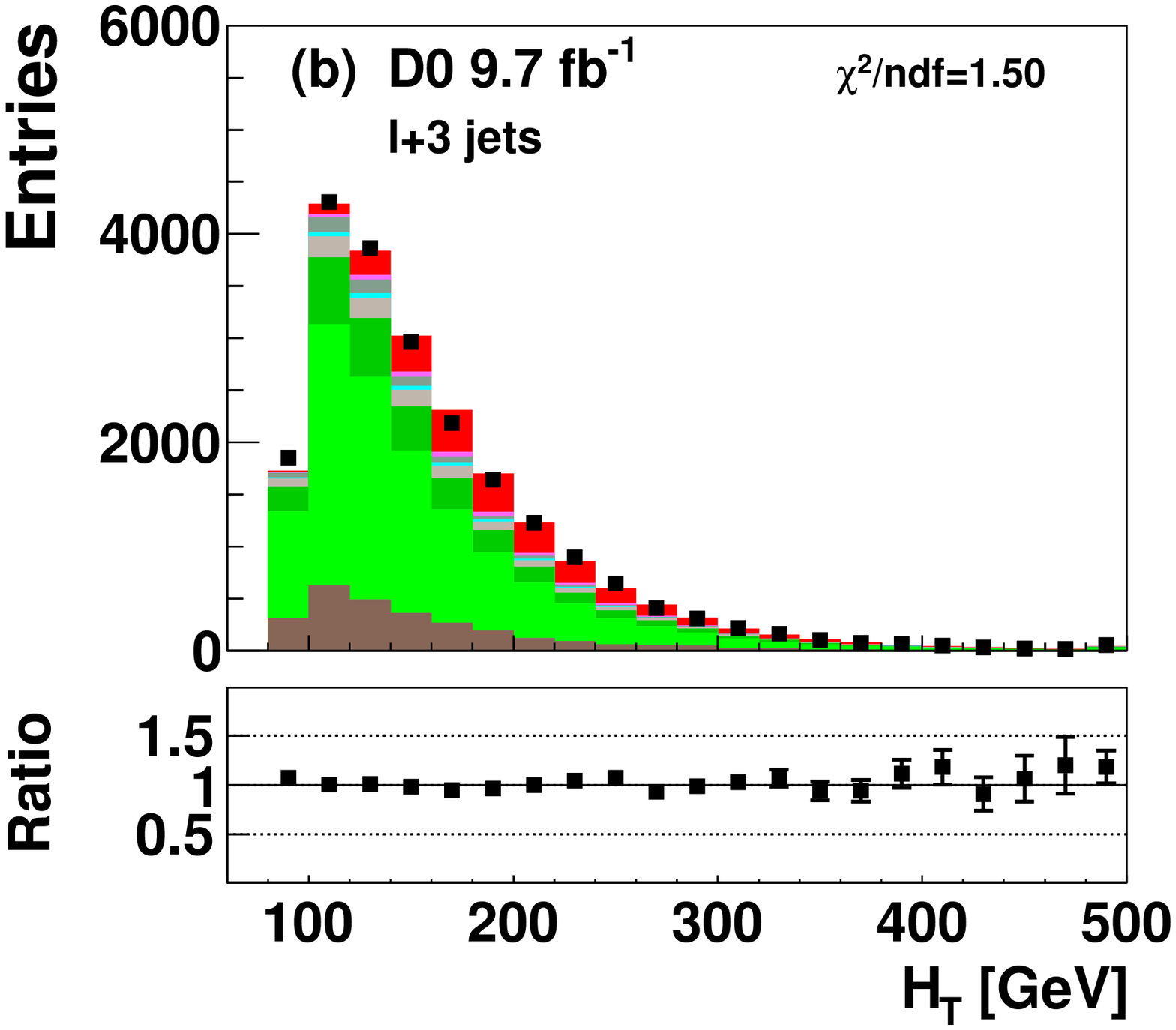}
   \includegraphics[width=0.675\columnwidth,angle=0]{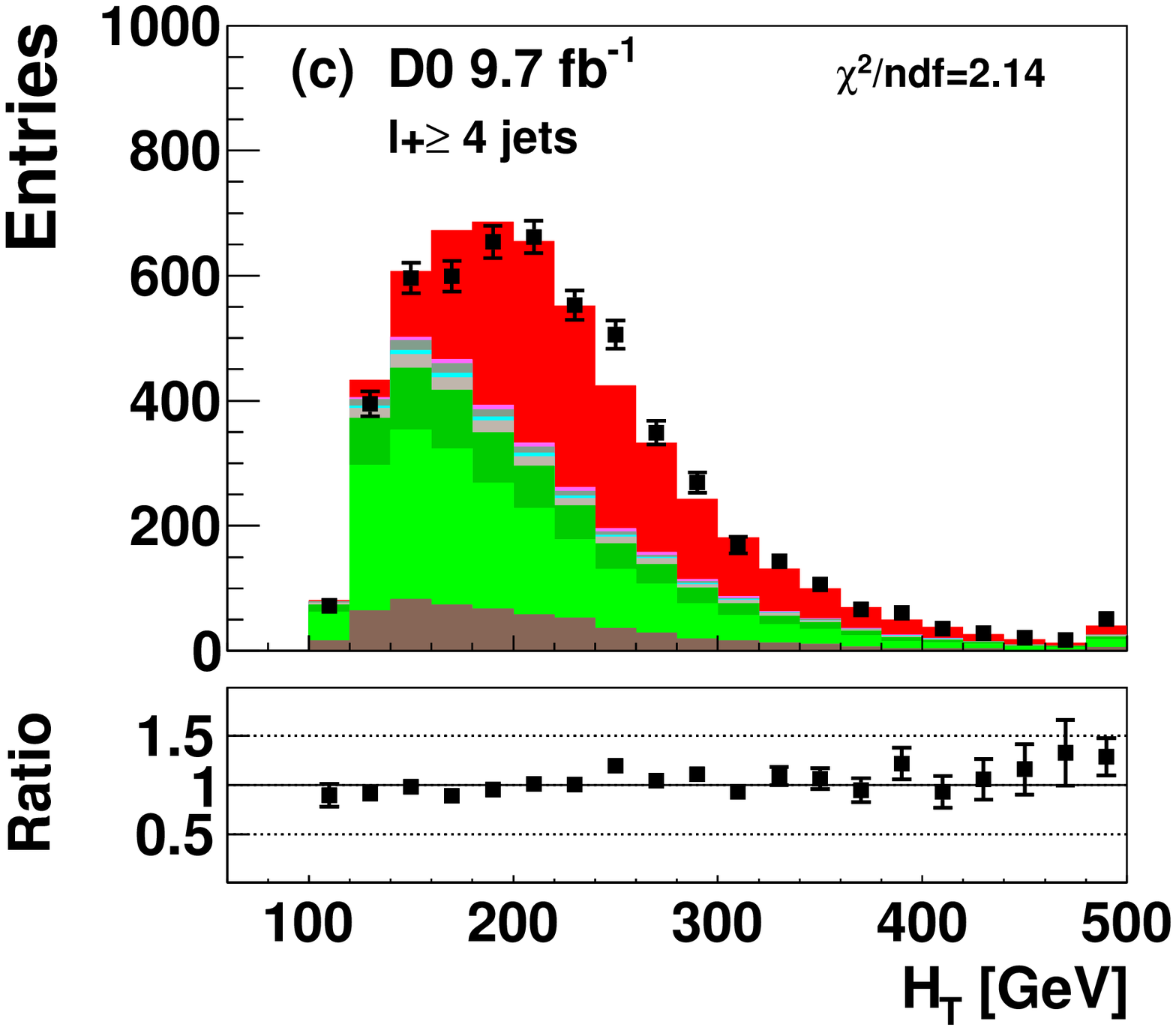}
   \includegraphics[width=0.675\columnwidth,angle=0]{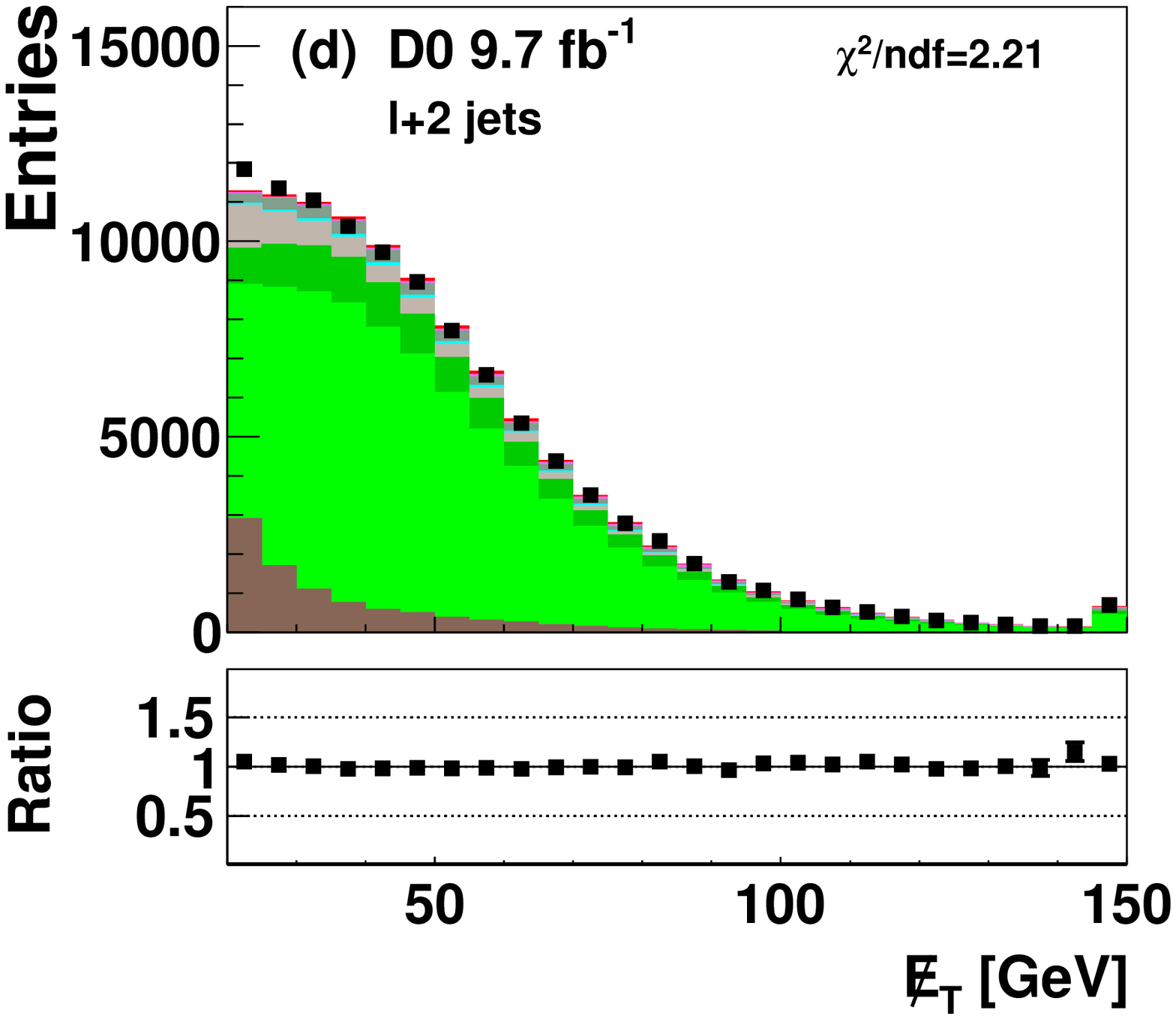}
   \includegraphics[width=0.675\columnwidth,angle=0]{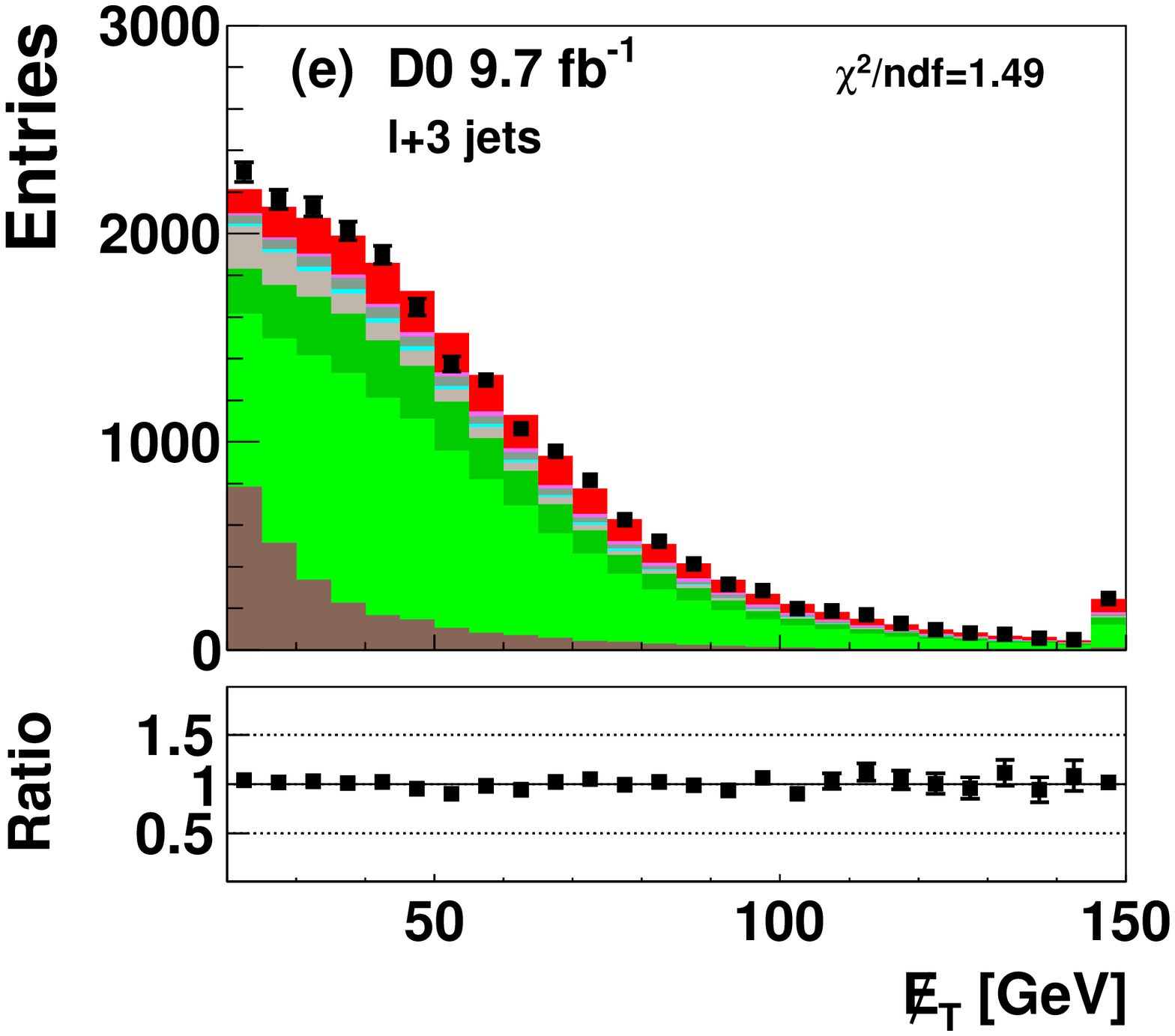}
   \includegraphics[width=0.675\columnwidth,angle=0]{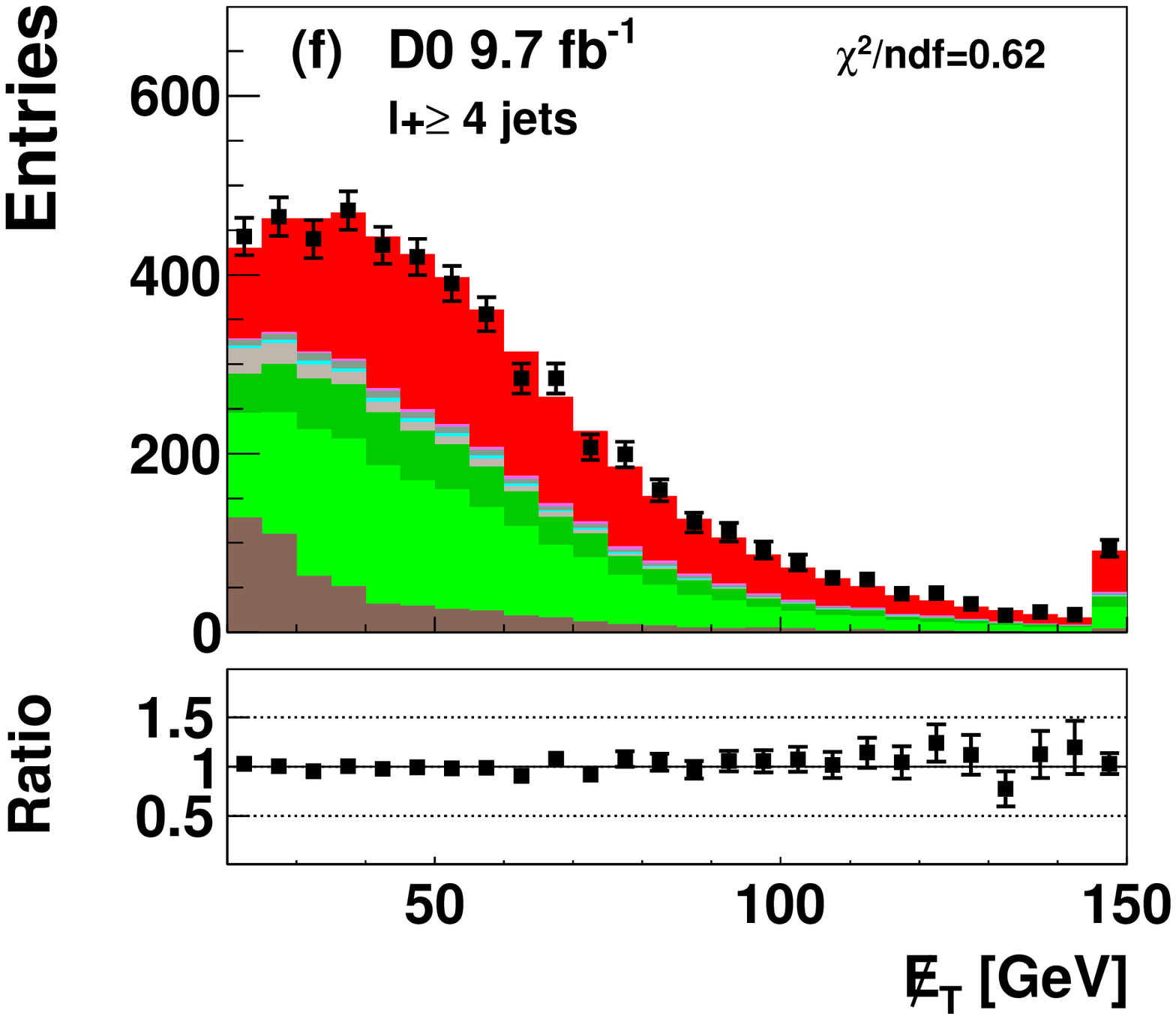}
   \includegraphics[width=0.675\columnwidth,angle=0]{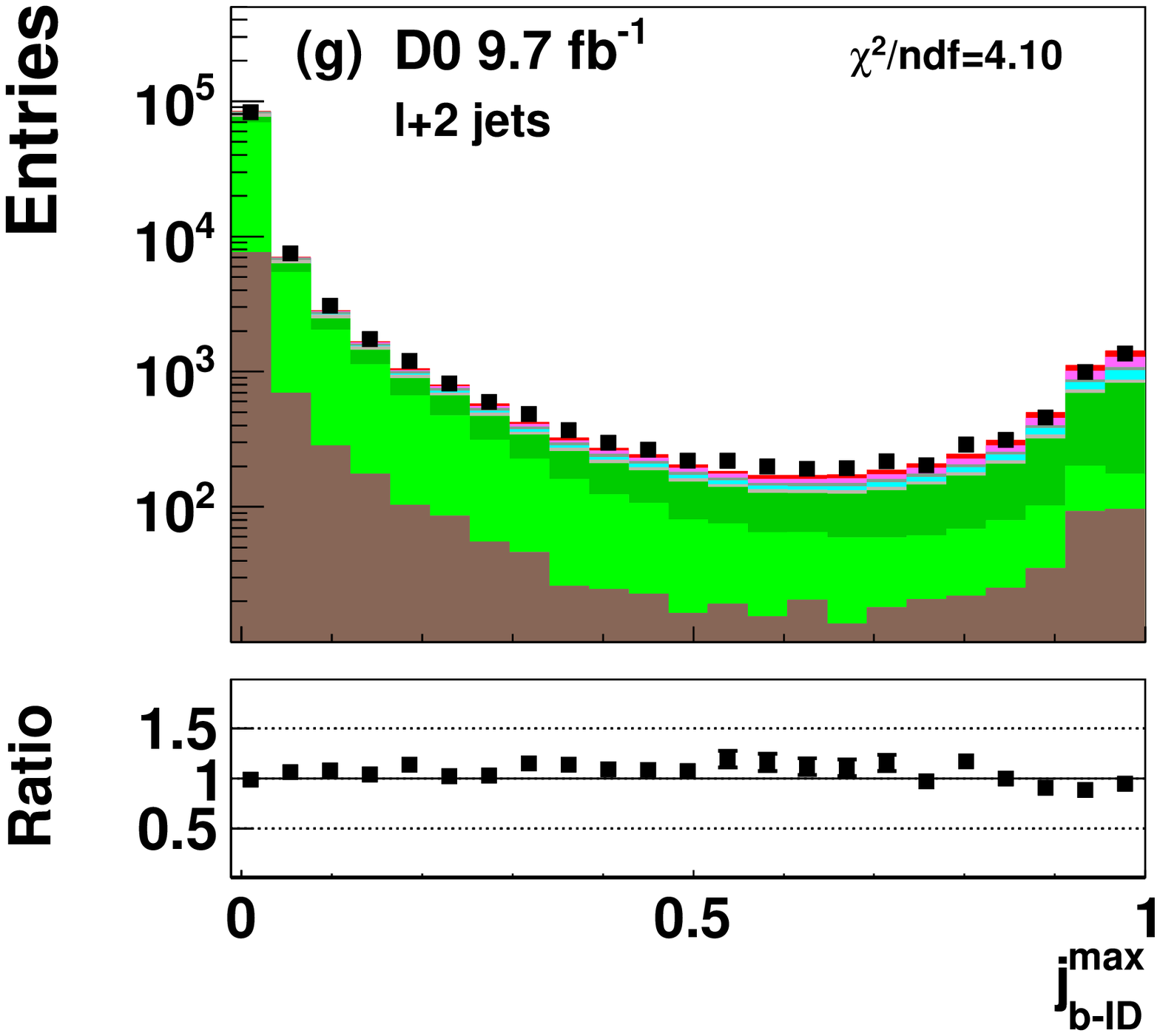}
   \includegraphics[width=0.675\columnwidth,angle=0]{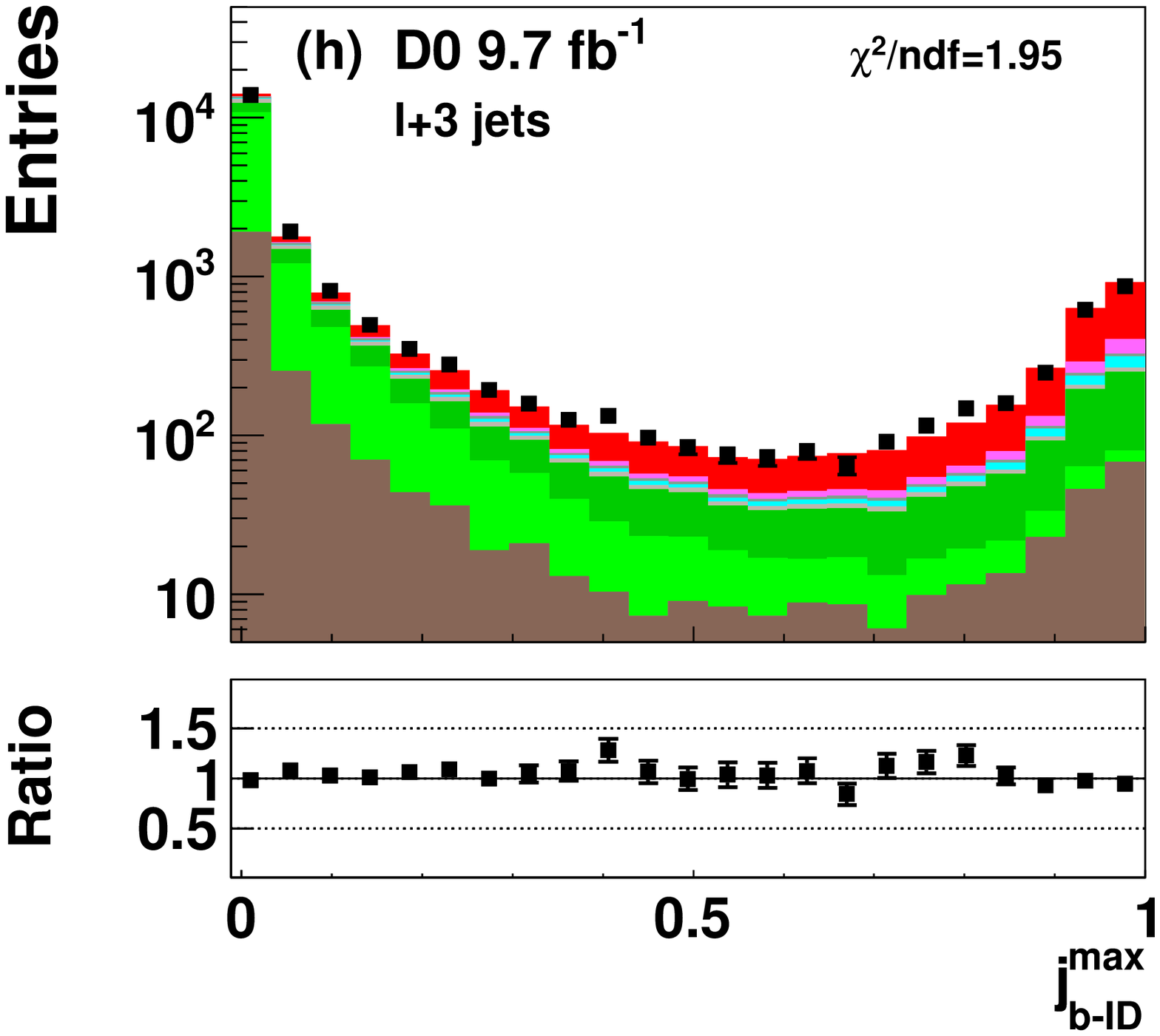}
   \includegraphics[width=0.675\columnwidth,angle=0]{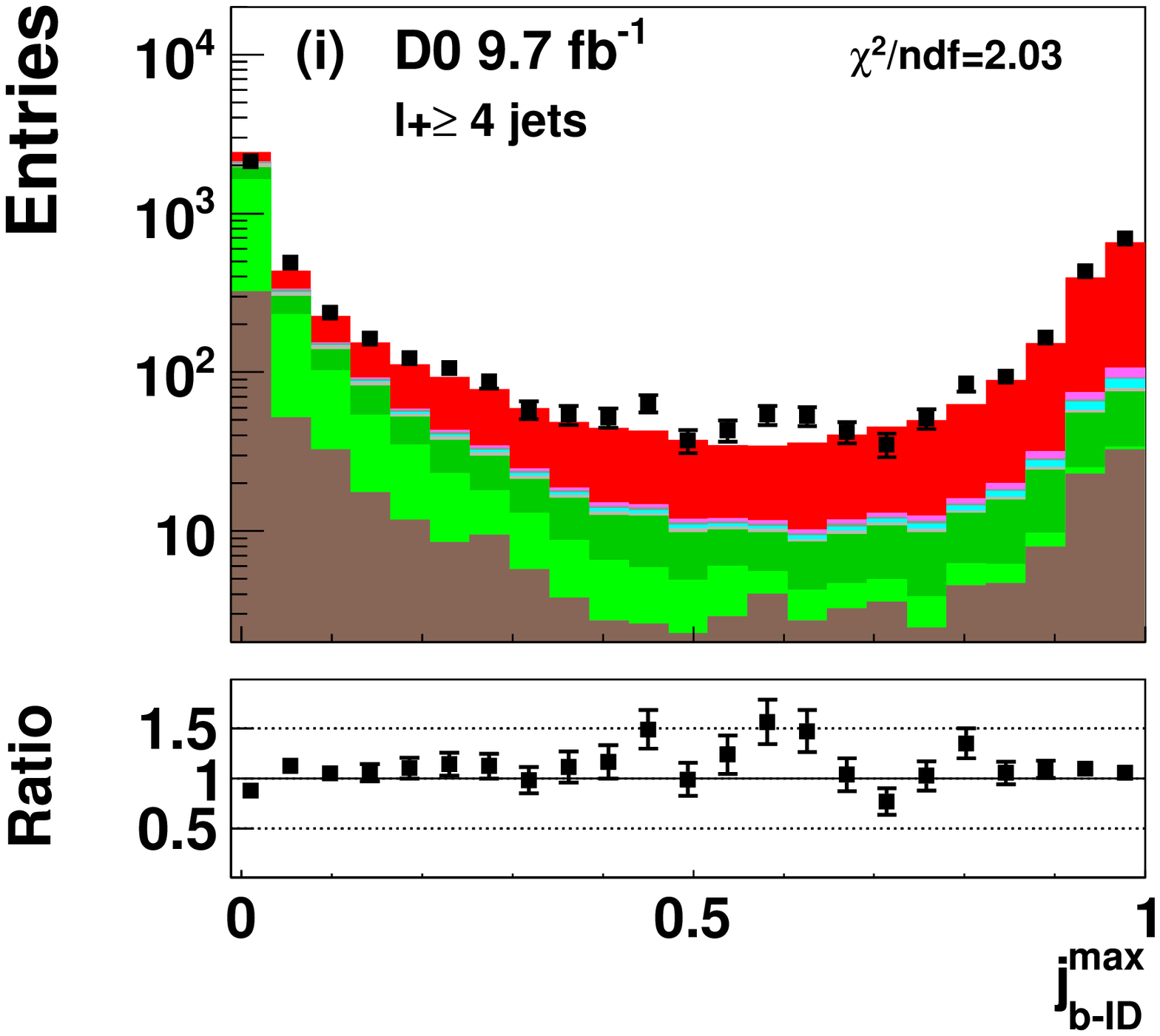}
\end{center}
    \caption{Distributions of (a)--(c) the scalar sum of the $p_T$ values of the lepton and jets, $H_T$, (d)--(f) $\met$ and (g)--(i) the maximal MVA $b$-ID value of all jets, \mmax, for events with a lepton and two, three or four or more jets. The data are compared to the sum of predicted contributions from signal and background processes, using the theoretical value of the inclusive $t\bar{t}$ cross section of $7.48~\mathrm{pb}$ \cite{mochUwer} and $m_t=172.5$ GeV. The highest bin in the histograms includes overflows. The ratios of data to the sum of the signal and all background contributions are shown in the panels below the individual distributions. Only statistical uncertainties of the data are shown and the $\chi^2/\mm{ndf}$ values only take statistical uncertainties into account.} \label{fig:ljets_control}
\end{figure*}

\begin{figure*}[ht]
  \begin{center}
   \includegraphics[width=0.675\columnwidth,angle=0]{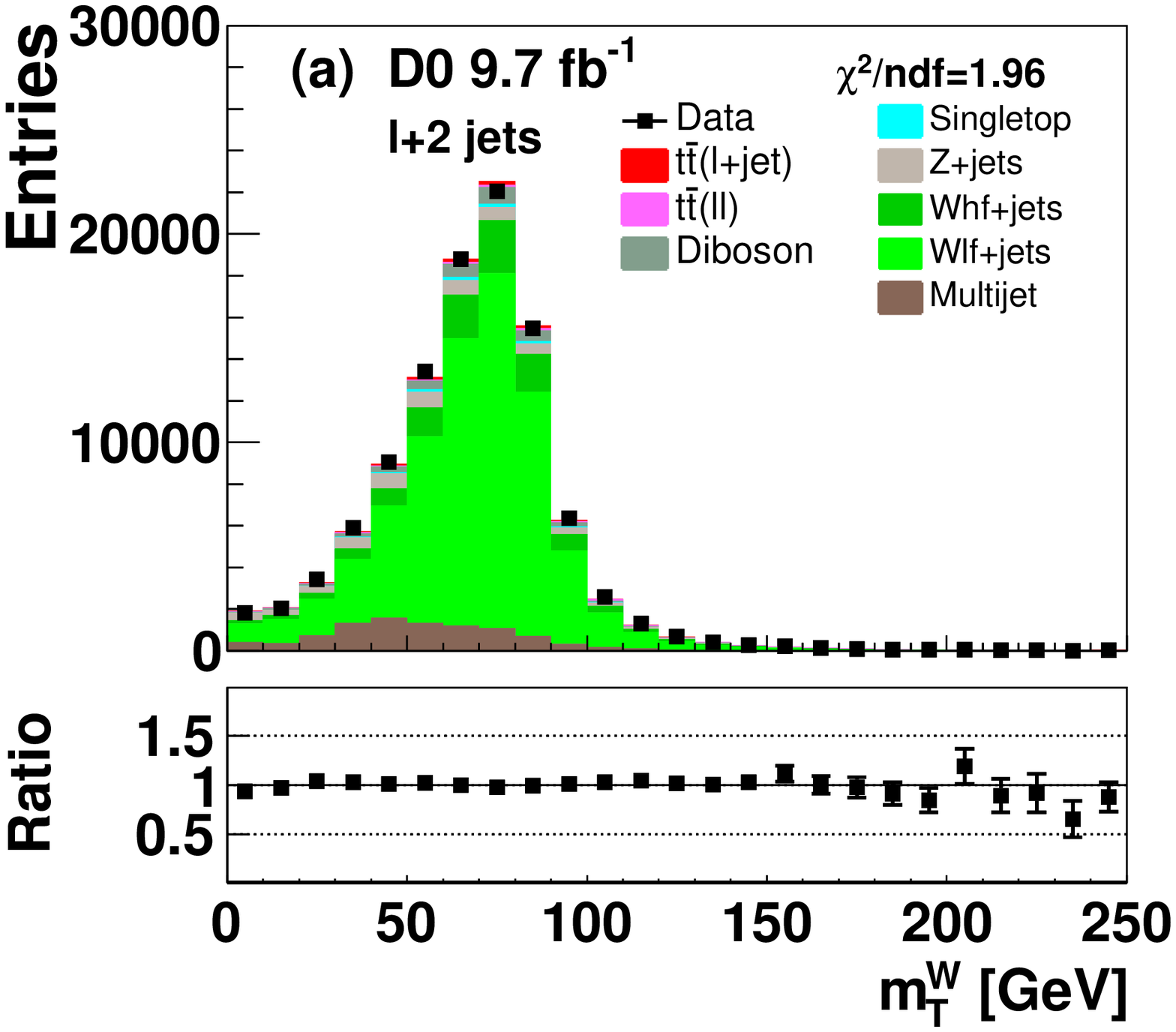}
   \includegraphics[width=0.675\columnwidth,angle=0]{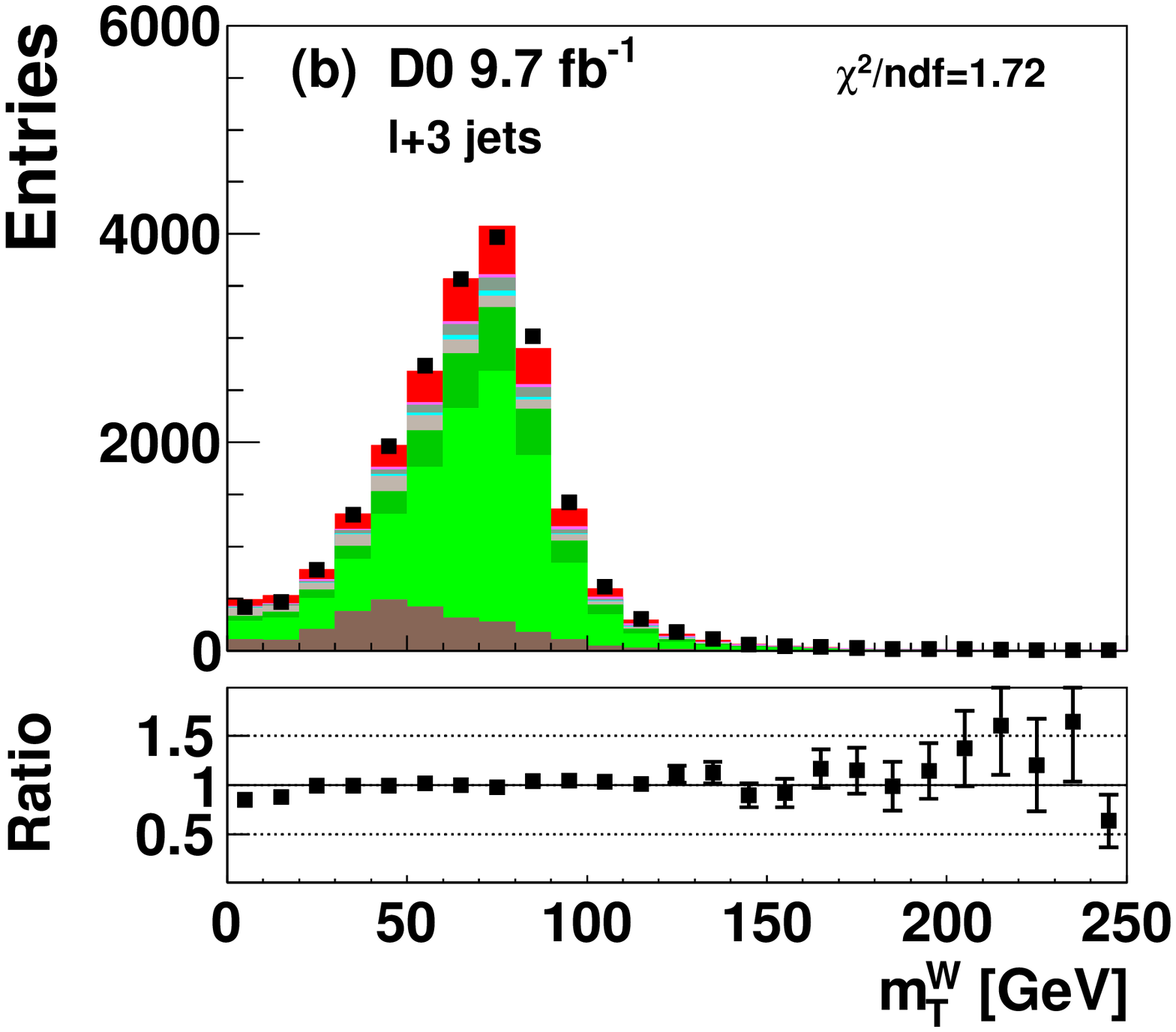}
   \includegraphics[width=0.675\columnwidth,angle=0]{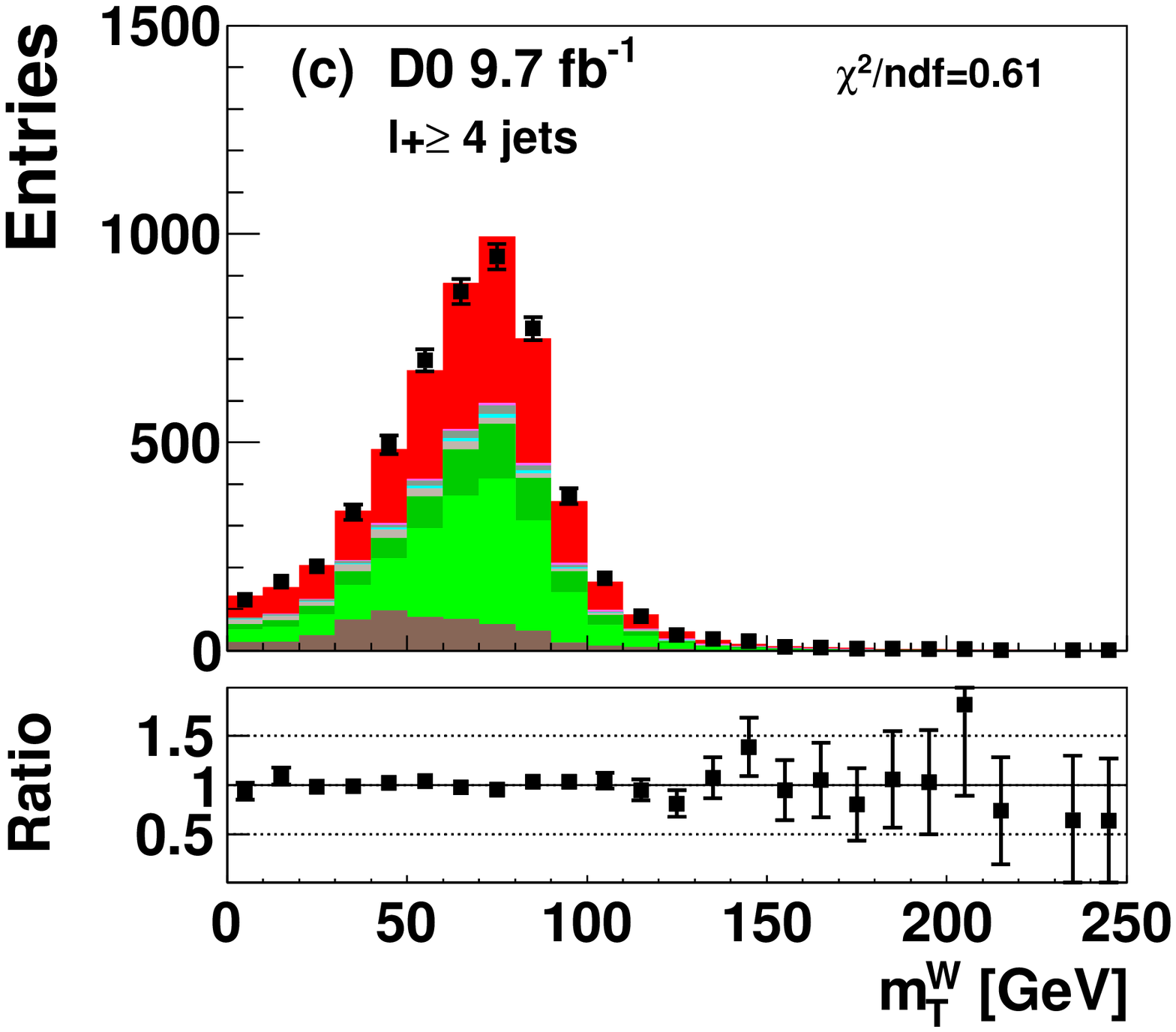}
   \includegraphics[width=0.675\columnwidth,angle=0]{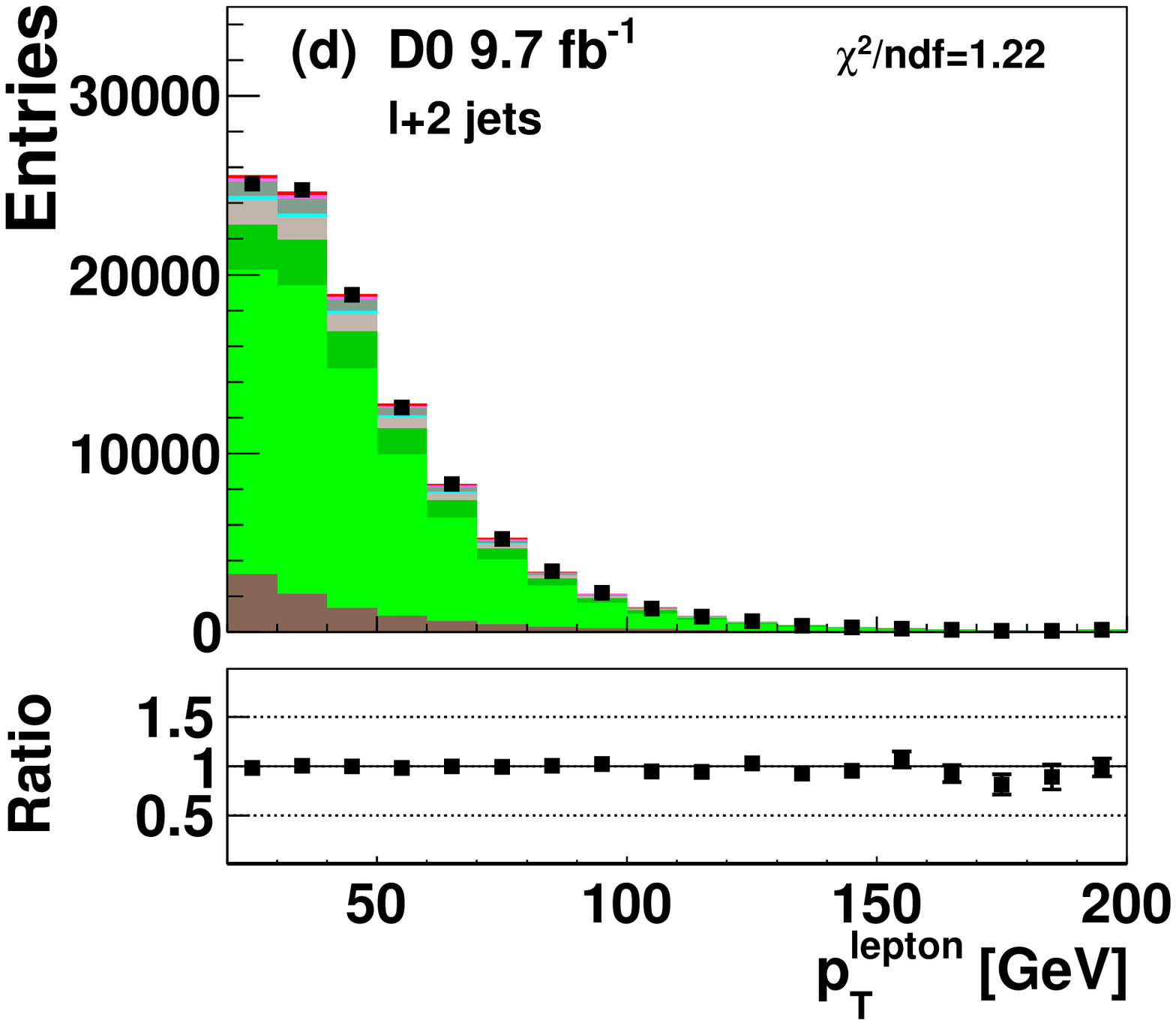}
   \includegraphics[width=0.675\columnwidth,angle=0]{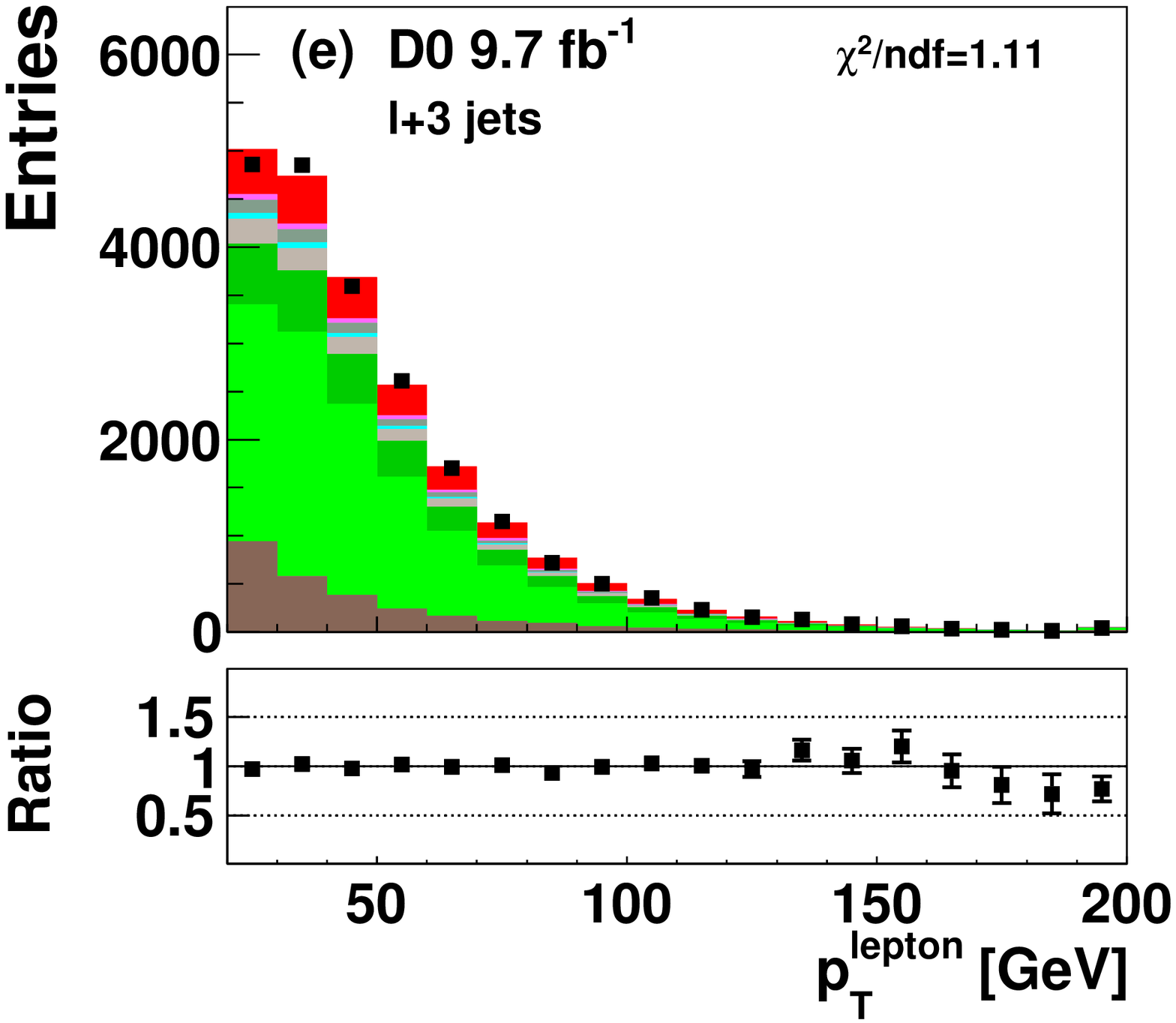}
   \includegraphics[width=0.675\columnwidth,angle=0]{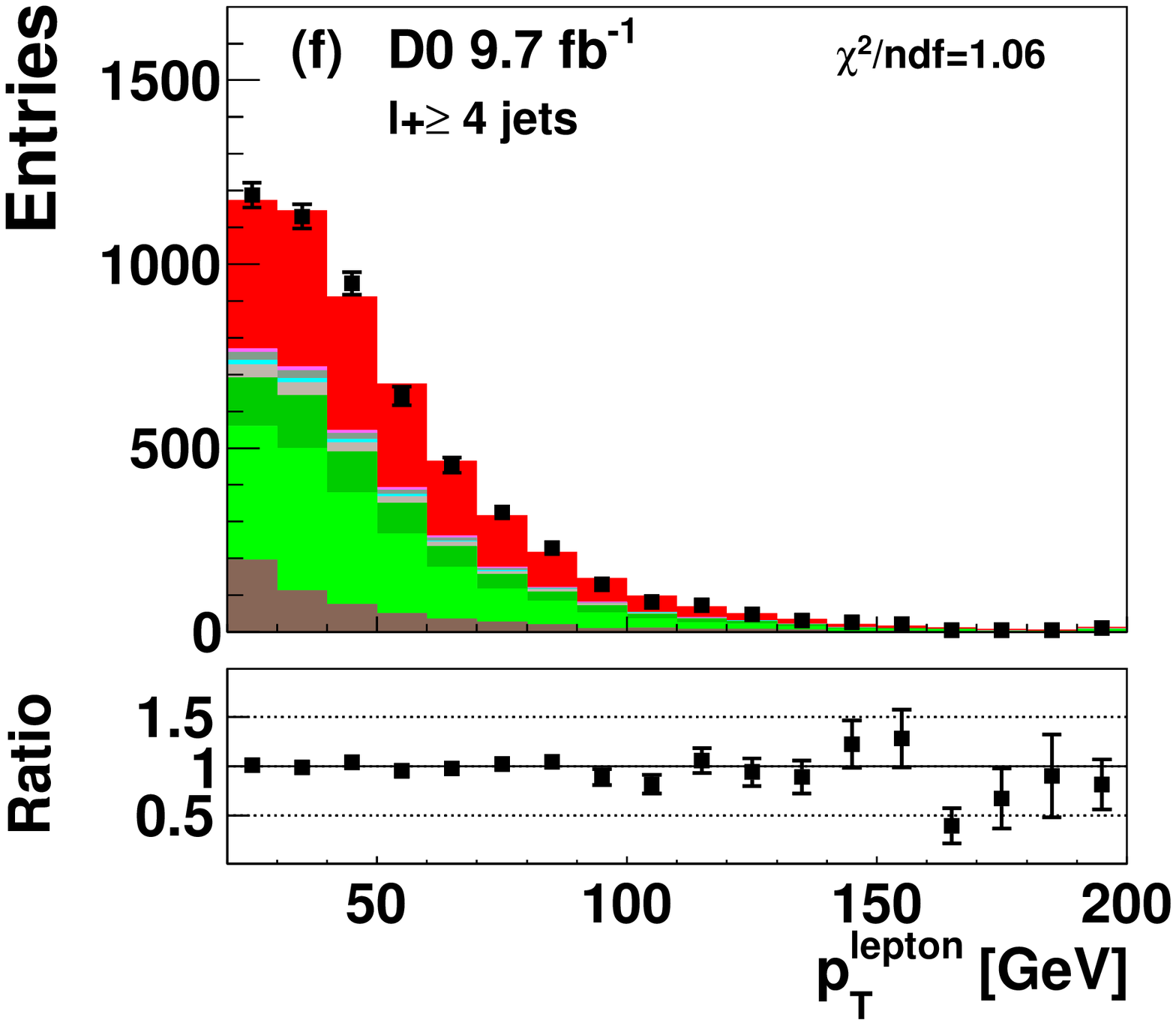}
  \end{center}
    \caption{Distributions of (a)--(c) the transverse mass of the lepton + \met system, $m_T^W$, and (d)--(f) $p_{T}^{\mm{lepton}}$ for events with a lepton and two, three or four or more jets. The data are compared to the sum of predicted contributions from signal and background processes, using the theoretical value of the inclusive $t\bar{t}$ cross section of $7.48~\mathrm{pb}$ \cite{mochUwer} and $m_t=172.5$ GeV. The highest bin in the histograms includes overflows. The ratios of data to the sum of the signal and all background contributions are shown in the panels below the distributions. Only statistical uncertainties of the data are shown and the $\chi^2/\mm{ndf}$ values only take statistical uncertainties into account.} \label{fig:ljets_controlO}
\end{figure*}

\subsection{Event selection in the $\bm{\ell+}$jets decay channel}
The selection requirements for the cross section measurement for the \ljets channel are very similar to the ones described in Ref.~\cite{diffXsecPaper} and are summarized briefly in the following:

\begin{enumerate}
\item \label{sel:triggerLJ} The trigger requirement is a logical ``OR" of the conditions for at least ``one lepton" and for at least ``a lepton plus a jet" in an event. Lepton trigger thresholds of 15 or 10 GeV were implemented for the single lepton trigger or the lepton plus jet trigger, respectively. 
\item \label{sel:leptonsLJ} Exactly one isolated lepton with a transverse momentum $p_{T}>20$~GeV and $\left|\eta\right|<1.1$ (for electrons) or $\left|\eta\right|<2$ (for muons) is required. Events with more than one lepton satisfying these criteria are rejected.
\item \label{sel:met} We require $\met>20$~GeV.
\item \label{sel:muonQual} For the \muplus sample we remove misreconstructed muons by requiring upper limits on the transverse mass of the muon + \met system, $M_T^W$, of $M_T^W < 250$ GeV and \met$<250$ GeV. To further remove such events, we employ an additional requirement on the significance of the track curvature described in more detail in Ref.~\cite{diffXsecPaper}.
\item \label{sel:fakes} To reduce multijet background we require a minimum separation between the direction of the lepton and the direction of the missing momentum \cite{diffXsecPaper}: \mbox{$\Delta \phi(e, $ \metNo$) > 2.2 - 0.045 \cdot \metNo/\mm{GeV}$} and \mbox{$\Delta \phi(\mu, $ \metNo$) > 2.1 - 0.035 \cdot \metNo/\mm{GeV}$}.
\item \label{sel:jetsLJ} At least two jets are required. To suppress jets from additional collisions, jets are required to contain at least two tracks with a closest approach in $z$ to the PV of less than 1 cm.

\end{enumerate}

\subsection{Event selection in the dilepton decay channel}
In addition to the general selection requirements discussed in the opening of this section, additional requirements specific to the dilepton channel are made. The selection requirements for this cross section measurement are very similar to those used for the leptonic asymmetry measurements in the dilepton channel published earlier~\cite{leptonicAfbdilepton}  and are summarized briefly in the following list.

\begin{enumerate}
\item \label{sel:globtrig} In the $e\mu$ channel, no explicit trigger requirement is applied, whereas in the $ee$ or $\mu\mu$ channels single lepton triggers with thresholds at 15 GeV are employed.
\item \label{sel:leptonsLL} Electrons are required to have a transverse momentum of $p_T> 15$~GeV and $|\eta|<2.5$. We exclude the $1.1 < |\eta| < 1.5$ region with poor resolution.
\item \label{sel:leptonsLLmu} Muons are selected with $p_T > 15$ GeV and $|\eta|<2$. To remove misreconstructed muons we require muons to have $p_T<200$~GeV for the dimuon channel.
\item \label{sel:tighte} For the $e\mu$ channel exactly one electron and one or more muons are required.
\item \label{sel:mu} For the $\mu\mu$ channel two or more muons are required.
\item \label{sel:die} For the $ee$ channel two or more electrons are required.
\item \label{sel:emupair} The two selected leptons must have opposite charges. If more than one oppositely charged lepton pair is found, the lepton pair with the largest $p_T$ scalar sum is chosen.
\item At least one jet is required in the $e\mu$ channel and at least two jets are required in the $\mu\mu$ and $ee$ channel. 
\item \label{sel:leptonsQualLL} Additional quality requirements are imposed to remove background from bremsstrahlung. 
\item \label{sel:topological} We further reduce background contributions by imposing the following topological requirements: in the $ee$ channel we require a \met significance of $\ge 5$, in the $\mu\mu$ channel we require $\met > 40$~GeV and a \met significance of $\ge 2.5$, and in the $e\mu$ channel we require $H_{T}>110$~GeV, where $H_{T} = p_{T}(\mm{leading~lepton}) + p_{T}(2~\mm{leading~jets})$. More details are described in Ref. \cite{leptonicAfbdilepton}.
\end{enumerate}

\begin{table*}[htp]
    \caption{Expected number of events in the \ljets channel with two, three or $\ge$ four jets. The sum of signal and background agrees well with the number of data events by construction; uncertainties are statistical and systematic added in quadrature (see Sec.\ \ref{toc:sys_sampleComp} for details). Events from \ttbar dilepton decays are treated as background and denoted as ``\ttbar $\ell \ell$."} \label{tab:yields_ljets}
\begin{ruledtabular}
\begin{tabular}{l d{-1} d{-1} d{-1} d{-1} d{-1} d{-1}}
\multicolumn{7}{c}{\ljets decay channel}\\
\multicolumn{1}{c}{Process} & \multicolumn{1}{c}{$e + 2$ jets} & \multicolumn{1}{c}{$e + 3$ jets} & \multicolumn{1}{c}{$e + \ge 4$ jets} & \multicolumn{1}{c}{$\mu + 2$ jets} & \multicolumn{1}{c}{$\mu + 3$ jets} & \multicolumn{1}{c}{$\mu + \ge 4$ jets} \\ \hline \T
Multijet	& 9160.2350 &  2266.550 & 464.120	& 1546.$\hphantom{0}$630 & 418.170 & 99.40\\  \T 
Single top	& 471.$\hphantom{00}$60 & 129.$\hphantom{0}$20 & 27.$\hphantom{00}$5 & 331.$\hphantom{00}$40	& 92.$\hphantom{0}$10 	& 20.$\hphantom{0}$3 \\  \T 
\wlp	        & 37937.\,\,^{1350}_{700} & 5544.\,\,^{200}_{100} & 850.\,\,\,\,^{30}_{20} & 32701.\,\,^{1150}_{600} & 5313.\,\,^{200}_{100} & 835.\,^{30}_{15} \\ \T 
\whf		& 6020.\,\,^{1000}_{1400} & 1502.\,\,^{250}_{350} & 329.\,\,\,\,^{60}_{80} & 4998.\,\,^{850}_{1150} & 1391.\,\,^{250}_{300} & 315.\,^{50}_{70} \\ \T 
\zlp	                 & 2031.$\hphantom{0}$400 & 390.$\hphantom{0}$80 & 57.$\hphantom{0}$10 & 2557.$\hphantom{0}$500 & 422.$\hphantom{0}$80 & 49.10 \\ \T 
\zhf		        & 369.$\hphantom{00}$70 & 114.$\hphantom{0}$20 & 24.$\hphantom{00}$5 & 485.$\hphantom{0}$100 & 120.$\hphantom{0}$20 & 21.$\hphantom{0}$5 \\ \T 
Diboson	        & 1926.$\hphantom{0}$140 & 338.$\hphantom{0}$20 	& 52.$\hphantom{00}$5 	& 1417.$\hphantom{0}$100 	& 249.$\hphantom{0}$20 	& 40.$\hphantom{0}$5 \\ \T 
\ttbar $\ell \ell$   & 566.$\hphantom{00}$30 & 182.$\hphantom{0}$10 & 31.$\hphantom{00}$5 & 345.$\hphantom{00}$20 & 118.$\hphantom{0}$10 & 22.$\hphantom{0}$5 \\ \hline \T 
$\sum\,\mm{background}$  & 58479.2900 & 10465.650 & 1834.140 & 44381.1650 & 8123.350 & 1402.80 \\ \T 
\ttbar \ljets        & 669.$\hphantom{00}$30 & 1460.$\hphantom{0}$70 & 1177.$\hphantom{0}$60 & 393.$\hphantom{00}$20 & 1002.$\hphantom{0}$50 & 909.50 \\  \hline \T
$\sum\, \mm{(signal + background)}$ & 59148.2900  & 11925.650  & 3011.140 & 44773.1650 & 9125.350 & 2310.80 \\  \T 
Data	                 & 59122  & 11905	&  3007 &  44736 &  9098 & 2325 \\
 \end{tabular}
\end{ruledtabular}
\end{table*}

%
%
%
%
\section{Sample Compositions}
\label{toc:sampleComp}
We distinguish between instrumental backgrounds and irreducible backgrounds from processes with final states identical to \ttbar. Instrumental backgrounds are due to multijet processes where one or more jets are misidentified as an electron, or where one or more muons originating from the semileptonic decay of a heavy hadron appear to be isolated, and hence fulfill all selection requirements of a lepton stemming from the decay of a top quark. Irreducible backgrounds are for example due to \wplus or \zplus processes with the same final state as the \ljets and \dilep top quark decay channel. Systematic uncertainties on the determination of the sample composition are discussed in Sec.\ \ref{toc:sys_sampleComp}. The following section describes the composition of the input or ``pre-fit" \ljets and \dilep samples, which are used to extract the \ttbar cross section as described in Sec.\ \ref{toc:fitting_collie}.

\subsection{Determination of the \ljets sample composition}
\label{toc:sampleCompLjets}
The irreducible background processes are estimated using MC simulations as described in Sec.\ \ref{toc:generators}. Compared to the \dilep channel, the \ljets channel has a larger background fraction, each with larger systematic uncertainties than in the \dilep channel. Since most of this background arises from \wplusTw production we estimate this contribution following the same approach as for the measurement of the differential \ttbar cross section \cite{diffXsecPaper}. The \wplus cross section is iteratively scaled for each jet multiplicity bin separately by a \wplus heavy-flavor scale factor $s_{\mm{fit}}^{\mm{WHF}}$ and \wplus light-flavor scale factor $s_{\mm{fit}}^{\mm{WLF}}$ to match the number of data events after subtraction of all other instrumental and irreducible background contributions as well as the signal contribution. This approach yields reasonable initial values for the log-likelihood profile fit (introduced in Sec.\ \ref{toc:fitting_collie}). The details of the estimation of systematic uncertainties are described in Sec.\ \ref{toc:sys_sampleComp}. 

Data-driven and MC methods are combined in the ``matrix method" \cite{matrixMethod, Publ54_xsec}, which is employed to model the instrumental background originating from multijet (MJ) processes in the \ljets channel. The MJ contribution is determined employing two samples of \ljets events: one applying the nominal lepton selection requirements and one with looser lepton selection requirements denoted ``loose." In addition, an orthogonal data sample is defined by requiring $\not\!\!E_T < 10~\mm{GeV}$ (the nominal requirement is $\not\!\!E_T > 20~\mm{GeV}$) and the above selection criteria for the signal sample. This data sample is enriched in MJ events and any contributions from isolated leptons, as expected from MC, are subtracted from all considered distributions. No real isolated leptons are assumed to be included. We determine the shape and absolute contribution (misidentification rate) of multijet events for different jet multiplicities by comparing this data sample with the data sample containing loose leptons but the same \met requirement.

Figures \ref{fig:ljets_control} and \ref{fig:ljets_controlO} show the modeling of the selected events in the \ljets sample with the background and signal contributions. The expected composition of the sample after the final selection is given in Table \ref{tab:yields_ljets}.

\begin{table*}[htp]
    \caption{Expected number of events in the $ee + \ge 2$ jets, $\mu\mu + \ge 2$ jets, $e\mu + 1$ jets and $e\mu + \ge 2$ jets channels due to each process; uncertainties are statistical and systematic added in quadrature (see Sec.\ \ref{toc:sys_sampleComp} for details).} \label{tab:yields_dilepton}
\begin{ruledtabular}
\begin{tabular}{l z{-1} z{-1} z{-1} z{-1}}
\multicolumn{5}{c}{dilepton decay channel}\\
\multicolumn{1}{c}{Process} & \multicolumn{1}{c}{$ee + \ge 2$ jets} & \multicolumn{1}{c}{$\mu\mu + \ge 2$ jets} & \multicolumn{1}{c}{$e\mu + 1$ jets} & \multicolumn{1}{c}{$e\mu + \ge 2$ jets} \\ \hline  \T
Multijet		         & 5.7,\,\,^{0.9}_{0.9} & 7.0,\,\,^{3.3}_{2.6} & 28.3,^{6.6}_{6.6} & 32.5,^{7.4}_{7.4} \\ \T
\zlllp		&66.6,^{17.9}_{17.2} & 107.6,^{22.1}_{22.0} & 74.6,^{15.8}_{15.8} & 57.5,^{13.8}_{13.4}\\ \T
Diboson	        	        & 9.9,\,\,^{2.4}_{2.2} & 12.6,\,\,^{2.8}_{3.0} & 38.5,^{4.6}_{4.2} & 14.7,^{3.7}_{3.5} \\ \hline \T
$\sum\,\mm{background}$   & 82.2,$\hphantom{0}$18 & 172.2,$\hphantom{0}$22 & 141.4,18 & 104.7,15 \\ \T 
\ttbar $\ell \ell$  	& 107.7,$\hphantom{0}$15 & 101.5,$\hphantom{0}$12 & 86.5,11 & 313.7,38 \\ \hline \T
$\sum\, \mm{(signal + background)}$ &190,$\hphantom{0}$23 & 229,$\hphantom{0}$25 & 228,21 & 418,42 \\ \T
Data	          & 215 &  242 & 236 & 465 \\
 \end{tabular}
\end{ruledtabular}
\end{table*}

\begin{figure*}[ht]
  \begin{center}
   \includegraphics[width=0.675\columnwidth,angle=0]{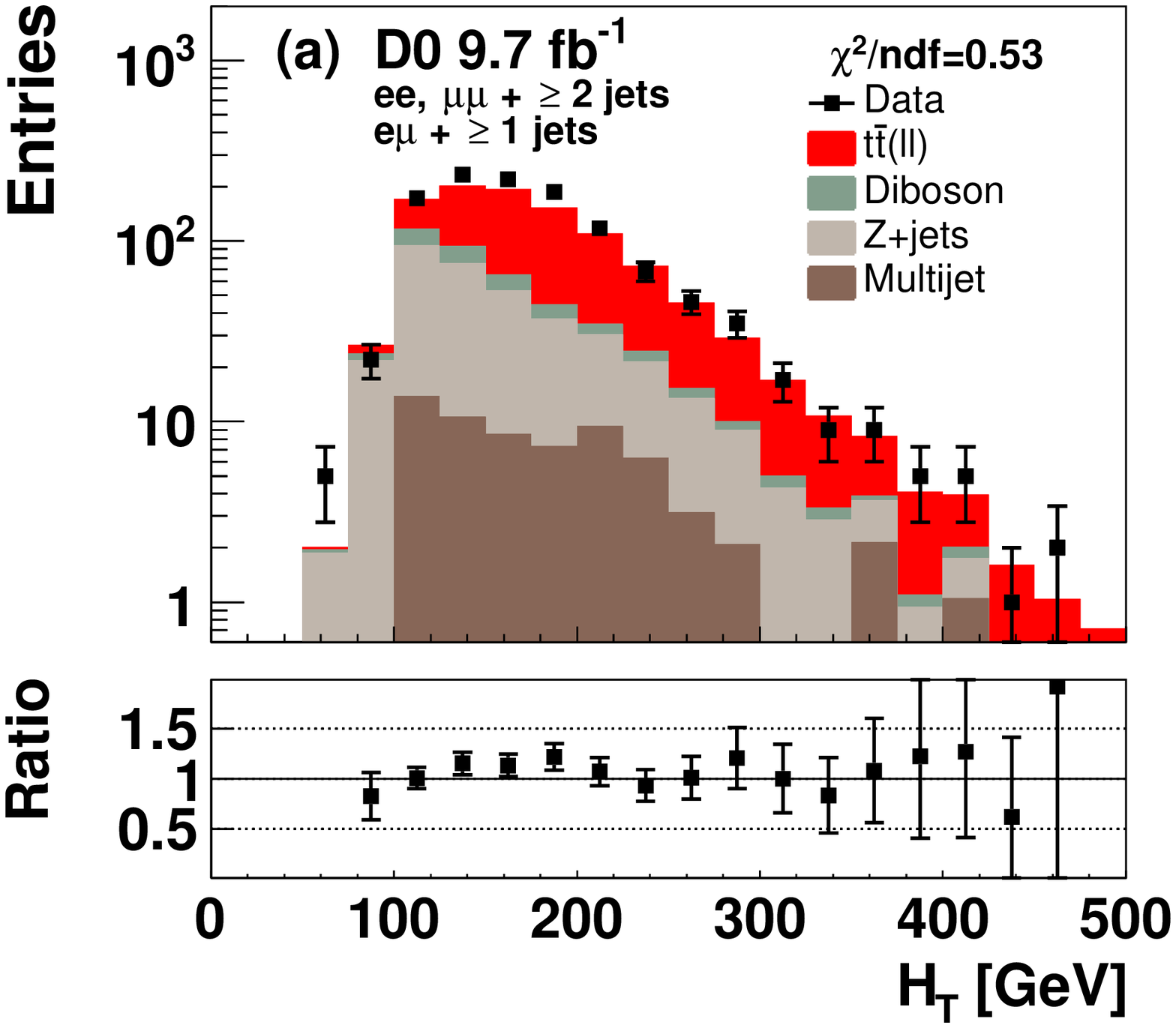}
   \includegraphics[width=0.675\columnwidth,angle=0]{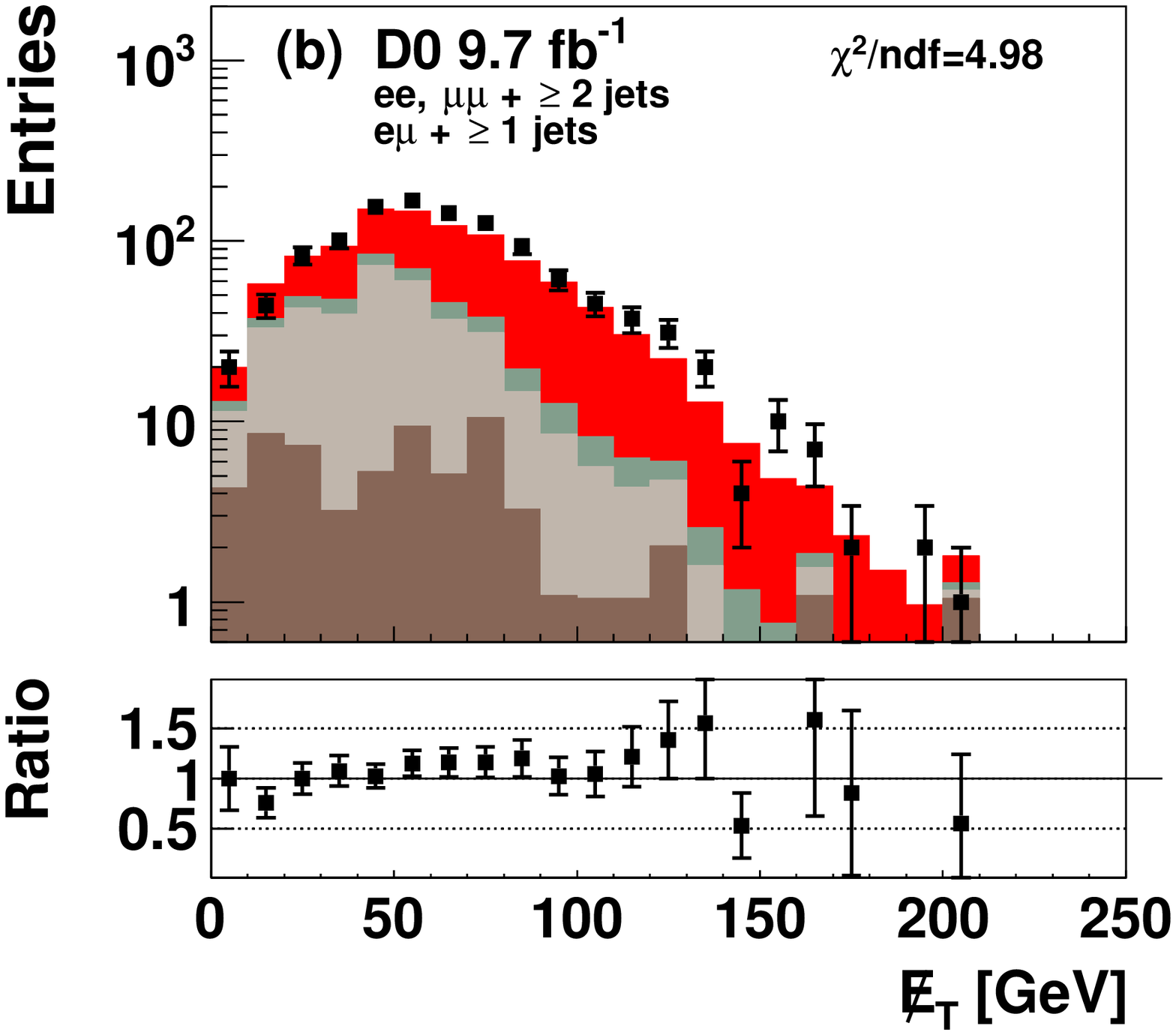}
   \includegraphics[width=0.675\columnwidth,angle=0]{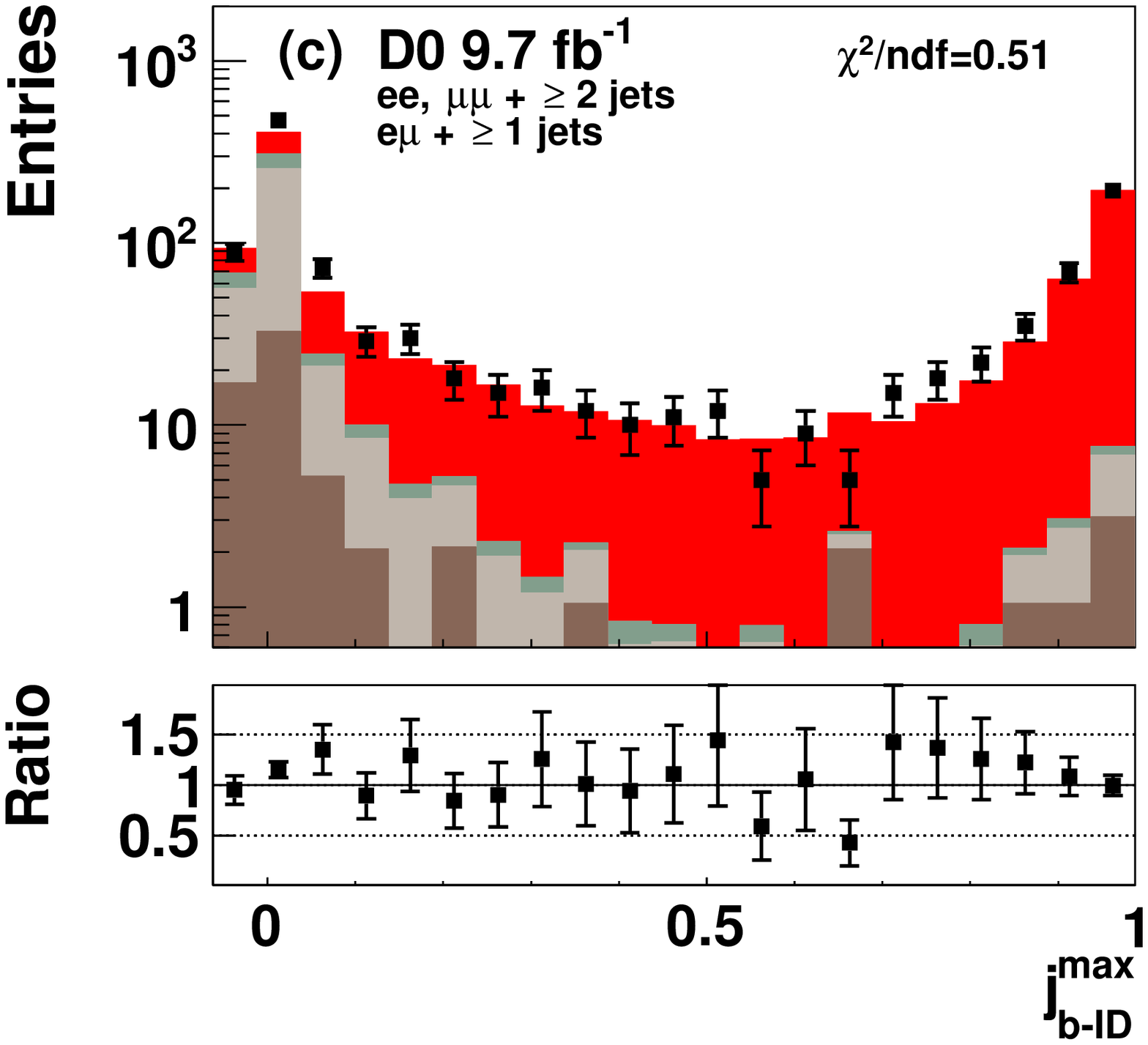}
   \includegraphics[width=0.675\columnwidth,angle=0]{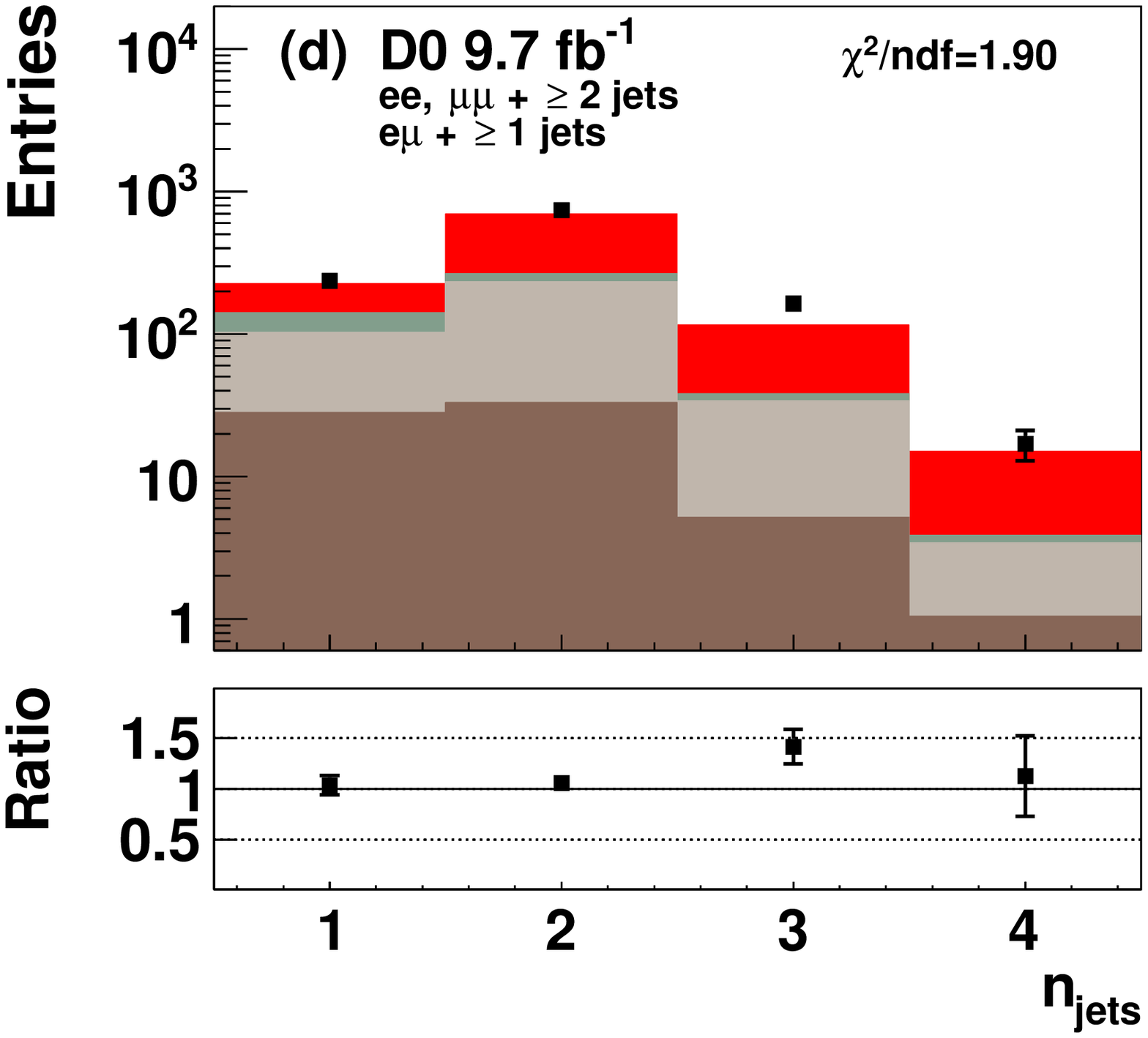}
   \includegraphics[width=0.675\columnwidth,angle=0]{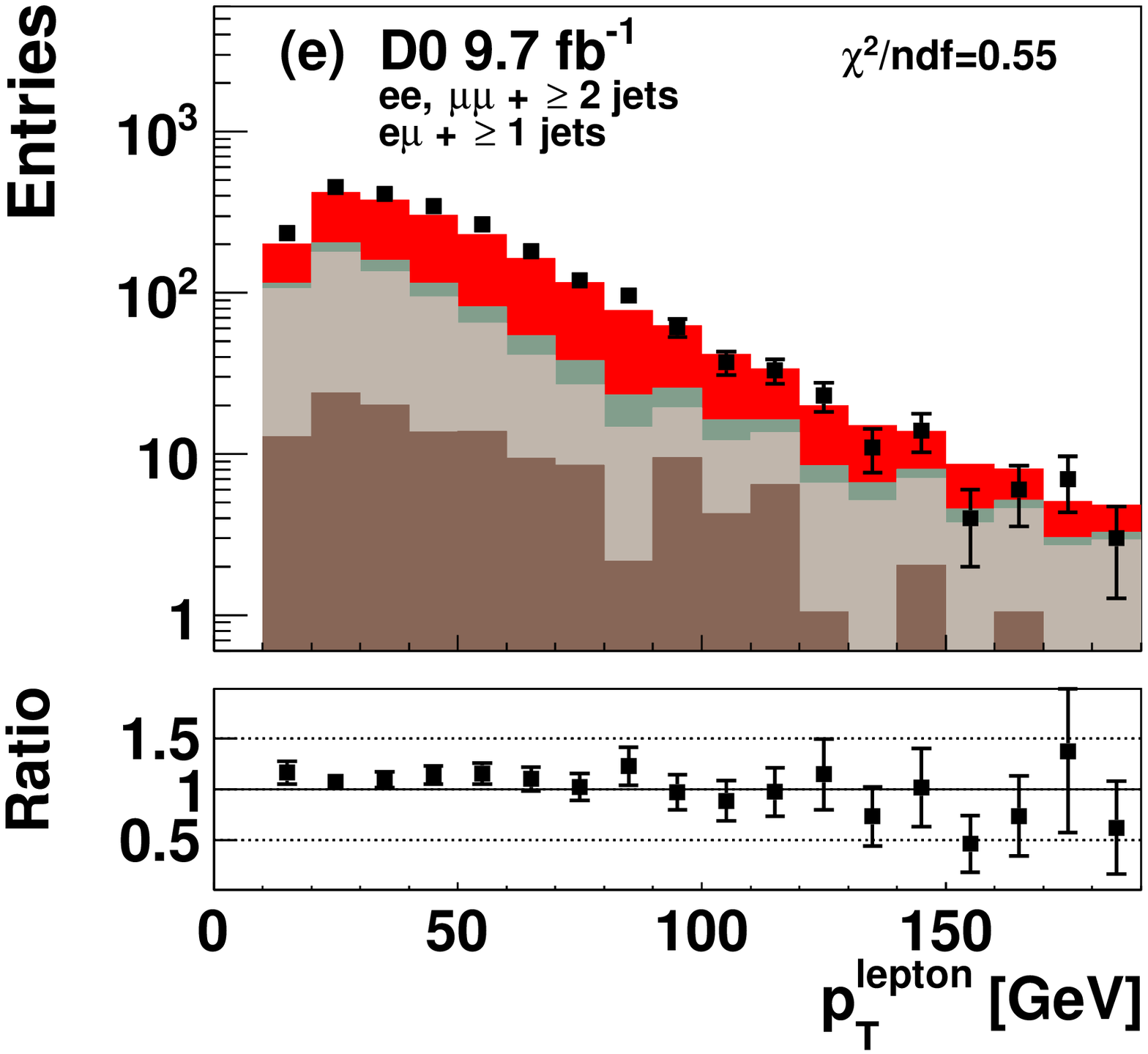}
  \end{center}
\caption{Distributions of (a) the scalar sum, $H_{T}$, of the $p_T$ values of the leading lepton and leading and second-leading jets, (b) $\met$, (c) \mmax, (d) the number of jets, and (e) lepton $p_T$ for \dilep final states with at least one jet in the $e\mu$ and at least two jets in the $ee$ and $\mu\mu$ channels. The data are compared to the sum of predicted contributions from signal and background processes, using the theoretical value of $7.48~\mathrm{pb}$ for the $t\bar{t}$ cross section and $m_t=172.5$ GeV. The highest bin in the histograms includes overflows. The ratios of data to the sum of the signal and all background contributions are shown in the panels below the distributions. Only statistical uncertainties of the data are shown and the $\chi^2/\mm{ndf}$ values only take statistical uncertainties into account.} \label{fig:dilepton_control}
\end{figure*}

\subsection{Determination of the $\bm{\ell \ell}$ sample composition}
The main backgrounds in the dilepton final state originate from $Z/\gamma^{*} \rightarrow$ \dilep instrumental backgrounds, and diboson production ($WW$, $WZ$, $ZZ$). The $Z/\gamma^{*} \rightarrow$ \dilep and diboson backgrounds are evaluated from MC as described in Sec.~\ref{toc:backgroundModel_ll}. We use a mixture of MC and data-driven approaches for the instrumental background determination. Similarly to the \ljets channel, the normalization of events with jets misidentified as electrons is estimated directly from data using the matrix method separately for the $ee$ and $e\mu$ channels \cite{leptonicAfbdilepton}. The estimation of the instrumental background from events with jets producing muons is based on events with two leptons of the same charge in the $\mu\mu$ and $e\mu$ channels, where for the latter the contribution from misidentified electrons is subtracted beforehand. The electron misidentification rate for the matrix method is derived from an orthogonal data sample by requiring that the two leptons have the same charge. This sample is selected applying the same selection criteria as for \ttbar events, but the final selection on $H_T$ is replaced with requiring \met significance $<15$ to avoid contribution from $W \rightarrow e\nu +$ jets events. The remaining contribution of the instrumental background is small, and we combine the MJ and \wjets components to reduce the statistical uncertainty on the background estimate. The shape of the MJ template is derived using a looser electron selection of only $e\mu$ events and employed for all \dilep channels. Since the MJ contribution in the $ee$ and $\mu\mu$ sample is small the difference in shape is not significant. 

The yields, after applying the described selection, are given in Table \ref{tab:yields_dilepton} for the individual channels.

Figure \ref{fig:dilepton_control} demonstrates the quality of the modeling of the selected events in the \dilep sample with the background and signal contributions, using a theoretical inclusive $t\bar{t}$ cross section of $7.48~\mathrm{pb}$ \cite{mochUwer} and $m_t = 172.5$ GeV. The slight disagreement in the \met distribution between data and the sum of signal and background contributions is covered by systematic uncertainties related to the JES and jet $p_T$ resolution.

%
%
%
%
\section{Multivariate measurement techniques}
\label{toc:methods_intro}
The inclusive \ttbar cross section \xsttbar is measured using the different MVA techniques introduced in Sec.\ \ref{toc:strategy}. We use different discriminant output distributions of decision trees to separate the signal from the background for the \lplus and \dilep final states. To construct the MVA, the event sample is subsequently split into smaller samples until each event is placed in one of a set of distinct nodes. At each splitting point the separation is optimized by employing training samples for the signal and background contributions. The output or discriminant value provides the probability of an event to be signal. In the case of the \topoM applied in the \lplus channel we use each individual background contribution in the training process and verify that there is no bias due to overtraining of the method. We employ a method called ``boosted decision trees with gradients" (BDTG) \cite{bdtgRef}. The BDTG implements additional weights to minimize classification errors in the training sample and improve signal to background separation. To measure the cross section we perform a log-likelihood profile fit of MC simulation templates to the data using a nuisance parameter for every source of systematic uncertainty as described in Sec.\ \ref{toc:fitting_collie}.  

\subsection{MVA methods in the $\bm{ \ell+}$jets channel}
\label{toc:methods_intro_topo}
Events in the \ljets channel are separated into six different samples according to the lepton type and the number of jets, $n_\mm{jet} = 2, 3, \geq 4$. We studied a further separation according to the number of $b$-tagged jets which gave increased systematic uncertainties and was not used for the final measurement. To build a discriminant, a total of 50 variables were analyzed. The individual distributions are verified to have a good modeling of the data by the MC by means of a Kolmogorov-Smirnov test \cite{KStest} and a $\chi^2$ test. We exclude all variables with poor modeling of the data. One variable among those studied, namely the maximal MVA $b$-ID value of all jets in the event, \mmax, shows only moderately good modeling of the data by MC [see Figs. \ref{fig:ljets_control}(g)--\ref{fig:ljets_control}(i)]. However, the data are reasonably described when considering systematic uncertainties related to $b$-tagging. The \mmax variable provides good separation power, and hence we use that variable in the discriminant.

Depending on the number of jets in an event, at least 24 variables are selected as input to the \topoM for each jet multiplicity and lepton type bin (see the Appendix for a detailed variable description). In particular we include various MVA $b$-ID discriminants into the combined MVA discriminant, which allows superior signal-to-background separation, as discussed in the following paragraph. All selected variables are defined in the Appendix. Adding more variables has a negligible effect on the signal-to-background separation of the discriminant. 

\begin{figure}[ht]
  \begin{center}
  \includegraphics[width=0.975\columnwidth,angle=0]{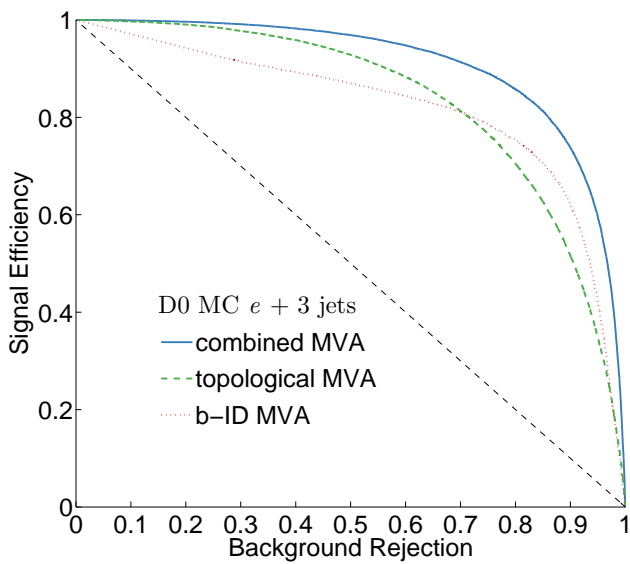}
  \end{center}
    \caption{Signal efficiency vs background rejection for different MVA choices for the $e + 3$ jets sample. For details, see text in Sec.\ \ref{toc:methods_intro}. The dashed line from $(0,1)$ to $(1,0)$ is shown for reference.} \label{fig:ROC_example}
\end{figure}

Figure \ref{fig:ROC_example} shows the separation of signal from background events in the \ljets decay channel using as an example events with exactly three jets in the \eplus channel. In this figure we show the performance of the \topoM discriminant compared to the performance of a similar MVA discriminant that does not include the \mmax input variable (\topoMnoB). For comparison we also show the performance of the MVA solely based on \mmax. The \bidM has a higher signal and lower background efficiency than the \topoMnoB for a signal efficiency less than about 80\%. Above this point the \topoMnoB surpasses the \bidM. Compared to these two MVA methods the \topoM shows superior behavior, with the area under the curve increased by $6-10$\%. 

\begin{figure*}[htb]
\begin{center}
    \includegraphics[width=0.675\columnwidth,angle=0]{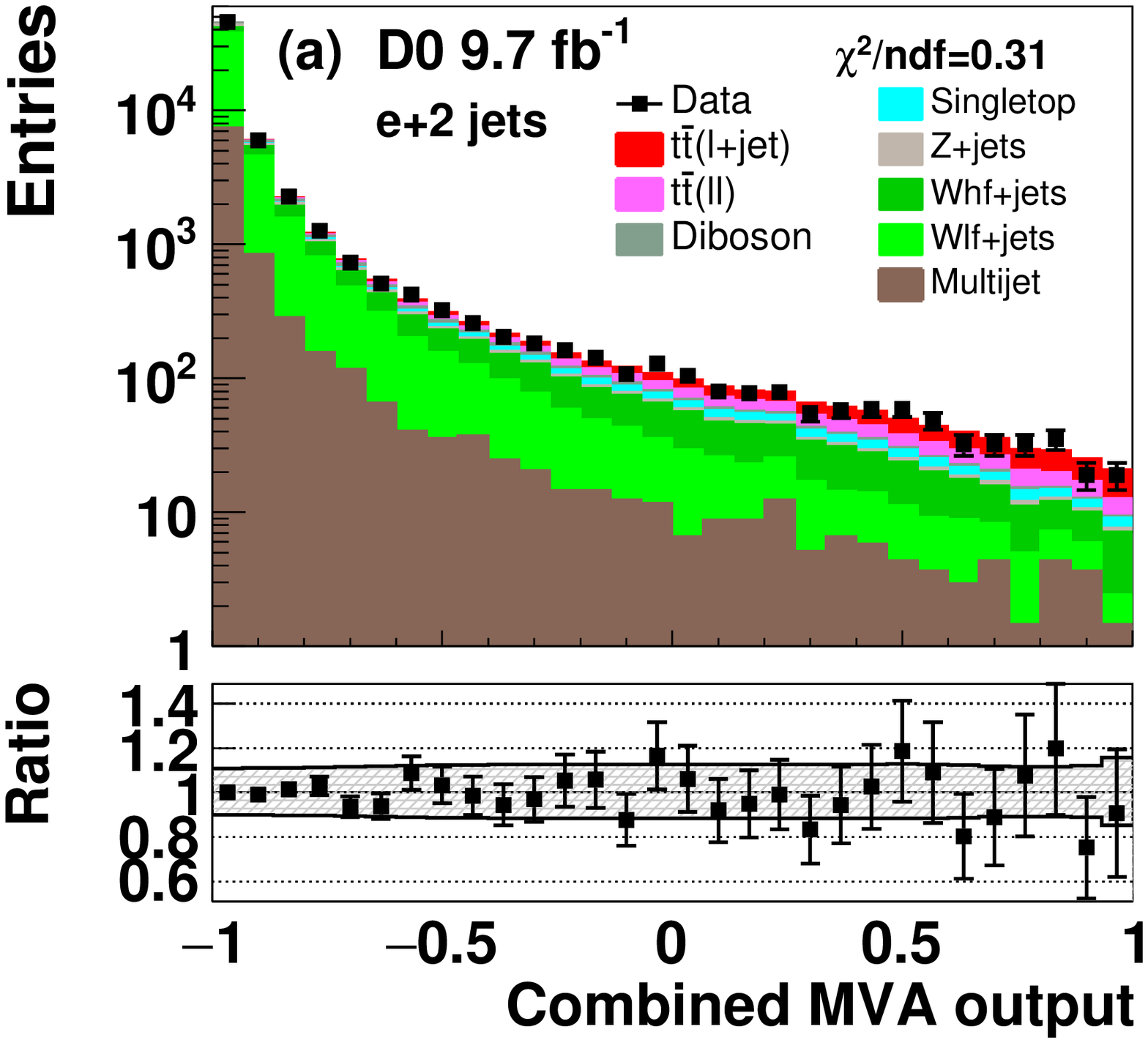}
    \includegraphics[width=0.675\columnwidth,angle=0]{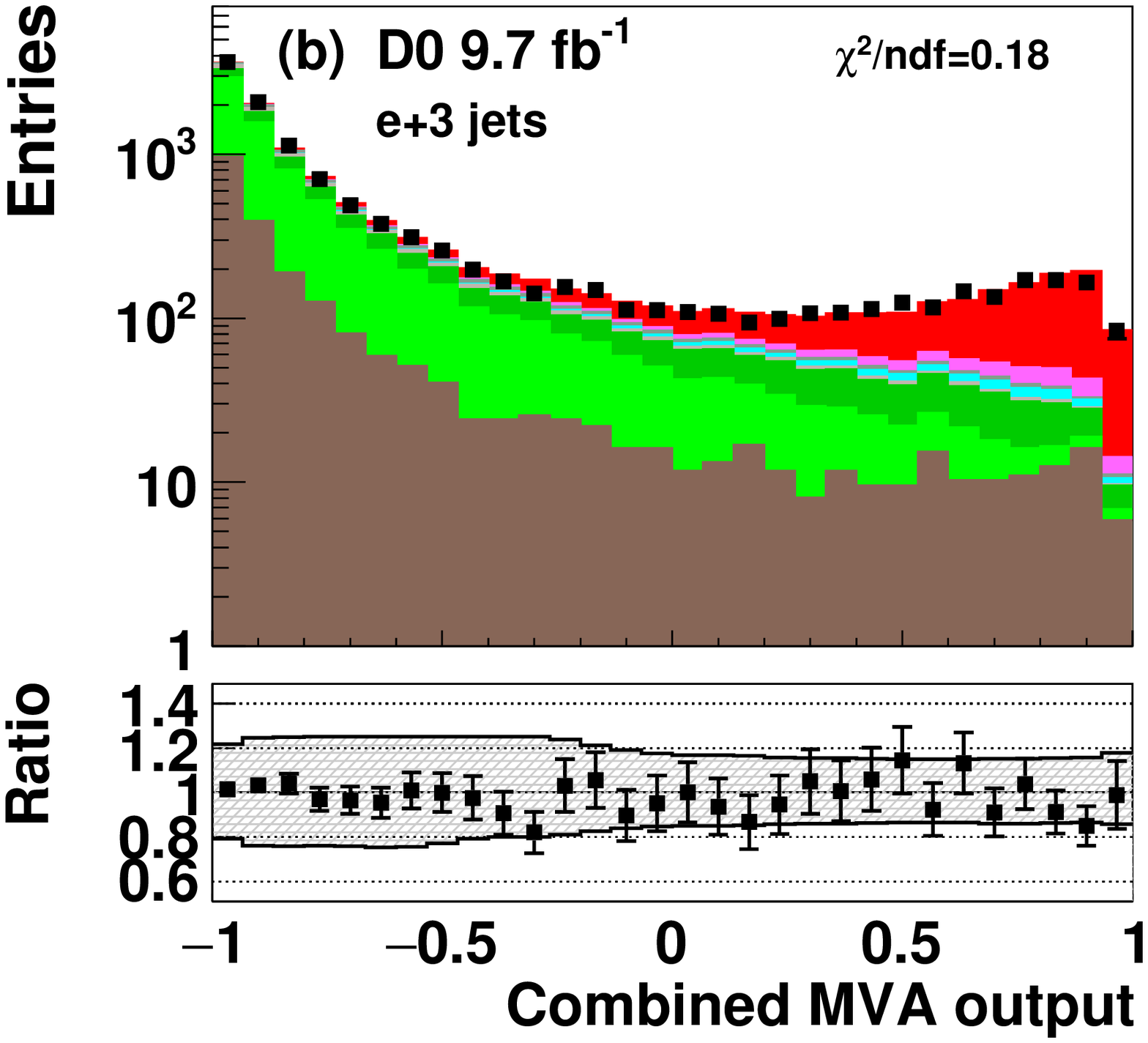}
    \includegraphics[width=0.675\columnwidth,angle=0]{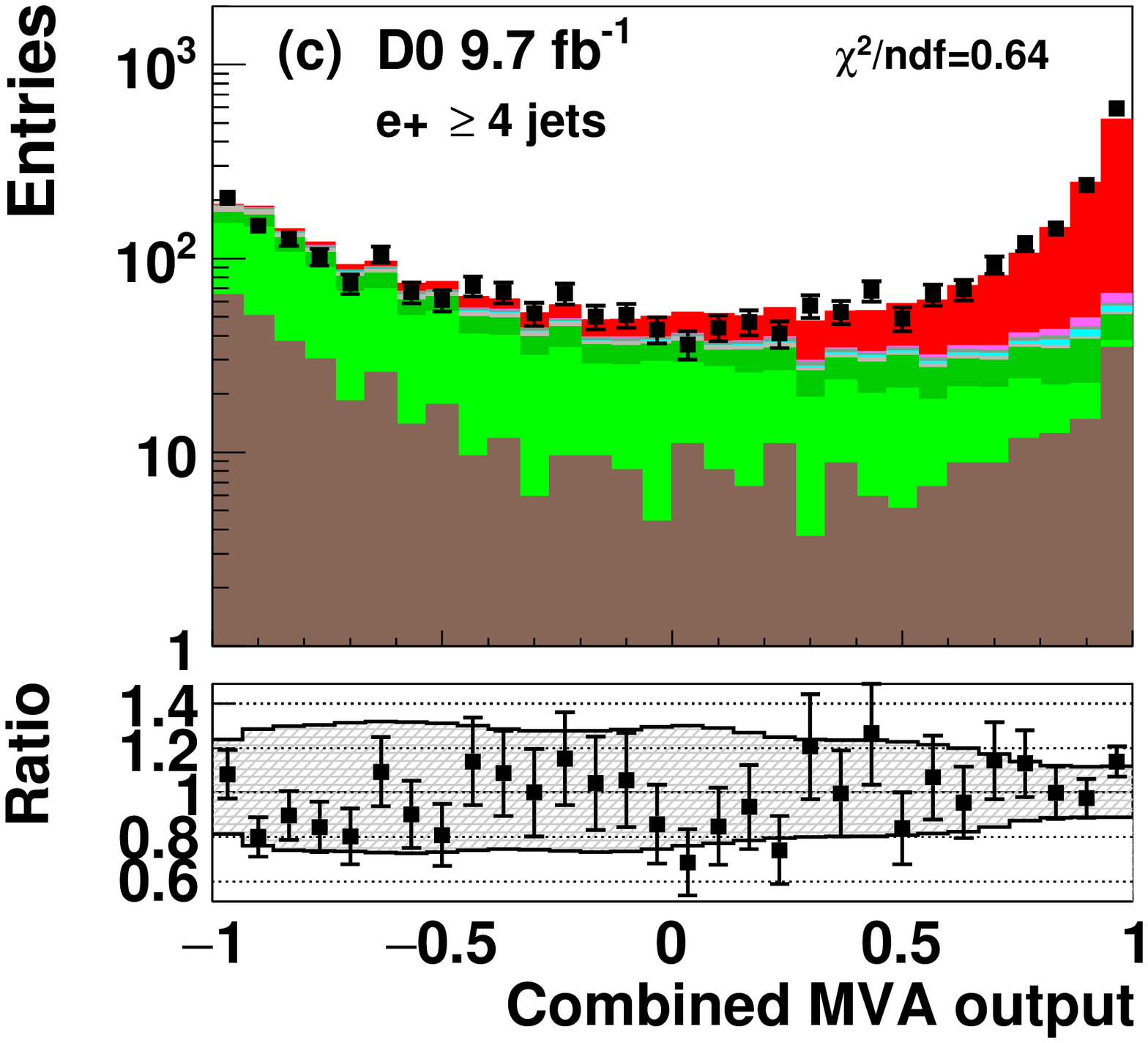}
    \includegraphics[width=0.675\columnwidth,angle=0]{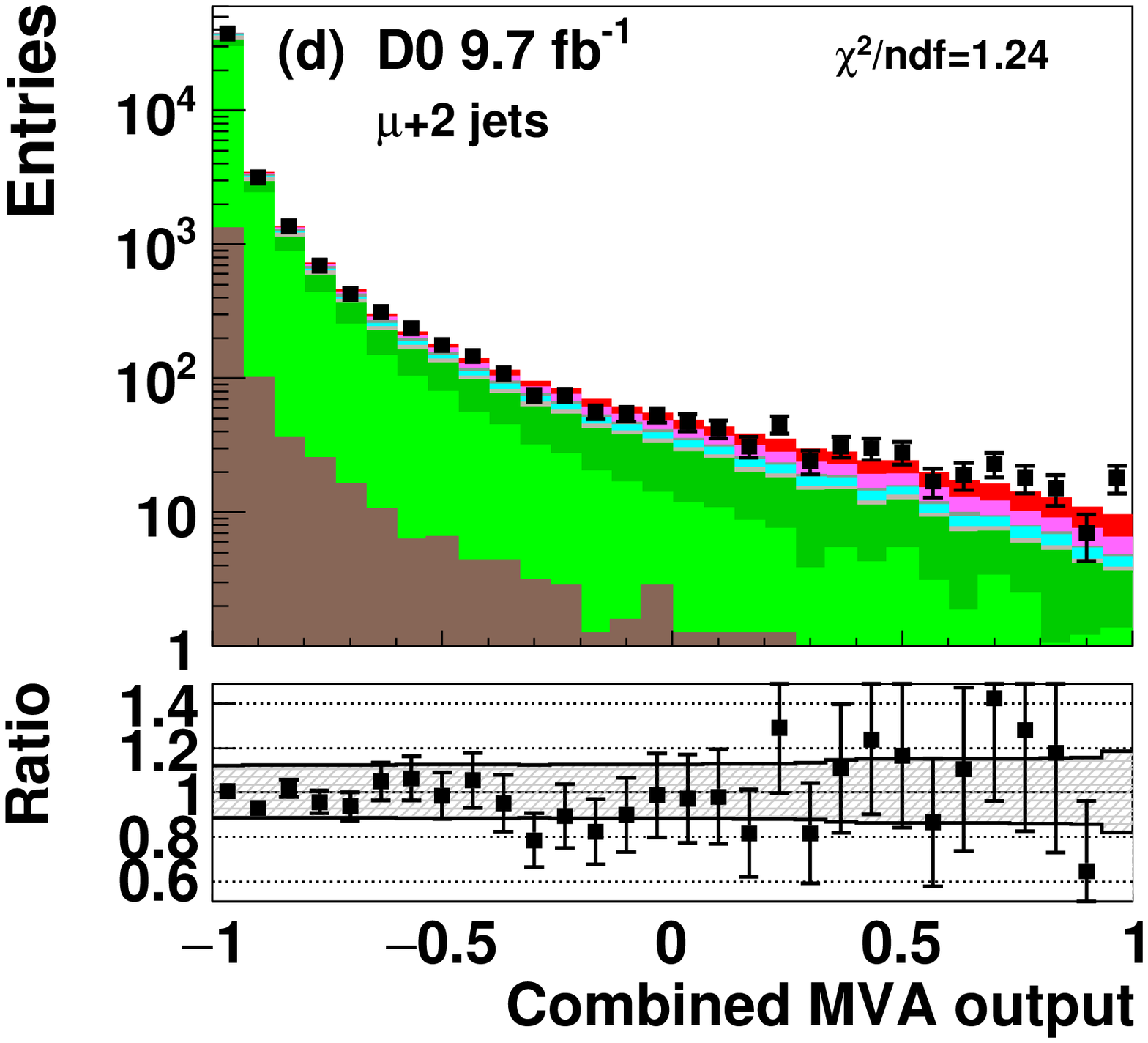}
    \includegraphics[width=0.675\columnwidth,angle=0]{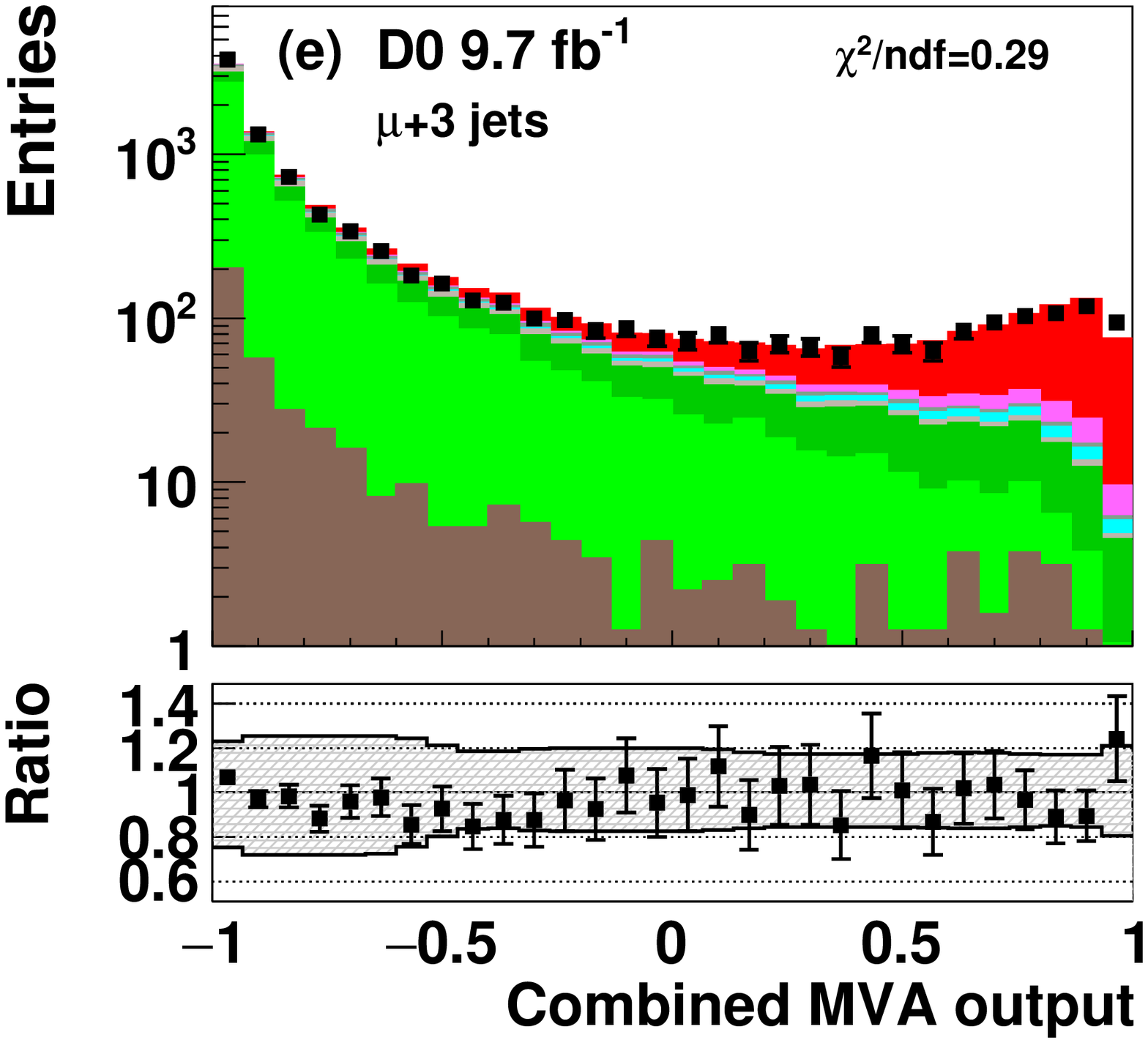}
    \includegraphics[width=0.675\columnwidth,angle=0]{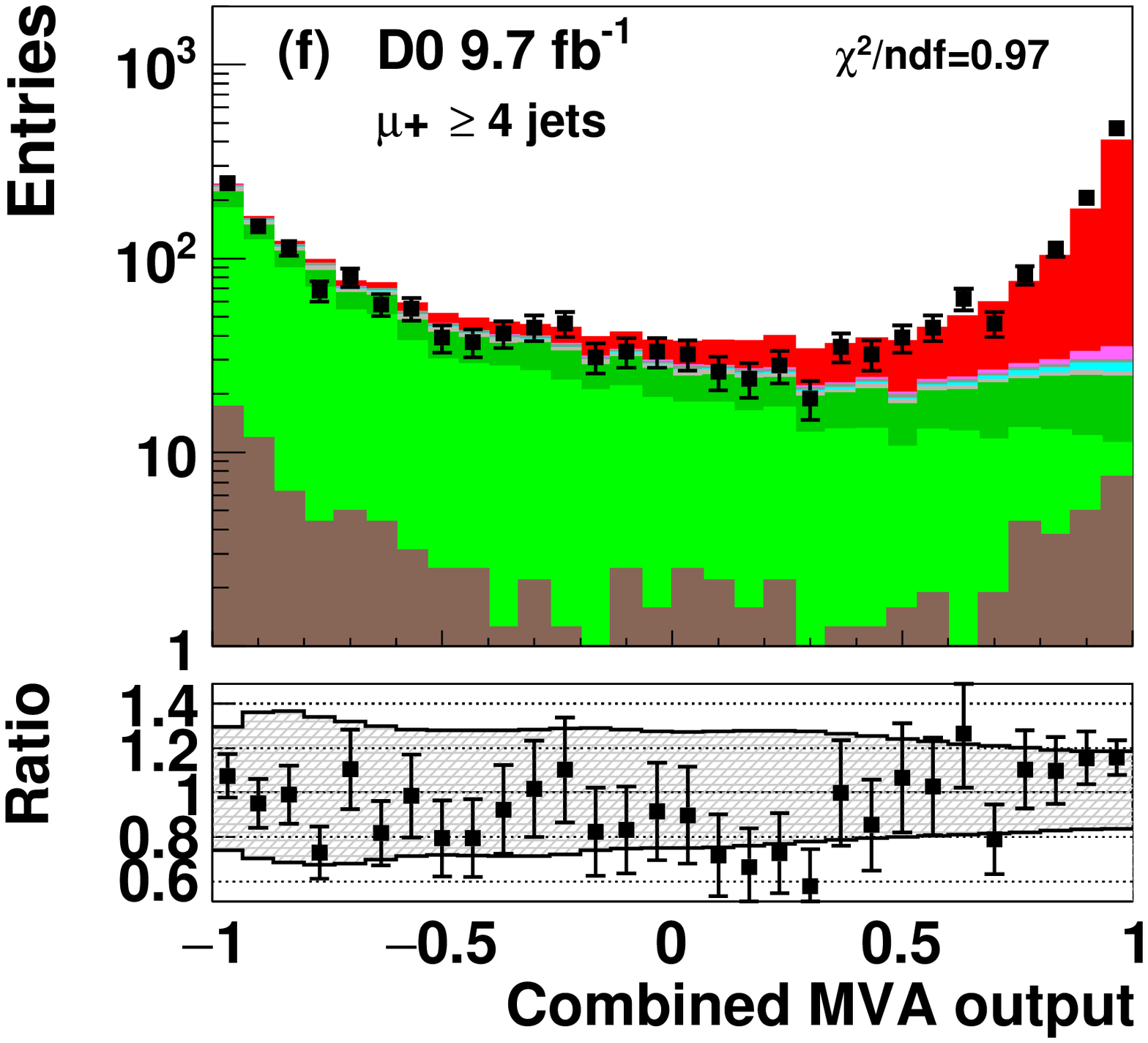}
\end{center}
\caption{Pre-fit output distributions of the combined MVA discriminant using the theoretical \ttbar cross section and $m_t=172.5$ GeV for the $e + $ two, three or four or more jets [panels (a) -- (c)], and for the $\mu + $ two, three or four or more jets [panels (d) -- (f)]. Statistical uncertainties of the data are shown and the pre-fit systematic uncertainties are indicated by the hashed band in the bottom panel of the histogram. The $\chi^2/\mm{ndf}$ values take statistical and systematic uncertainties into account.}
\label{fig:prefit_topo_emujets}
\end{figure*}

%
%
%
%
Figure \ref{fig:prefit_topo_emujets} shows the pre-fit MVA output distributions of the combined MVA method using a theoretical \ttbar cross section of $7.48$ pb. 

\subsection{MVA ${\bm b}$-jet method in the $\bm{ \ell \ell}$ channel}
\label{toc:methods_intro_mvab}
We measure the \ttbar production cross section in the \dilep channel using \mmax to separate signal from background. Events in the dilepton channel are separated into samples according to the lepton type and the number of jets. Due to the small background contribution and the size of the signal contribution in the dilepton channel, the separation provided by the \mmax is sufficiently good and the \topoM was not employed for the \dilep channel.

\begin{figure*}[htb]
\begin{center}
    \includegraphics[width=0.675\columnwidth,angle=0]{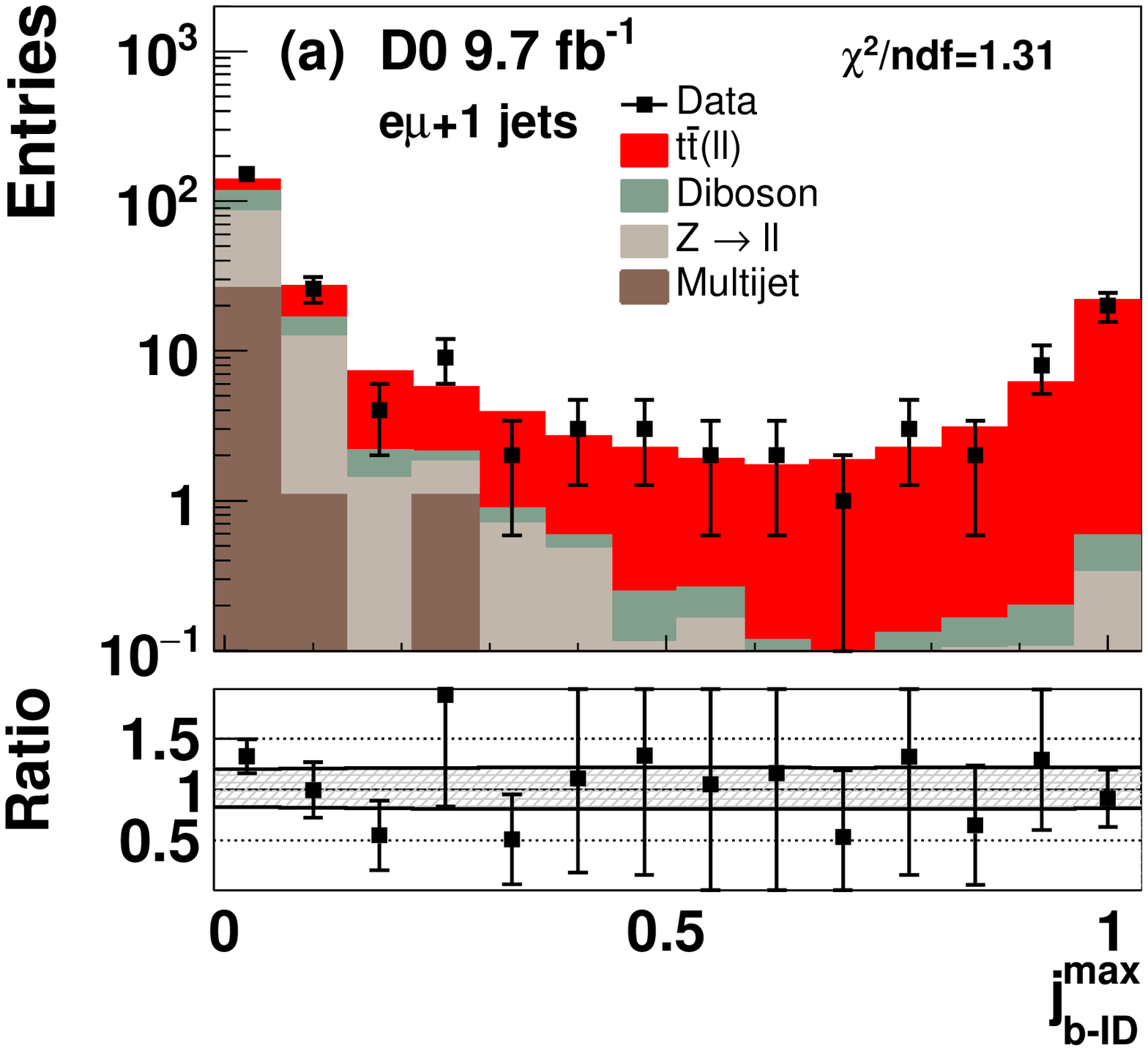}
    \includegraphics[width=0.675\columnwidth,angle=0]{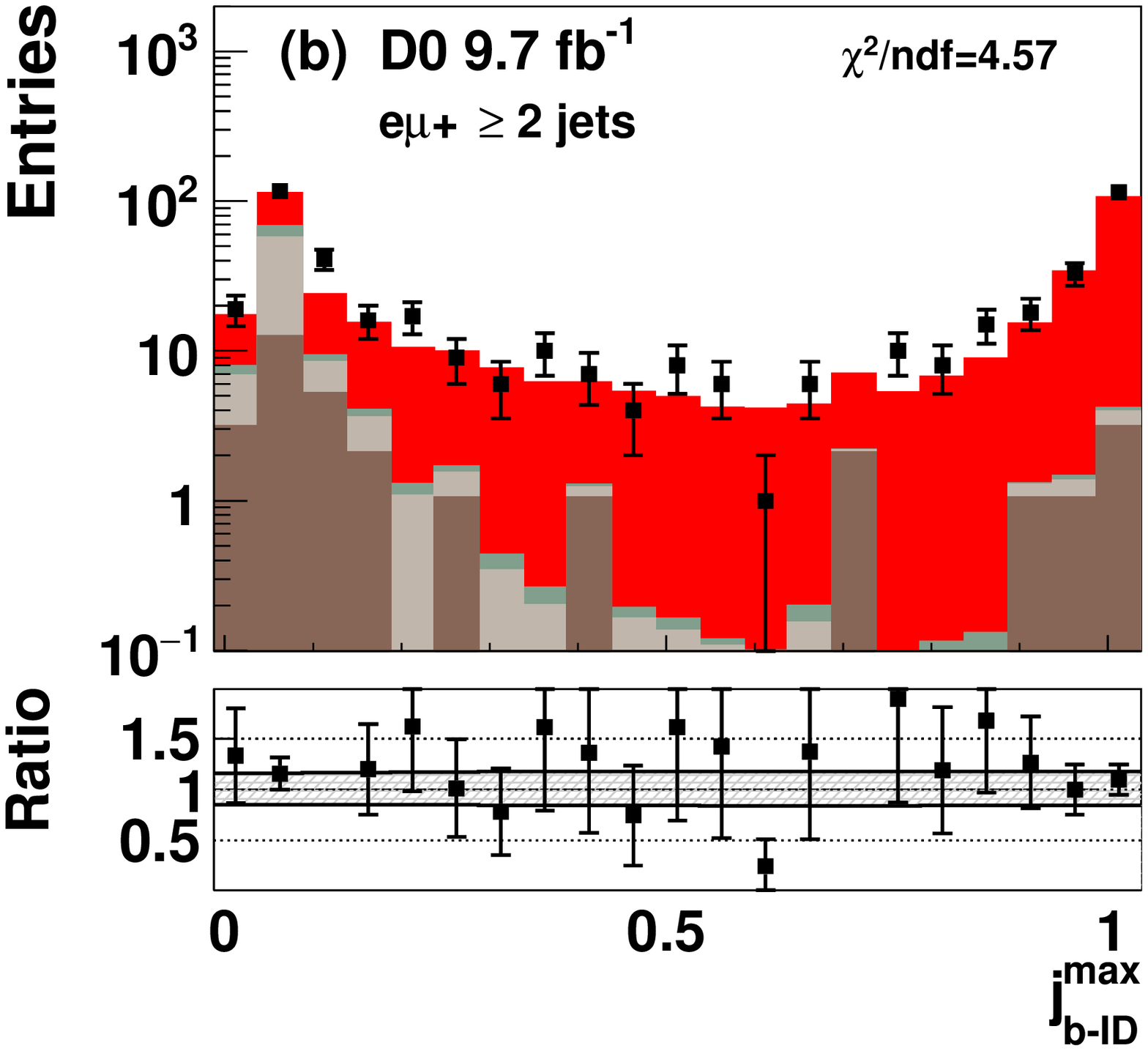}
    \includegraphics[width=0.675\columnwidth,angle=0]{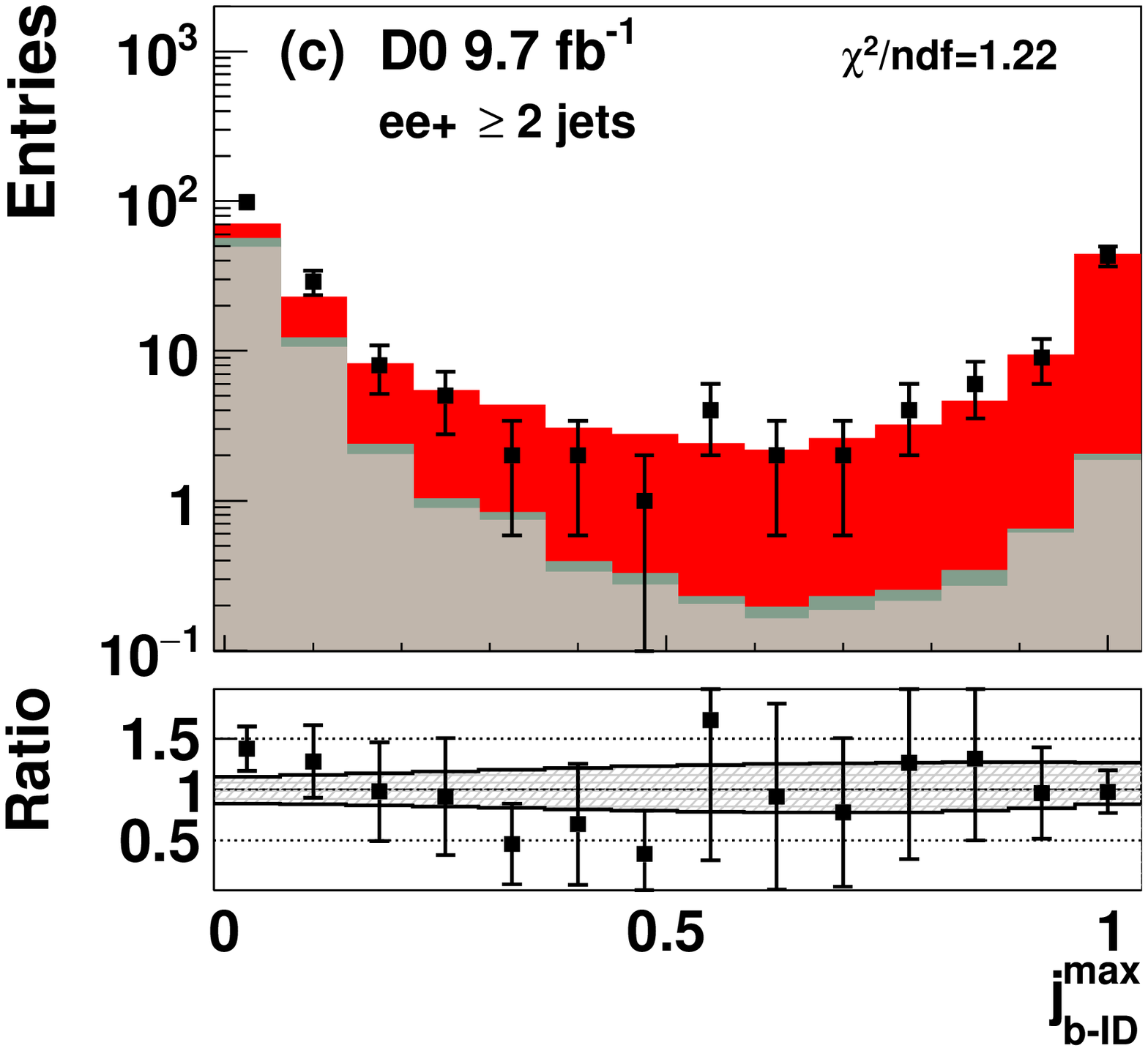}
    \includegraphics[width=0.675\columnwidth,angle=0]{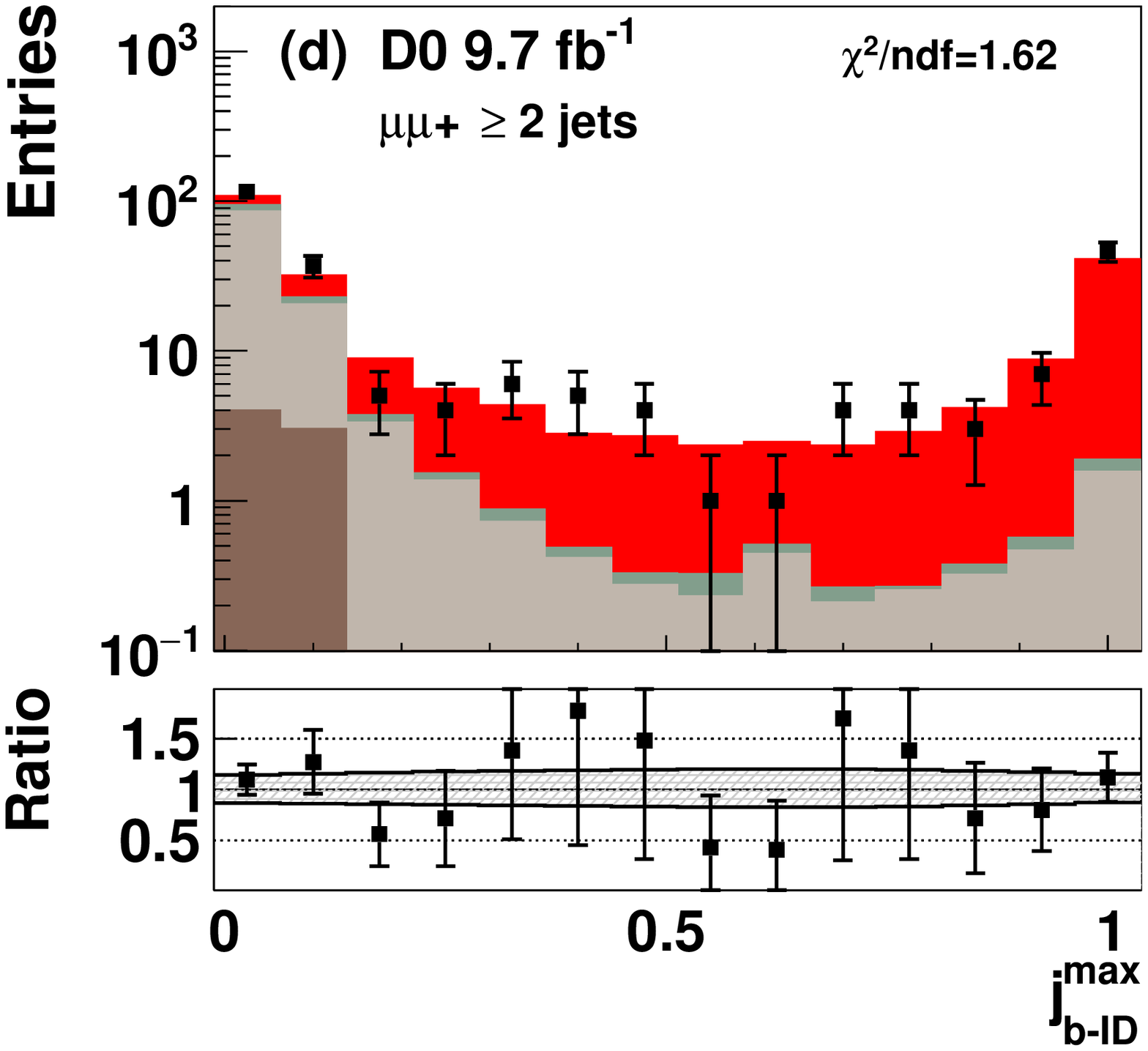}
\end{center}
\caption{The pre-fit MVA output distributions (using the theoretical \ttbar cross section and $m_t=172.5$ GeV) of the \bidM given by the jet with the maximum value, \mmax, for (a) $e\mu$ events with exactly one jet, (b) $e\mu$ events with at least two jets, (c) $ee$ events with at least two jets, and (d) $\mu\mu$ events with at least two jets. Statistical uncertainties of the data are shown and the pre-fit systematic uncertainties are indicated by the hashed band in the bottom panel of the histogram. The $\chi^2/\mm{ndf}$ values take statistical and systematic uncertainties into account.}
\label{fig:prefit_mva_dilep}
\end{figure*}

The MVA output distributions in the dilepton channel allow one to distinguish between \ttbar events dominantly located at high output values and the most dominant \zplus background contribution typically located at low output values. For the $e\mu$ channel we split the sample into sub-samples with exactly one and $\ge$ two jets, whereas for $ee$ and $\mu\mu$ only events with two or more jets are used. The \mmax distributions are shown in Fig.\ \ref{fig:prefit_mva_dilep}. A theoretical \ttbar cross section of 7.48 pb is used \cite{mochUwer}.

%
%
%
%
%
\section{Cross section determination}
\label{toc:fitting_collie}
As shown in Figs. \ref{fig:prefit_topo_emujets} and \ref{fig:prefit_mva_dilep}, the MVA output distributions for the \ljets and \dilep channels allow the discrimination of the \ttbar signal contribution and the most dominant background sources. We perform a simultaneous fit of MC templates to the data using the software package \collie (A Confidence Level Limit Evaluator) \cite{collie} to determine the inclusive \ttbar cross section \xsttbar. 

The combined likelihood includes prior probability densities on systematic uncertainties $\pi(\vec{\theta})$, and is based on the product of likelihoods for the individual channels, each of which is a product over bins of histograms of a particular analysis channel:
\begin{equation}
 \label{eqn:xsecDef}
{\cal L}(\vec{s},\vec{b}|\vec{n},\vec{\theta}) \times \pi(\vec{\theta}) = \prod_{i=1}^{N_C} \prod_{j=1}^{N_{\mm{bins}}} \mu_{ij}^{n_{ij}} \dfrac{e^{-\mu_{ij}}}{n_{ij}!} \times \prod_{k=1}^{n_\mm{sys}} e^{-\theta^{2}_{k}/2}.
\end{equation}
The first product is over the number of channels ($N_C$) and the second product is over histogram bins containing $n_{ij}$ events, binned in ranges of the final discriminants used for the individual analyses. The predictions for the bin contents are $\mu_{ij} = s_{ij}(\vec{\theta}) + b_{ij}(\vec{\theta})$ for channel $i$ and histogram bin $j$, where $s_{ij}$ and $b_{ij}$ represent the expected signal and background in the bin. The predictions $\mu_{ij}$ include effects from limited detector resolution and efficiency, including those from trigger and selection efficiencies and for the kinematic and geometric acceptance.

Systematic uncertainties are parametrized by the dependence of $s_{ij}$ and $b_{ij}$ on $\vec{\theta}$. Each of the $n_\mm{sys}$ components of $\vec{\theta}$, $\theta_k$, corresponds to a single independent source of systematic uncertainty scaled by its standard deviation, and each parameter may affect the predictions of several sources of signal and background in different channels, thus accounting for correlations. For the combination of the \topoM in the \ljets channel with the \bidM in the \dilep channel using \collie systematic uncertainties are either assumed to be fully correlated or not correlated (see Sec.\ \ref{toc:xsec_sys}).

\collie models nuisance parameters using a Gaussian prior probability density function specified by $\pm1$ standard deviation of the systematic uncertainty in question [see Eq.\ (\ref{eqn:xsecDef})]. For asymmetric uncertainties, two half-Gaussian functions model separately the positive and negative parts of the nuisance parameters. In the log-likelihood profile fit, the nuisance parameters and the cross section are simultaneously fitted. Hence, sources of systematic uncertainties not only contribute to the final cross section uncertainty, but also shift the fitted cross section value. Since the simultaneous fit is constrained by data it also provides a reduction of the impact of the different systematic uncertainty sources.

The central value of the \collie fit provides a scale factor that is applied to the expected number of signal events using the theoretical \ttbar cross section. The scaled number of signal events, $N^{\mm{signal}}$, is preferred by the data and the systematic uncertainties. To determine $\sigma_{t\bar{t}}$ for the full phase space of \ttbar production we correct $N^{\mm{signal}}$ for the detector efficiency and acceptance, the branching ratio and the integrated luminosity.

%
%
%
%
\section{Systematic Uncertainties}
\label{toc:xsec_sys}
Systematic uncertainties are assessed by varying the values of a specific parameter in the modeling of the data, and determining the effect on the distributions or MC templates of the \topoM or the \bidM. Compared to the earlier D0 measurement \cite{Publ54_xsec} we employ a more refined strategy for systematic uncertainties including the newly added hadronization uncertainty. Unless otherwise stated, the magnitude of the parameter modifications is obtained from alternative calibrations of the MC simulation. Each of the modified MVA distributions is used to determine the effect of systematic uncertainties. As described in Sec.\ \ref{toc:fitting_collie} all nuisance parameters are fitted simultaneously with the nominal MVA distributions to measure the \ttbar production cross section. Systematic uncertainties are constrained by the data and are minimized since we use the full shape information of the MVA templates. A further reduction of correlated systematic uncertainties is achieved when combining the \ljets and \dilep decay channels, since systematic uncertainties are then cross-calibrated.

In total we assign 39 (\ljets channel), 37 (\dilep channel), and 53 (combination) individual systematic uncertainties as discussed below for the decay channels. The pre-fit systematic uncertainties are summarized in Table \ref{tab:pre_syst_uncorr}, whereas the post-fit effects of the systematic uncertainties are summarized in Table \ref{tab:syst_uncorr}. We group systematic uncertainties addressing a similar object, e.g.\ jet-related ones, into a combined source of systematic uncertainty.

\begin{table}[tp]%
\begin{center}
\caption {\label{tab:pre_syst_uncorr} Sources of grouped pre-fit systematic uncertainties for the \ttbar cross section measurement assuming the theoretical \ttbar cross section of 7.48 pb \cite{mochUwer} and $m_t = 172.5$ GeV. The systematic uncertainty in pb from each source on the inclusive cross section is given for the \ljets and the \dilep channels. The column denoted as ``Type" refers to a systematic uncertainty affecting the shape and normalization $S$ or only the normalization $N$ of a MVA distribution. The numbers presented for shape-dependent uncertainties represent averages across the entire set of distributions and all samples. The term ``n.a." is used where a systematic uncertainty is not applicable for a given decay channel.}
\end{center}
\begin{ruledtabular}
\begin {tabular}{llclc}
\multicolumn{1}{l}{Source of uncertainty} &
\multicolumn{1}{c}{$\delta_{\ell \mm{+ jets}}$, pb} & \multicolumn{1}{c}{Type} & 
\multicolumn{1}{c}{$\delta_{\ell \ell}$, pb} & \multicolumn{1}{c}{Type} \\ \hline

\textit{Signal modeling} 				&				& \T \\
~~~Signal generator & 						     $\pm 0.86$	 	& $S$ & $\pm 0.28$     &  $S$  \T \\
~~~Hadronization & 				         $\pm 0.59$		 & $S$ & $\pm 0.29$     &  $S$ \T \\
~~~Color reconnection & 			         $\pm 0.21$	 	& $S$ & $\pm 0.08$     &  $S$ \T \\
~~~ISR/FSR variation &				$\pm 0.07$		 & $S$ & $\pm 0.04$     &  $S$ \T \\
\textit{PDF} & 						$\pm 0.20$		 & $S$ & $\pm 0.07$     &  $S$ \T \\
\textit{Detector modeling}	&	                  &                \T\\
~~~Jet modeling \& ID & 			         $\pm 0.33$		 & $S$ & $\pm 0.34$	&  $S$ \T \\
~~~$b$-jet modeling \& ID & 			$\pm 0.19$		 & $S$ & $\pm 0.56$     &  $S$ \T \\
~~~Lepton modeling \& ID	 &		         $\pm 0.23$	 	& $S$ & $\pm 0.31$     &  $N$ \T \\
~~~Trigger efficiency& 			         $\pm 0.06$		 & $N$ & $\pm 0.07$     &  $N$ \T \\
~~~Luminosity\footnote{To prevent constraining the luminosity uncertainty by data, we do not assign the luminosity uncertainty to the \wjets and multijet contribution.} & 				         $\pm 0.32$	 	& $N$ & $\pm 0.32$   &  $N$ \T \\
\textit{Sample Composition} 		&		&  		&  &\T \\
~~~MC cross sections &		                $\pm 0.02$ 	 & $N$ & $\pm 0.12$     &  $N$ \T \\
~~~$Z / W$ $p_T$ reweighting &		$\pm 0.03$	 	& $S$ & $\pm 0.28$    &  $S$ \T \\
~~~Multijet contribution &			          $\pm 0.26$  	& $S$ & $\pm 0.13$& $S$ \T \\
~~~\zplus SF			& 			$\pm 0.09$ 	 & $S$ &$\pm 0.12$ &$S$ \T \\
~~~\wplus HF SF & 					$\pm 0.41$ 	 & $S$  &   $\mm{n.a.}$ & $\mm{n.a.}$\T \\
~~~\wplus LP SF &  					$\pm 0.16$ & $S$  &  $\mm{n.a.}$ & $\mm{n.a.}$\T \\
\textit{MC statistics} 				&	$\pm 0.01$ &  $\mm{S}$ & $\pm 0.10$  &  $S$ \T \\ 
\end {tabular}
\end{ruledtabular}
\end {table}

\begin{table*}[tp]%
\caption {\label{tab:syst_uncorr} Sources of grouped post-fit systematic uncertainties for the \ttbar cross section measurement. Owing to the complexity of the correlations among the systematic uncertainties, we only show here the symmetrized uncertainties. For the top quark signal and background contributions we assume $m_t=172.5$ GeV. For each group, the systematic uncertainty on the inclusive cross section is given for the \ljets, \dilep, and combined measurement.  The last column shows the shift in pb in the combined inclusive cross section due to a particular group. A shift of $0.00$ pb indicates a shift of $0.004$ pb or less. The total uncertainty is provided by the nominal fit, when including all individual sources of systematic uncertainties, and denoted as ``central \collie". For comparison only we also provide the combined systematic uncertainty (quadratic sum) of the grouped post-fit systematic uncertainties. Due to correlations between the systematic uncertainties, that value differs from the total systematic uncertainty of the nominal fit.}
\begin{ruledtabular}
\begin {tabular}{lcccc}
\multicolumn{1}{l}{Source of uncertainty} & 
                  \multicolumn{1}{c}{$\delta_{\ell \mm{+ jets}}$, pb} & \multicolumn{1}{c}{$\delta_{\ell \ell}$, pb} & \multicolumn{1}{c}{$\delta_{\mm{comb,}}$, pb} & \multicolumn{1}{c}{Shift, pb} \\ \hline
  \textit{Signal modeling} 	&	            	    & & 	       &\\
  ~~~Signal generator &		 		$\pm 0.21$ &$\pm0.05$& $\pm 0.17$	 &  $+0.08$ \\ 
  ~~~Hadronization & 				$\pm 0.26$ &$\pm0.33$& $\pm 0.25$	 &  $+0.12$ \\ 
  ~~~Color reconnection & 			$\pm 0.08$ &$\pm0.05$& $\pm 0.09$	 &  $+0.02$ \\
  ~~~ISR/FSR variation &				$\pm 0.08$ &$\pm0.04$& $\pm 0.06$	 &  $-0.05$ \\ 
  \textit{PDF} & 						$\pm 0.04$ &$\pm0.03$& $\pm 0.02$	 &  $-0.01$ \\ 
  \textit{Detector modeling}	&	            	    & & 	       &\\
  ~~~Jet modeling \& ID & 		$\pm 0.11$ &$\pm0.08$& $\pm 0.04$	 &  $+0.07$ \\ 
  ~~~$b$-jet modeling \& ID & 	$\pm 0.27$ &$\pm0.26$& $\pm 0.23$	 &  $-0.15$ \\ 
  ~~~Lepton modeling \& ID&	$\pm 0.20$ &$\pm0.26$& $\pm 0.17$	 &  $-0.11$ \\ 
  ~~~Trigger efficiency & 				$\pm 0.32$ &$\pm0.08$& $\pm 0.16$	 &  $+0.01$ \\ 
  ~~~Luminosity & 					$\pm 0.30$ &$\pm0.30$& $\pm 0.27$	 &  $+0.10$ \\ 
  \textit{Sample Composition} 		&	& & 	    	   & \\
  ~~~MC cross sections 				&$\pm 0.07$ &$\pm0.13$& $\pm 0.09$	 &  $+0.01$ \\ 
  ~~~Multijet contribution &			$\pm 0.11$ &$\pm0.02$& $\pm 0.10$	 &  $+0.10$ \\ 
  ~~~\wplus scale factor & 			$\pm 0.21$ &$\pm0.01$& $\pm 0.15$	 &  $-0.50$ \\ 
  ~~~\zplus scale factor &  			$\pm 0.07$ &$\pm0.11$& $\pm 0.12$	 &  $+0.12$ \\ 
  \textit{MC statistics} 		&	        $\pm 0.01$ &$\pm0.01$& $\pm 0.02$	 &  $+0.00$ \\ \hline 
  Total systematic uncertainty (quadratic sum) 	& 		$\pm 0.70$ &$\pm 0.64$&$\pm 0.60$ & \\  
  Total systematic uncertainty (central \collie) & $\pm 0.67$ &$\pm 0.73$&$\pm 0.55$ & \T 
\end {tabular}
\end{ruledtabular}
\end {table*}

\subsection{The $\bm{\ell+}$jets channel}
In the following we describe the sources of systematic uncertainties studied in the \ljets channel. As discussed above, each source of systematic uncertainty yields a modified discriminant distribution, which is parametrized with a nuisance parameter (see Sec.\ \ref{toc:fitting_collie}). We assign an uncertainty on the shape, but not on the normalization, of the \wplus and multijet contribution (see Sec.\ \ref{toc:sampleCompLjets}). In particular the trigger and luminosity uncertainties affecting the normalization are not assigned to the \wplus and multijet contribution, and consequently the luminosity uncertainty cannot be constrained by data.

\subsubsection{Signal modeling}
The effect of an alternative signal model for \ttbar production is estimated by comparing \ttbar events generated with \mcherwig to those from \alpher. Comparing \alppyt to \alpher, we estimate the effect of hadronization uncertainties. Additional uncertainties on signal arise from color reconnection (CR), and initial- and final-state radiation (ISR/FSR) producing additional jets. The effect of CR is determined by comparing identical \alpgen events interfaced to \pythia with two different tunes, Perugia 2011 and Perugia 2011NOCR \cite{perugia}, which either include color reconnection effects (Perugia 2011) or not (Perugia 2011NOCR). The effect of ISR/FSR is determined by modifying the factorization and renormalization scale implemented in the MC. More details can be found in Ref.\ \cite{ljetsMassPRD}.

\subsubsection{Parton distribution functions} 
The uncertainty on the cross sections due to the uncertainty on PDFs is estimated following the procedure of Ref.\ \cite{cteq6m} by reweighting the MC simulation according to each of the 20 pairs of error eigenvectors of the CTEQ6M PDF.

\subsubsection{Detector modeling}
Uncertainties on the modeling of the detector include uncertainties on trigger efficiency, lepton identification, and $b$-quark identification. The identification efficiencies for $b$, $c$, light quarks ($u,d,s$), and gluons in MC simulations are calibrated using dijet data \cite{topmass_sdc}, and variations within the calibration uncertainty are used to determine the systematic uncertainty due to $b$-quark identification. Additional uncertainties arise from track multiplicity requirements on the selected jets in the identification of $b$ quarks.

The measurement of the \ttbar cross section and the subsequent extraction of the top quark pole mass relies on a precise knowledge of normalization uncertainties. Hence, this measurement is the first measurement in D0 employing the reduced systematic uncertainty on the luminosity measurement of 4.3\% \cite{lumi_nim,lumiTM}. We use an auxiliary data sample where no cut is made on the primary vertex position in $z$ to verify that negligible uncertainty arises for the $|z_{\mm{PV}}|< 60$ cm requirement used in this analysis. Other instrumental uncertainties from modeling the detector arise from the calibration of the jet energy, resolution, and efficiency.

\subsubsection{Sample composition} 
\label{toc:sys_sampleComp}
Uncertainties in the composition of the selected events arise from $s_{\mm{fit}}^{\mm{WHF}}$ and $s_{\mm{fit}}^{\mm{WLF}}$ used for \wplus events, the assumed \ttbar cross section, single top quark and diboson cross sections, and the estimate of the contributions from misidentified leptons. As introduced in Sec.\ \ref{toc:sampleComp}, we determine an initial sample composition from a simultaneous fit to the MVA distribution in the \lplustw, \lplusth and \lplusgefo samples. For this initial sample composition we fit $s_{\mm{fit}}^{\mm{WHF}}$ and $s_{\mm{fit}}^{\mm{WLF}}$ assuming an uncertainty of 5\% on the normalization of the \ttbar processes. This initial sample composition is only used to determine a systematic uncertainty on the contribution of \wjets processes. From the fit we derive a systematic uncertainty of $^{+3.5}_{-1.8}$\% on the normalization of the \wlp and $^{+17}_{-23}$\% on the normalization of the $Wc\bar{c}+\mm{jets}$ and $Wb\bar{b}+\mm{jets}$ processes. The statistical uncertainties on these processes are negligible. An uncertainty of 25\% on the \zplus cross section is assigned. The uncertainty on the single top quark cross sections is 12.6\%, taken from varying the factorization and renormalization scales simultaneously by factors of $2$ and $0.5$. An uncertainty of 7\% on the diboson cross sections is assigned, corresponding to half the difference between the LO and NLO predictions. The uncertainties on the single top quark and diboson contributions are labeled ``MC cross sections" in the corresponding tables. The  uncertainties on the data-driven method of estimating MJ background and its kinematic dependences, mostly due to the uncertainties on the selection rates of true and false lepton candidates, are 40\% in the \muplus and 25\% in the \eplus sample (including statistical components). These uncertainties are estimated by varying the contribution of $Wc\bar{c}+\mm{jets}$, $Wb\bar{b}+\mm{jets}$, $Zc\bar{c}+\mm{jets}$ and $Zb\bar{b}+\mm{jets}$ by $\pm 20\%$, the \ttbar contribution by $\pm 10\%$, and then comparing the fake and true signal rates in different variables (quoting the largest difference as additional parametrization uncertainty) \cite{diffXsecPaper}.

\begin{figure*}[htb]
\begin{center}
    \includegraphics[width=0.675\columnwidth,angle=0]{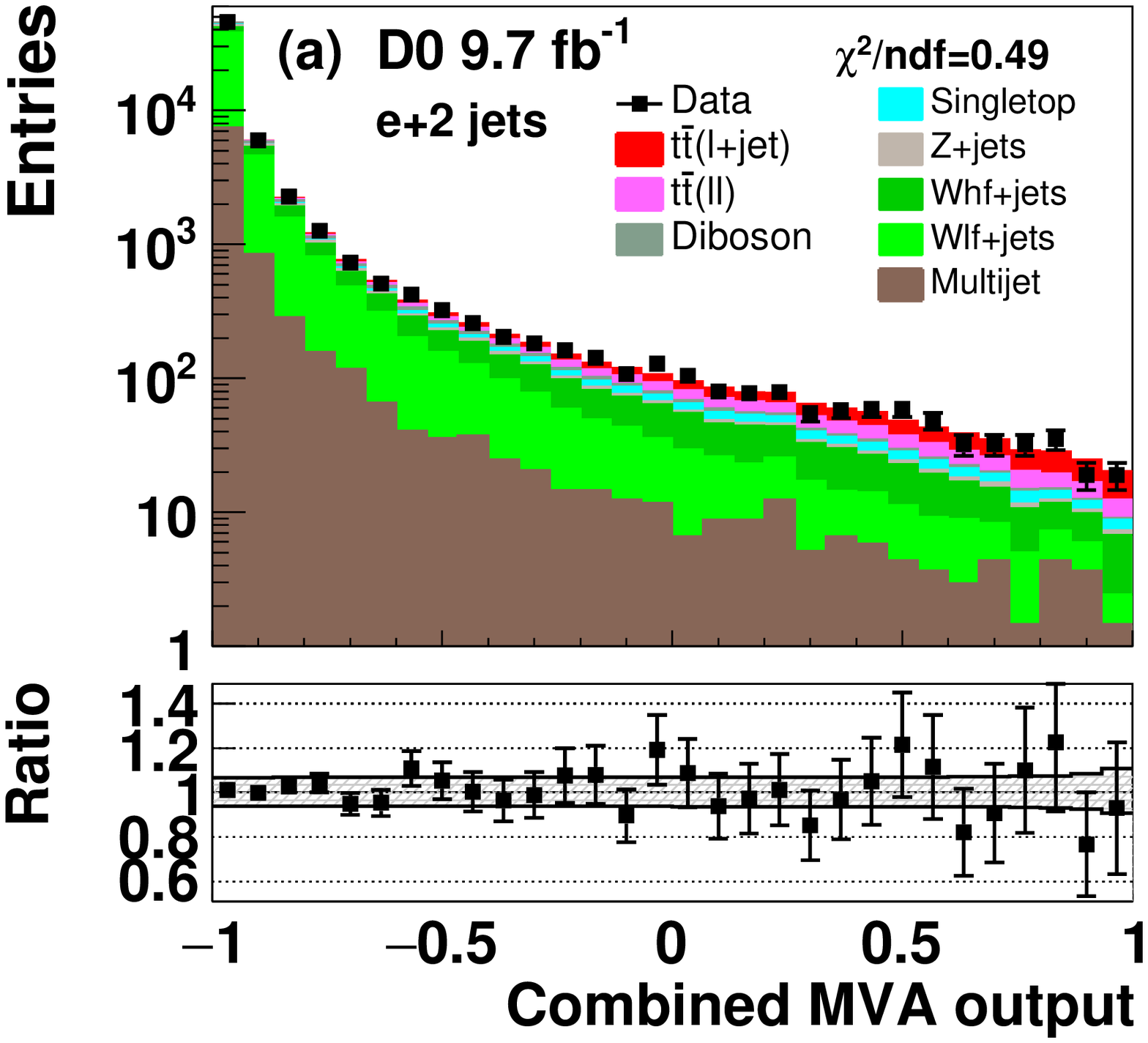}
    \includegraphics[width=0.675\columnwidth,angle=0]{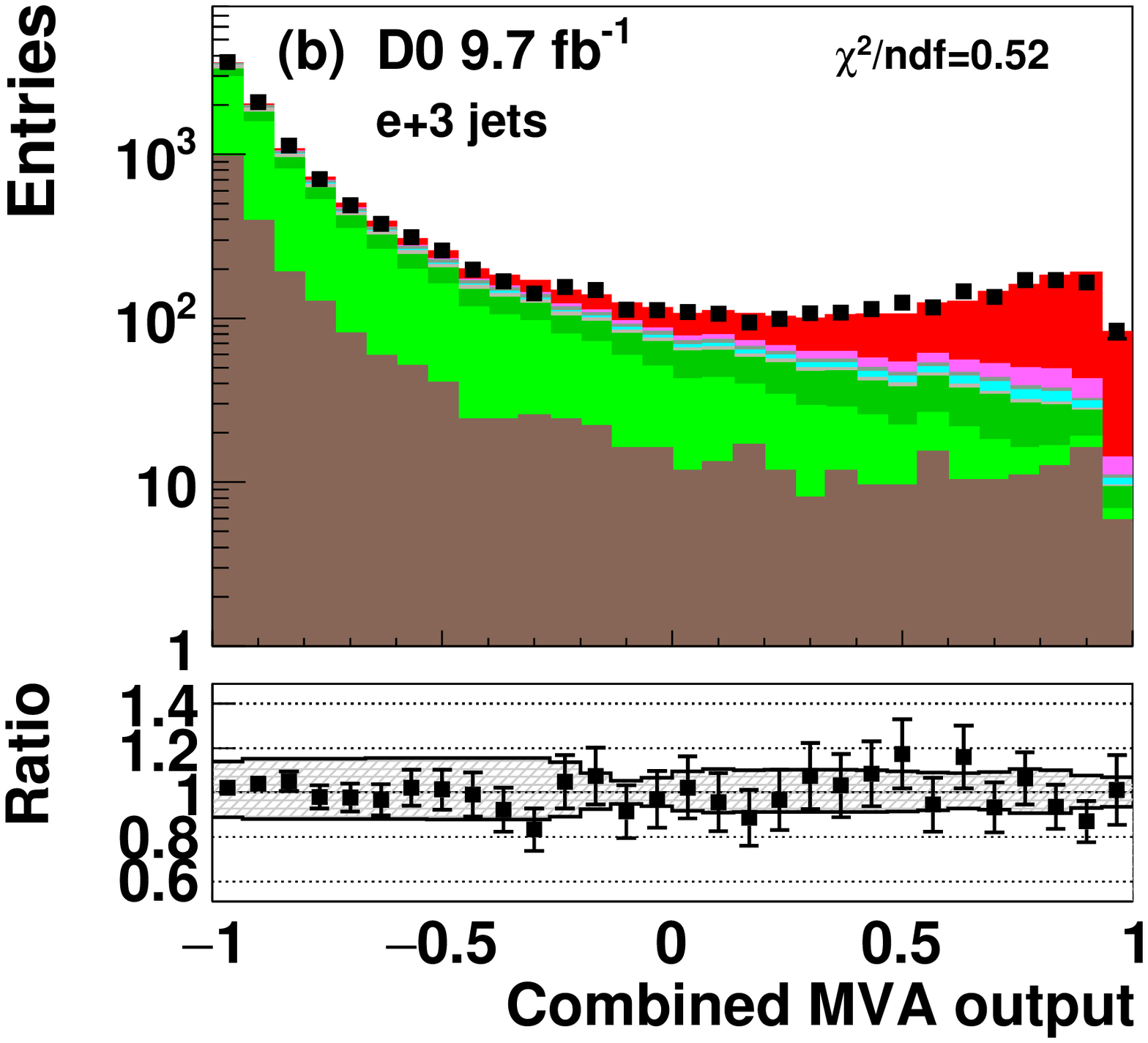}
    \includegraphics[width=0.675\columnwidth,angle=0]{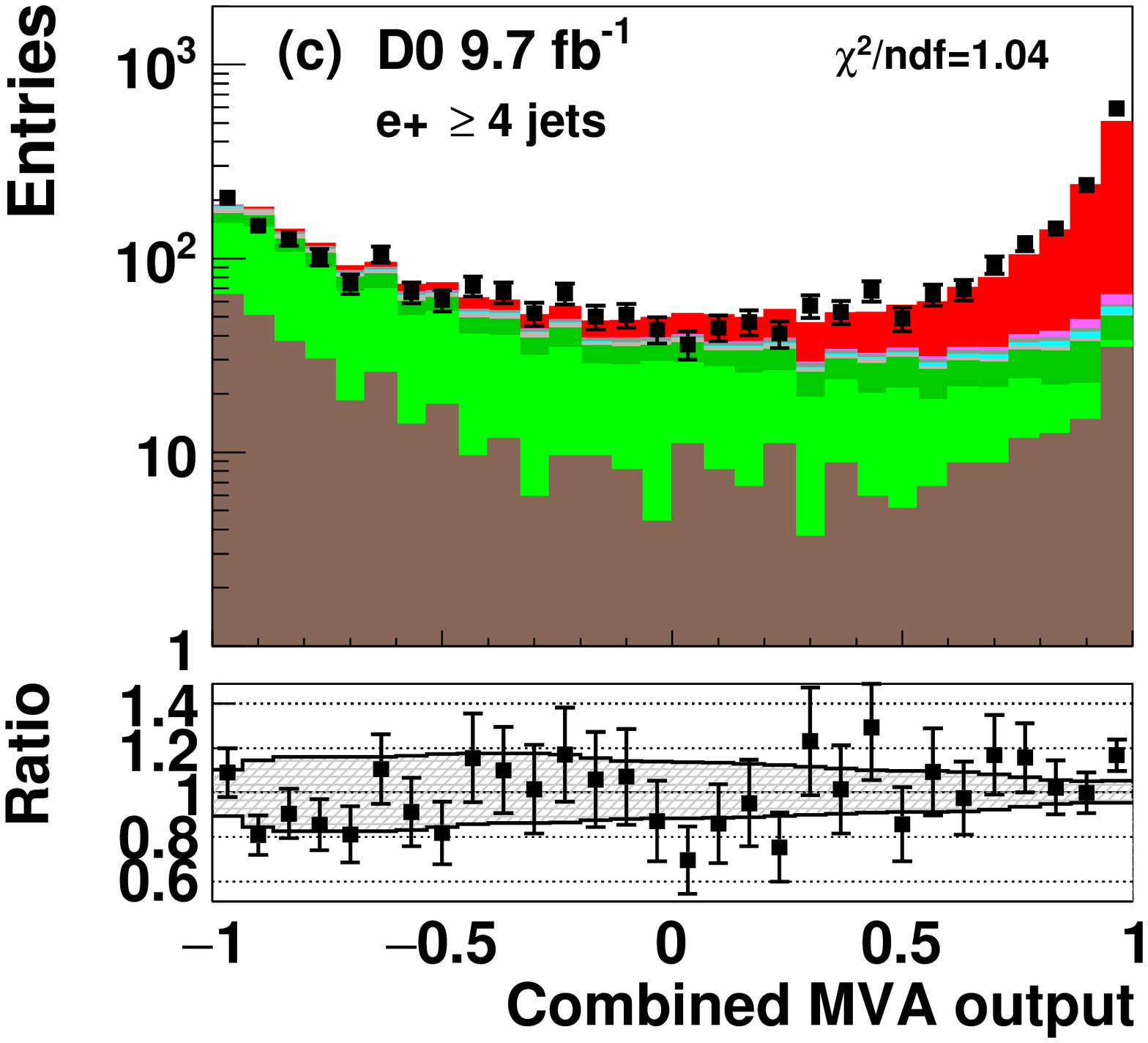}
    \includegraphics[width=0.675\columnwidth,angle=0]{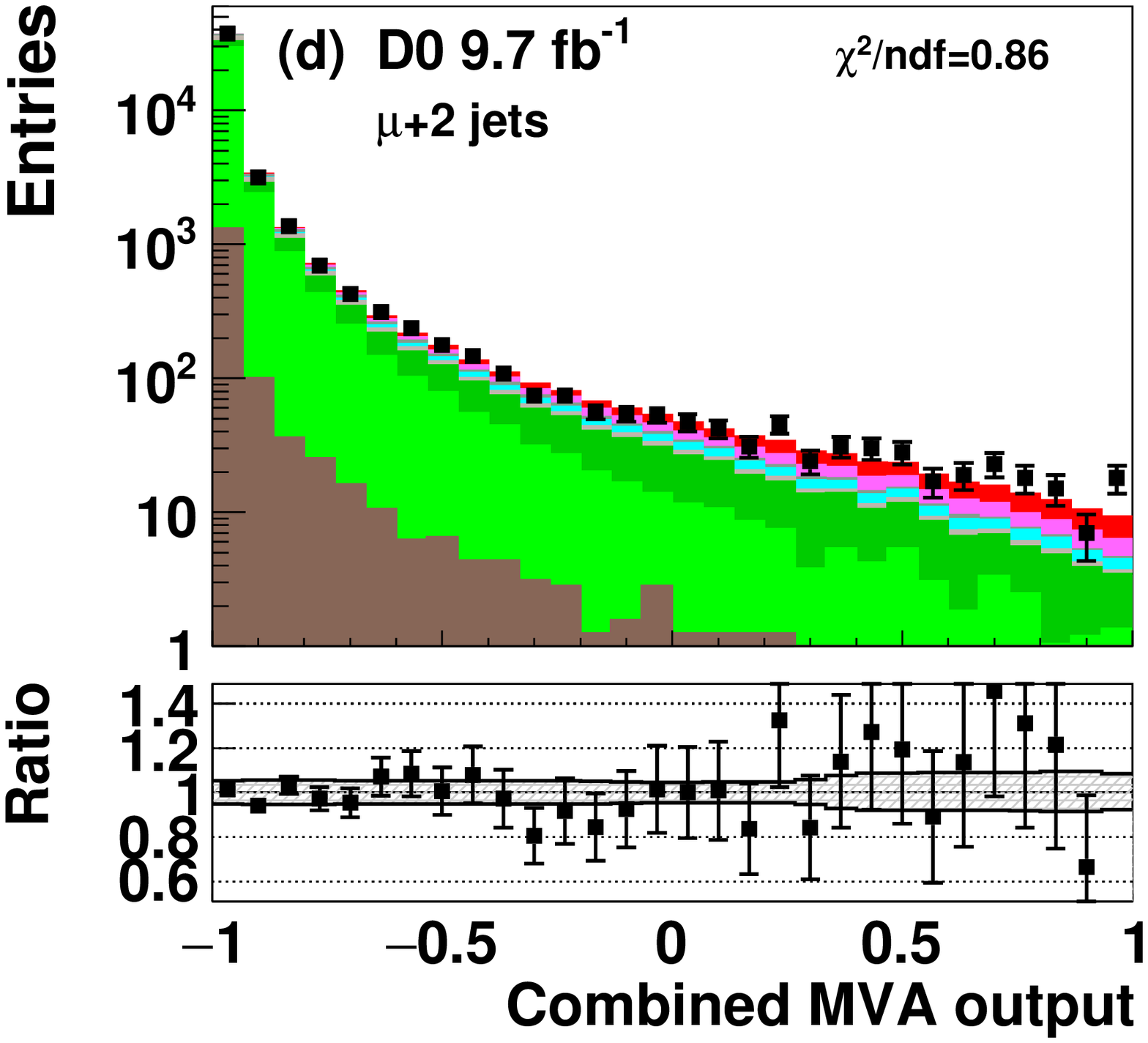}
    \includegraphics[width=0.675\columnwidth,angle=0]{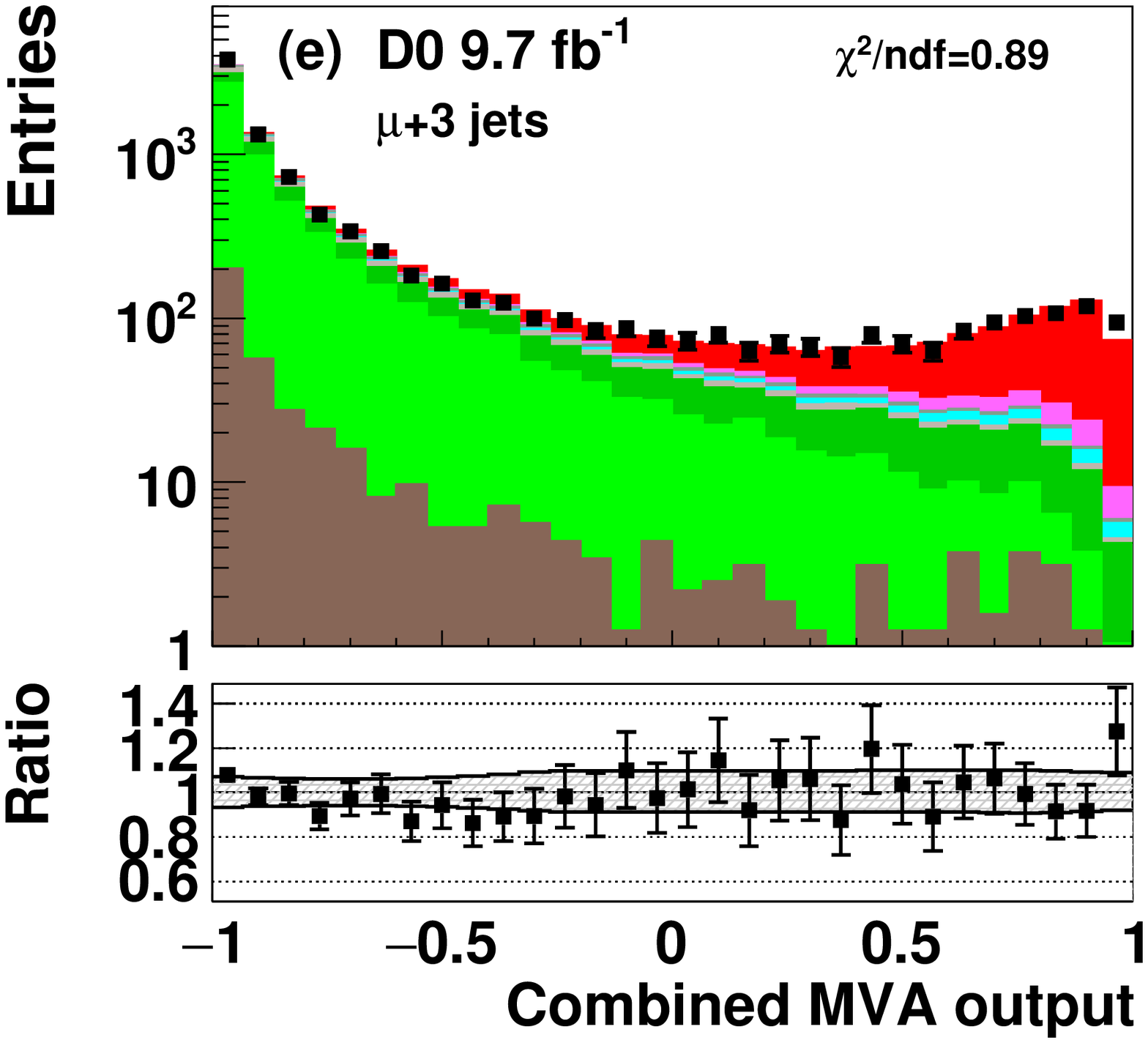}
    \includegraphics[width=0.675\columnwidth,angle=0]{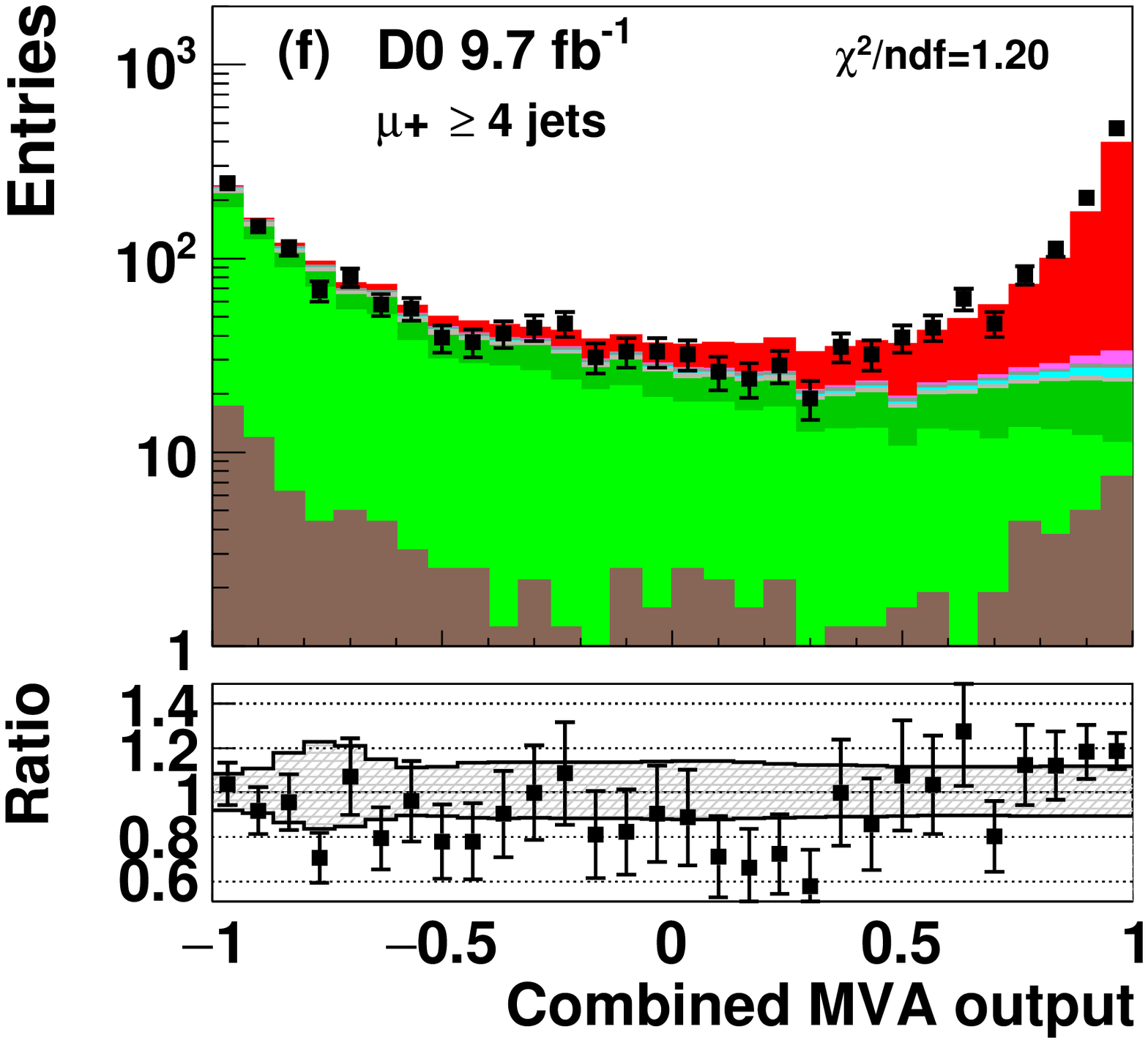}
\end{center}
\caption{Post-fit output distributions of the combined MVA discriminant using the measured combined \ttbar cross section and $m_t=172.5$ GeV for the $e + $ two, three or four or more jets [panels (a) -- (c)], and for the $\mu + $ two, three or four or more jets [panels (d) -- (f)]. Statistical uncertainties of the data are shown and the post-fit systematic uncertainties are indicated by the hashed band in the bottom panel of the histogram. The $\chi^2/\mm{ndf}$ values take statistical and systematic uncertainties into account.}
\label{fig:postfit_topo_emujets}
\end{figure*}


\begin{figure*}[htb]
\begin{center}
     \includegraphics[width=0.675\columnwidth,angle=0]{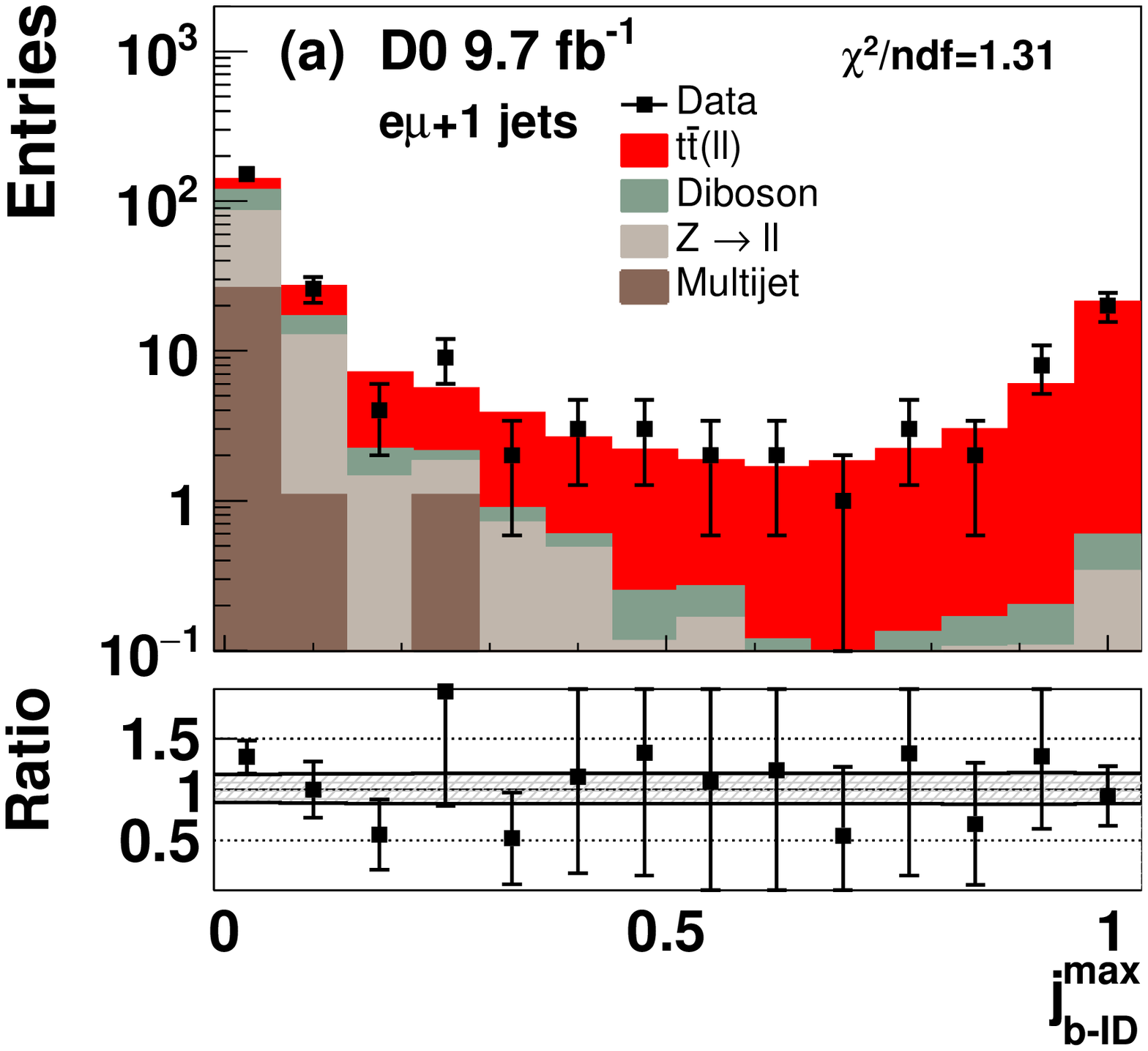}
     \includegraphics[width=0.675\columnwidth,angle=0]{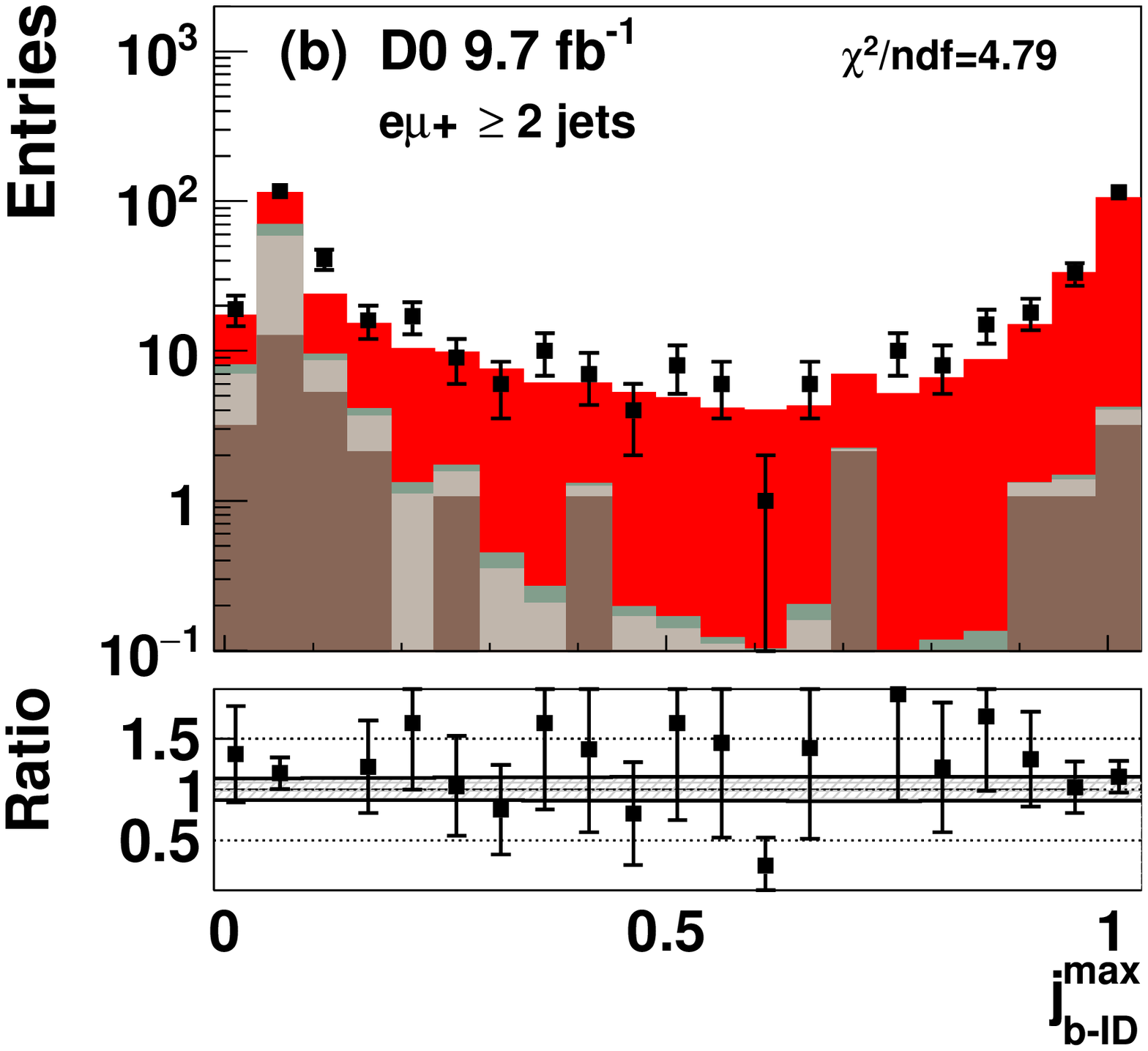}
     \includegraphics[width=0.675\columnwidth,angle=0]{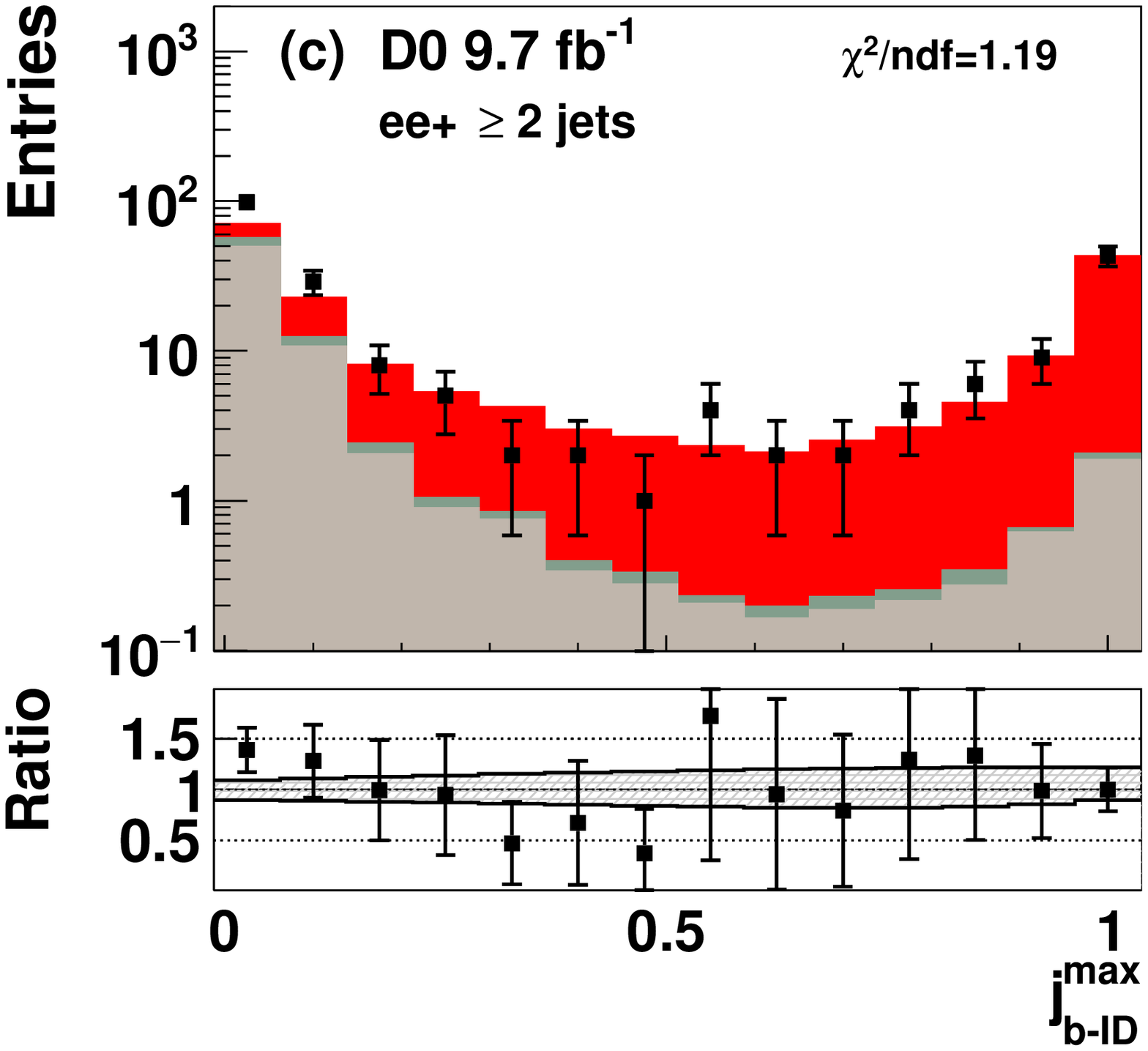}
     \includegraphics[width=0.675\columnwidth,angle=0]{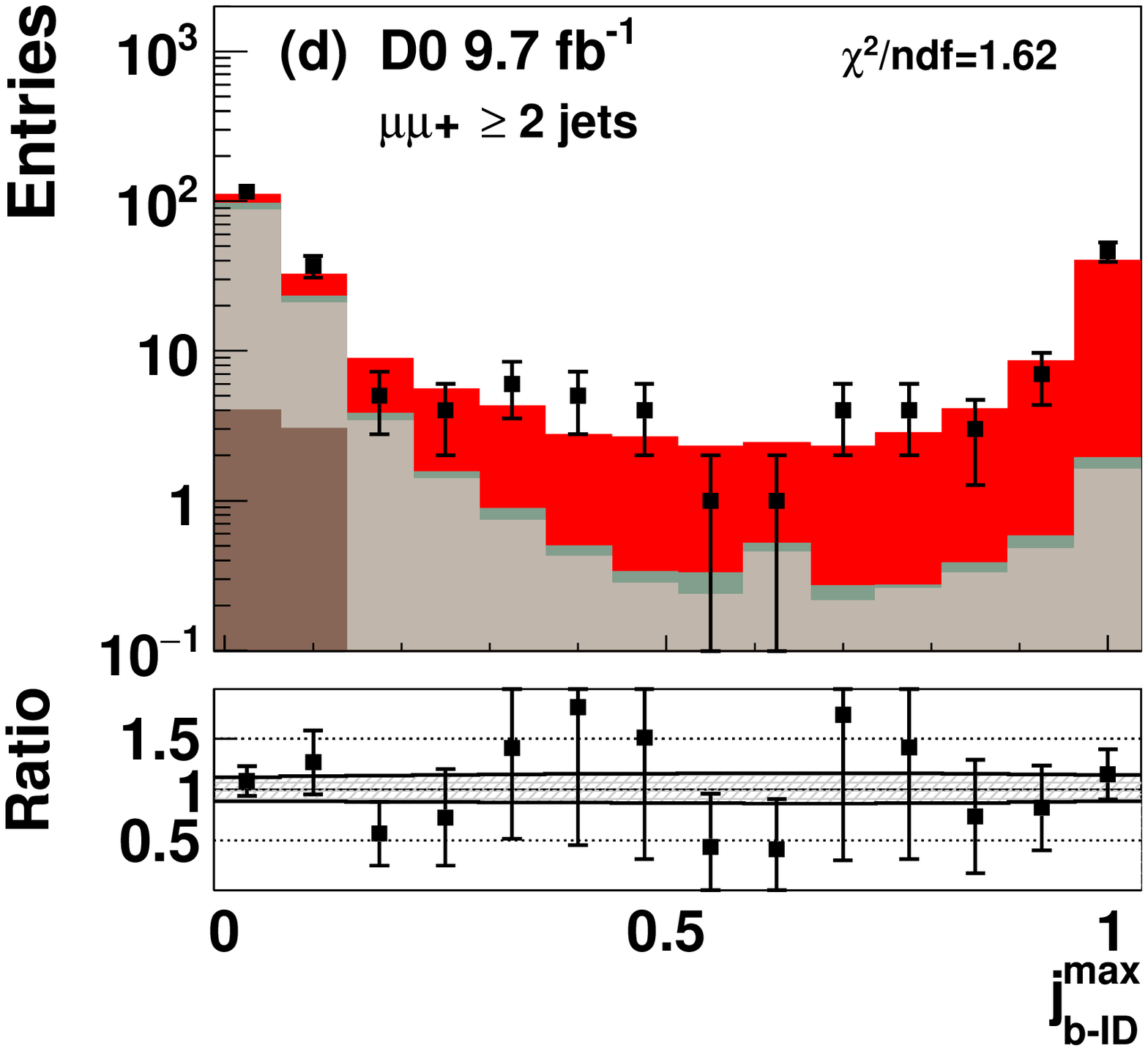}
\end{center}
\caption{The post-fit MVA output distributions (assuming $m_t=172.5$ GeV) of \mmax for (a) $e\mu$ events with exactly one jet, (b) $e\mu$ events with at least two jets, (c) $ee$ events with at least two jets, and (d) $\mu\mu$ events with at least two jets are shown. Statistical uncertainties of the data are shown and the post-fit systematic uncertainties are indicated by the hashed band in the bottom panel of the histogram. The $\chi^2/\mm{ndf}$ values take statistical and systematic uncertainties into account.}
\label{fig:postfit_mva_dilep}
\end{figure*}

\subsection{The $\bm{ \ell \ell}$ channel}
In the following we describe the sources of systematic uncertainties studied in the \dilep channel, which are mostly similar to those in the \ljets channel. As discussed above, each source of systematic uncertainty yields a modified discriminant distribution of the \bidM and a nuisance parameter is used in the fit to determine the \ttbar production cross section in the \dilep channel. Uncertainties on the sample composition only apply to certain sample contributions. Uncertainties due to common sources are assumed to be 100\% correlated between the \ljets channel and the \dilep channel unless otherwise specified.

\subsubsection{Signal modeling}
The same sources of systematic uncertainties for the modeling of the signal as in the \ljets decay channel are considered for the \dilep channel as well. 

\subsubsection{Parton distribution functions}
The uncertainty on the cross sections due to the uncertainty on PDFs is estimated following the same procedure as in the \ljets case. 

\subsubsection{Detector modeling}
The assigned uncertainties related to the modeling of the detector are the same as the ones assigned in the \ljets channel and include uncertainties on the efficiencies of electron and muon identification, uncertainties on trigger efficiencies, the uncertainty in jet energy scale, jet energy resolution, jet identification efficiency, and $b$-quark jet tagging efficiency. The \ljets and \dilep event selection ensures that the samples are dominantly orthogonal. Hence, we assume the uncertainties arising from modeling of the trigger to be not correlated between the \ljets and \dilep channel. 

\subsubsection{Sample composition}
We estimate the uncertainty on the instrumental background contribution in the \dilep channel by changing the normalization of that background by $\pm 1$ standard deviation of its uncertainty. It includes both the statistical uncertainty on the sample used to derive the normalization and the systematic uncertainty in the lepton misidentification rate. Uncertainties from \zplus and diboson production are taken into account with the same assumptions as in the \ljets case. Uncertainties arising from the determination of the MJ background are assumed to be not correlated between the \ljets and \dilep channels.

%
%
%
%
\section{Cross Section Results}
\label{toc:results}
The result of the measurement in the \ljets channel using the combined MVA method is\\
\begin{center}
\xsecLJ pb,
\end{center} with a relative total uncertainty of 9.2\%. For the $\ell\ell$ decay channel we employ the MVA $b$-jet method and measure\\
\begin{center}
\xsecDL pb,
\end{center} with a relative total uncertainty of 9.6\%. 

The combination of the \ttbar cross section is carried out by a simultaneous \collie fit of the combined MVA and the MVA $b$-jet discriminant distributions in the \ljets and \dilep channels. For a top quark mass of 172.5 GeV we measure\\
\begin{center}
\xsecComb pb,
\end{center} which corresponds to a relative total uncertainty of $7.6\%$. For the combined \ljets and \dilep \ttbar cross section measurement we profile the systematic uncertainties by employing the MVA discriminant distributions simultaneously in both channels. The combined \ttbar cross section does not coincide with the weighted average of the individual \ljets and \dilep results, which is attributed to the effect of correlations of systematic uncertainties between both channels.

Table \ref{tab:syst_uncorr} summarizes the post-fit systematic uncertainties on the \ttbar cross section in the \ljets and \dilep decay channels and for the combination. The impact of these sources is estimated by removing or including the corresponding ``group" of individual sources from the fit. The total uncertainty is provided by the nominal fit, when including all individual sources of systematic uncertainties, and denoted as ``central \collie'' in Table \ref{tab:syst_uncorr}. For comparison only we also provide the quadratic sum of the groups of systematic uncertainties. Due to correlations being different between the groups and all the individual systematic uncertainties, that value differs from the total systematic uncertainty of the nominal fit. In addition, we provide the ``Shift" in units of pb, which refers to shifts on the combined inclusive cross section due to a particular source of systematic uncertainty relative to the central value of the combined \ttbar cross section.

Figure \ref{fig:postfit_topo_emujets} shows the post-fit MVA combined discriminant distributions for the \lplus channel using the combined \ttbar cross section. Similarly Fig.~\ref{fig:postfit_mva_dilep} shows the post-fit MVA $b$-ID discriminant distribution for the \dilep channel using the combined \ttbar cross section. This result is consistent with and supersedes our earlier measurement using $5.3~\mm{fb^{-1}}$ of data \cite{Publ54_xsec}. The inclusive \ttbar production cross section is in agreement with the fully resummed next-to-next-to-leading logarithm at NNLO QCD calculation (see Sec.\ \ref{toc:generators}) of $\sigma_{\mathrm{tot}}^{\mathrm{res}} = 7.35 ^{+0.23}_{-0.27}\thinspace(\mathrm{scale} + \mathrm{pdf})$ pb.

%
%
%
%
\section{Top Quark Pole Mass}
\label{toc:tmass}
Table \ref{tab:mt_values} presents the measured combined inclusive \ttbar cross section as a function of the top quark mass. For each top quark mass point shown a separate combined log-likelihood fit of the \ljets and \dilep channel MVA discriminant inputs was performed, as was done for the mass point of 172.5 GeV. The measured \ttbar cross section only changes by 0.7\% for a change of 1 GeV in the assumed top quark mass. Systematic uncertainties of the \ttbar contribution are taken from the 172.5 GeV case and assigned to other masses as a relative systematic uncertainty of the same size. 

\begin{table}[htp]
    \caption{The measured combined inclusive \ttbar cross section as a function of the top quark MC mass with statistical and systematic uncertainties given separately.} \label{tab:mt_values}
\begin{center}
\begin{ruledtabular}
\begin{tabular}{c l}
Top quark mass [GeV] & Cross section \xsttbar [pb]\\ \hline \T
$150$ & $9.70 \pm 0.16\,(\mm{stat.}) ^{+0.73}_{-0.67}\,(\mm{syst.})$  \\ \T 
$160$ & $8.25 \pm 0.14\,(\mm{stat.}) ^{+0.63}_{-0.57}\,(\mm{syst.})$  \\ \T
$165$ & $7.46 \pm 0.13\,(\mm{stat.}) ^{+0.58}_{-0.51}\,(\mm{syst.})$  \\ \T
$170$ & $7.55 \pm 0.13\,(\mm{stat.}) ^{+0.58}_{-0.55}\,(\mm{syst.})$ \\ \T
$172.5$ &$7.26 \pm 0.12\,(\mm{stat.}) ^{+0.57}_{-0.50}\,(\mm{syst.})$  \\ \T
$175$ &$7.28 \pm 0.12\,(\mm{stat.}) ^{+0.54}_{-0.49}\,(\mm{syst.})$ \\ \T
$180$ &$6.86 \pm 0.12\,(\mm{stat.}) ^{+0.53}_{-0.47}\,(\mm{syst.})$ \\ \T
$185$ &$6.50 \pm 0.11\,(\mm{stat.}) ^{+0.50}_{-0.43}\,(\mm{syst.})$ \\ \T
$190$ &$6.70 \pm 0.11\,(\mm{stat.}) ^{+0.60}_{-0.47}\,(\mm{syst.})$ 
 \end{tabular}
\end{ruledtabular}
 \end{center}
\end{table}

\begin{figure}[htb]
\begin{center}
    \includegraphics[width=0.975\columnwidth,angle=0]{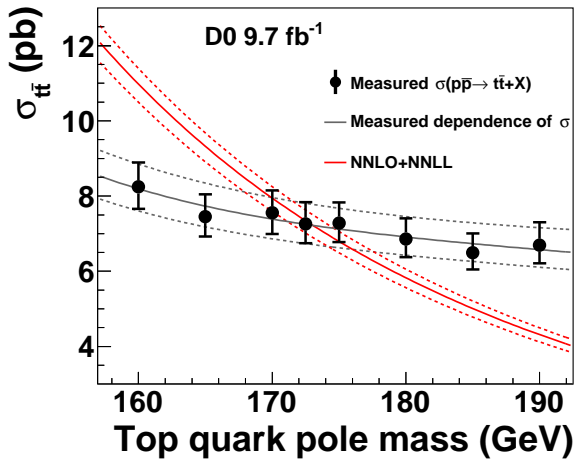}
\end{center}
\caption{The measured \ttbar production cross section dependence on the top quark mass (points) parametrized by a quartic function (solid black line) and compared to the dependence provided by the NNLO pQCD calculation \toppp \cite{toppp_prg}, which implements pQCD calculations according to Ref.~\cite{nnloInclXsec}.}
\label{fig:topMass_vs_topxsec_NNLO}
\end{figure}

Figure \ref{fig:topMass_vs_topxsec_NNLO} shows the measured and theoretical mass dependence of the inclusive \ttbar production cross section as provided by \toppp \cite{toppp_prg}. We parametrize the experimentally measured dependence with a fourth-order polynomial function  to the individual cross section measurements at the mass points reported in Table \ref{tab:mt_values}. There is negligible change if a cubic function is chosen. Uncertainties on the measured values include the statistical and systematic uncertainties discussed in Sec.\ \ref{toc:xsec_sys} and are indicated by the dotted lines. Theoretical uncertainties include those from the variation of the renormalization and factorization scales by a factor of 2 and PDF uncertainties \cite{mstw2008nnlo} taken from the MSTW2008 NNLO PDF set. These are added in quadrature and indicated by the dotted lines surrounding the central theoretical prediction.
 
To determine the top quark pole mass from the inclusive \ttbar cross section  following the method in Ref.~\cite{d0_polemass_5fb}, we extract the most probable $m_t$ value and uncertainty by employing a normalized joint-likelihood function, which takes into account the total experimental uncertainty, the theoretical uncertainties on the renormalization and factorization scales, and the PDF uncertainties. Employing the quartic parametrization and the theory predictions at NNLO pQCD we obtain
\begin{eqnarray*}
m_t &=& 172.8 \pm 1.1\,(\mm{theo.})\,^{+3.3}_{-3.1}\,(\mm{exp.})\,\mm{GeV}, \mm{or}\\
m_t &=& 172.8 ^{+3.4}_{-3.2}\,(\mm{tot.})\,\mm{GeV}
\end{eqnarray*}
The experimental uncertainties dominate the precision of the determination. The precision of this determination is 1.9\%, and represents the most precise determination of the top quark pole mass from the inclusive \ttbar cross section at the Tevatron. This supersedes our previous determination which had a precision of 3\% \cite{d0_polemass_5fb}.

%
%
%
%
\section{Conclusions}
\label{toc:conclusion}
The inclusive \ttbar production cross section has been measured combining the lepton+jets and dilepton top quark decay channels based on the full Tevatron data set at $\sqrt{s} = 1.96$ TeV. We performed a simultaneous log-likelihood fit to profile systematic uncertainties and, for a top quark mass of 172.5 GeV, we measured a combined \ttbar cross section of
\begin{center}
\xsecComb pb,
\end{center} which corresponds to a relative uncertainty of $7.6\%$. This result and the measured inclusive cross sections per decay channel are in good agreement with predictions by QCD. 

We employed the dependence of the theoretical cross section on the top mass, to determine a pole mass of the top quark of 
\begin{center}
 $m_t = 172.8 ^{+3.4}_{-3.2}\,(\mm{tot.})\,\mm{GeV}$.
 \end{center} 
The uncertainty corresponds to a precision of 1.9\% and represents the most precise determination of the top quark pole mass at the Tevatron.

\section{Acknowledgments}

We thank Vladimir Shiltsev for enlightening discussions. We thank the staffs at Fermilab and collaborating institutions,
and acknowledge support from the
Department of Energy and National Science Foundation (United States of America);
Alternative Energies and Atomic Energy Commission and
National Center for Scientific Research/National Institute of Nuclear and Particle Physics  (France);
Ministry of Education and Science of the Russian Federation, 
National Research Center ``Kurchatov Institute" of the Russian Federation, and 
Russian Foundation for Basic Research  (Russia);
National Council for the Development of Science and Technology and
Carlos Chagas Filho Foundation for the Support of Research in the State of Rio de Janeiro (Brazil);
Department of Atomic Energy and Department of Science and Technology (India);
Administrative Department of Science, Technology and Innovation (Colombia);
National Council of Science and Technology (Mexico);
National Research Foundation of Korea (Korea);
Foundation for Fundamental Research on Matter (The Netherlands);
Science and Technology Facilities Council and The Royal Society (United Kingdom);
Ministry of Education, Youth and Sports (Czech Republic);
Bundesministerium f\"{u}r Bildung und Forschung (Federal Ministry of Education and Research) and 
Deutsche Forschungsgemeinschaft (German Research Foundation) (Germany);
Science Foundation Ireland (Ireland);
Swedish Research Council (Sweden);
China Academy of Sciences and National Natural Science Foundation of China (China);
and
Ministry of Education and Science of Ukraine (Ukraine).
%

%
%
%
%

%
%
%
%
\appendix

\section*{Appendix: Variables selected for the combined MVA discriminant}
\label{toc:appendix_vars}
Depending on the number of jets in an event, we select at least 24 variables for the \topoM to measure the \ttbar production cross section in the \ljets decay channel. The list given below is sorted according to the ranking in terms of separation power as provided by the BDTG method. 

\begin{itemize}
\item \mmax: The maximum output value of the MVA $b$-jet discriminant of all jets in an event.

\item $H_T^3$: The scalar sum of transverse momenta of jets excluding the leading and subleading jets.

\item $H_T^{2.0}$: The scalar sum of transverse momenta of jets satisfying $|\eta| < 2.0$.

\item $j_{b}^{1}$: The $b-\mm{ID}\,\,\mm{MVA}$ output value of the leading jet.

\item Centrality $\cal{C}$: Ratio of the scalar sum of the transverse momentum of all jets to the energy of all jets.

\item $H_T$: The scalar sum of the transverse momenta of all jets, the lepton and \met.

\item $p_T^{\mm{j}_i}$: The transverse momenta of the individual \mbox{jets $i$}.

\item $j_{b}^{\mm{2}}$: The $b-\mm{ID}\,\,\mm{MVA}$ value of the second leading jet.

\item $H_T^{\ell}$: The scalar sum of the transverse momenta of all jets and the lepton.

\item Sphericity $\cal{S}$: Diagonalizing the normalized quadratic momentum tensor $\cal{M}$ yields three eigenvalues $\lambda_i$ \cite{Publ54_xsec}, with $\lambda_1 > \lambda_2 > \lambda_3$. The sphericity is defined as ${\cal S} = \frac{3}{2}(\lambda_2 + \lambda_3)$ and reflects the degree of isotropy of an event.

\item \mTT: The invariant mass of the \ttbar pair. The energy of the neutrino is determined by constraining the invariant mass of the lepton and vector \met (as the neutrino) to the mass of the $W$ boson. Of the two possible solutions for the longitudinal momentum of the neutrino, we use the one with the smallest absolute value.

\item $\eta^{\mm{j}_1}$: The rapidity of the leading jet.

\item $\Delta R(j^1, j^2)$: The separation in the distance $R$ between the leading and second leading jet.

\item $p_T^W$: The transverse momentum of the reconstructed $W$ boson which decays hadronically.

\item $M_T^{\mm{j}_2\nu\ell}$: The transverse mass of the system consisting of the second leading jet, the neutrino, and the lepton.

\item Aplanarity ${\cal A}$: The aplanarity is defined as $3/2$ times the momentum tensor eigenvalue $\lambda_3$.

\item $\Delta R(j^1, j^{i\ge3})$: The separation in the distance $R$ between the leading and each jet beyond the second leading jet.

\item $m_{\mm{jet}}$: The invariant mass of the jets.

\item $M_T^{\mm{jet}}$: The transverse mass of the first two leading jets.

\item $M_{\mm{j}_2\nu\ell}$: The invariant mass of the system consisting of the second leading jet, the neutrino, and the lepton.

\item $\Delta \phi(\ell, \met)$: The separation in azimuth between the lepton and the direction of \met.

\item \met: Missing transverse momentum.

\item $\eta^{\mm{lepton}}$: The rapidity of the lepton.

\item $\Delta \phi(j^1, j^i)$: The minimum separation in azimuth between the leading and any other jet.

\item \metNo$^{\mm{perp}}$: Component of the missing transverse momentum perpendicular to the direction of the lepton.
\end{itemize}

\end{document}